\documentclass[10pt]{article}
\usepackage[utf8]{inputenc}
\pdfoutput=1
\usepackage{natbib}
\usepackage[english]{babel}
\usepackage{graphicx,xspace,float,afterpage}
\usepackage[pdftex]{color}
\usepackage[labelsep=period,labelfont=bf]{caption}
\usepackage{geometry}
\usepackage{latexsym,amsmath,amsfonts,amssymb,textcomp,calc,units,booktabs,url}
\usepackage{graphicx}
\usepackage{rotating}
\usepackage{placeins}
\usepackage{afterpage}
\newcommand{\dpow}[2]{\ensuremath{#1\cdot 10^{#2}}}  
\newcommand{\chem}[3][]{\ensuremath{\mathrm{{#2}}_{#3}^{#1}}}

\begin{document}
\thispagestyle{empty}
\begin{center}
\Large Thomas Ruedas\textsuperscript{1,2}\\[2ex]
Doris Breuer\textsuperscript{2}\\[5ex]
\textbf{``Isocrater'' impacts: Conditions and mantle dynamical responses for different impactor types}\\[5ex]
final version\\[5ex]
7 February 2018\\[10ex]
published in:\\
\textit{Icarus} 306, pp.~94--115 (2018)\\[15ex]
\normalsize
\textsuperscript{1}Institute of Planetology, Westf\"alische Wilhelms-Universit\"at, M\"unster, Germany\\[5ex]
\textsuperscript{2}Institute of Planetary Research, German Aerospace Center (DLR), Berlin, Germany
\rule{0pt}{12pt}
\end{center}
\vfill
\footnotesize The version of record is available at \url{http://dx.doi.org/10.1016/j.icarus.2018.02.005}.\\
This author post-print version is shared under the Creative Commons Attribution Non-Commercial No Derivatives License (CC BY-NC-ND 4.0).
\normalsize
\newpage

\title{``Isocrater'' impacts: Conditions and mantle dynamical responses for different impactor types}
\author{Thomas Ruedas\thanks{Corresponding author: T. Ruedas, Institute of Planetology, Westf\"alische Wilhelms-Universit\"at, M\"unster, Germany (t.ruedas@uni-muenster.de)}\\{\footnotesize Institute of Planetology, Westf\"alische Wilhelms-Universit\"at, M\"unster, Germany}\\{\footnotesize Institute of Planetary Research, German Aerospace Center (DLR), Berlin, Germany}\\[2ex]
Doris Breuer\\{\footnotesize Institute of Planetary Research, German Aerospace Center (DLR), Berlin, Germany}}
\date{}
\maketitle
\textbf{Highlights}
\begin{itemize}
\item
Impactors of different type/size/velocity can produce craters of the same diameter.
\item
Conditions for such ``isocraters'' are derived from scaling laws, modeled numerically.
\item
Response of interior to isocrater impacts varies strongly between impactor types.
\item
Responses to similar impactors vary strongly with planetary structure.
\item
Observed geophysical anomalies may allow to resolve non-uniqueness of impactor.
\end{itemize}
\begin{abstract}
Impactors of different types and sizes can produce a final crater of the same diameter on a planet under certain conditions. We derive the condition for such ``isocrater impacts'' from scaling laws, as well as relations that describe how the different impactors affect the interior of the target planet; these relations are also valid for impacts that are too small to affect the mantle. The analysis reveals that in a given isocrater impact, asteroidal impactors produce anomalies in the interior of smaller spatial extent than cometary or similar impactors. The differences in the interior could be useful for characterizing the projectile that formed a given crater on the basis of geophysical observations and potentially offer a possibility to help constrain the demographics of the ancient impactor population. A series of numerical models of basin-forming impacts on Mercury, Venus, the Moon, and Mars illustrates the dynamical effects of the different impactor types on different planets. It shows that the signature of large impacts may be preserved to the present in Mars, the Moon, and Mercury, where convection is less vigorous and much of the anomaly merges with the growing lid. On the other hand, their signature will long have been destroyed in Venus, whose vigorous convection and recurring lithospheric instabilities obliterate larger coherent anomalies.
\end{abstract}
\begin{flushleft}
Impact processes; Terrestrial planets; Thermal histories; Interiors
\end{flushleft}

\section{Introduction}
The cratered surfaces of planetary bodies in the solar system offer abundant evidence that meteorite impacts have been an important geological factor in their evolution, especially in the first few hundreds of millions of years. However, the dynamics of the impact process itself as well as later degradation make it difficult to reconstruct the physical properties of the impactor unambiguously \citep[e.g.,][]{HeHy17}, because the impactor is usually obliterated during the impact, even though numerical models of central peak formation in complex craters suggest that a certain fraction of the impactor material may be preserved to some extent, especially in slow or oblique impacts \citep[e.g.,][]{ZYue:etal13,SvSh15}. The principal features of the remaining crater, i.e., its geometrical characteristics, depend on a combination of several parameters that characterize the impactor and the target. While it is in principle possible to determine the target properties reasonably well, the combination of properties of the projectile (diameter, velocity, density) can generally only be inferred on the basis of morphological studies of the crater and the ejecta \citep[e.g.,][]{OKeAh82,PHScCr16} or statistical information on candidate impactor types, especially if the crater or basin is very old and its ejecta are obliterated.\par
There are several candidate impactor classes, which belong to two general groups, namely asteroids and comets; in principle, larger trans-neptunian objects (TNO) which for some reason acquired very eccentric orbits that brought them into realms of the solar system closer to the Sun are a third group, although no such objects are known to exist presently. The asteroids are divided into various classes, of which especially the rocky S-types and the less dense C-types are important due to the relatively large mass and number fractions of the total asteroid population they constitute \citep{DeMeCa13}. These two major classes are expected to travel at similar velocities, but differ markedly in their average densities \citep{Carry12}. There is even more uncertainty about the density of comets, but it seems to be clear that they are, on average, less dense by at least a factor of two than even C-type asteroids. In the inner Solar System, their velocities cover a wide range at any given distance from the Sun, as they originate from different reservoirs at very different distances from the center.\par
Direct statistical information about their relative abundance is essentially limited to results from observations of present-day populations. While total meteorite fluxes can be deduced to some extent from cratering statistics, such statistics provide no direct information about the proportions of the different impactor types and their possible temporal evolution. In an attempt to establish a more stringent link with the past, \citet{BAIvan:etal02} compared the observed size--frequency distribution of the modern main-belt asteroid population with that of the impactors on the Moon as derived from the lunar cratering record and impact scaling laws and found that both have a similar shape. These authors hence argue that at least for the past $\sim 4$\,Gy, most impacts in the inner Solar System were caused by main-belt asteroids, or more generally, collisionally evolved bodies. Similar conclusions were also reached by other authors on the basis of cratering statistics and dynamical simulations \citep[e.g.,][]{Stro:etal05,Rick:etal17}. Still, many assumptions made about these issues rely heavily on knowledge of the current state and on models of long-term Solar System evolution including phenomena such as orbital shifts of planets \citep[e.g.,][]{Gome:etal05}, all of which are still poorly known. Furthermore, the earliest stage of impact history is not constrained by a similar argument, and the case has been made that different impactor populations have existed at different times during the history of the Solar System and are recorded on Mars and on the farside of the Moon, respectively \citep{Bott:etal17}.\par
A frequently made assumption in models of individual impacts on planets and/or their effects on their interiors is that the impactor is a body of the most frequent class, usually assumed to be an S-type asteroid traveling at the average impact velocity corresponding to the target planet and striking at an angle of 45\textdegree. Given the need to limit the scope of such studies, this focus on the most frequent and hence most likely category is perfectly legitimate, but it may lead to impacts of other bodies being forgotten. However, all of those other categories are not so rare that they can safely be considered insignificant: the contribution of comets has been estimated to lie between a few per cent and a few tens of per cents, increasing with proximity of the target planet to the Sun \citep[e.g.,][]{Chyba87,Olsson-Steel87}. It seems therefore rather unlikely that even all of the dozens of large impact basins expected to have existed in the inner Solar System were formed only by S-type asteroids; for instance, it has been suggested that the South Polar--Aitken basin on the Moon or the crater Eminescu on Mercury were formed by cometary impacts \citep{Shev:etal07,PHSchultz17}. Geochemical arguments, especially isotopic studies of water and nitrogen for the Earth, the Moon, and Mars, point to a strong dominance of asteroidal impactors but also confirm the existence of a contribution of cometary impactors on the order of a few percent \citep[e.g.,][]{JJBarn:etal16}; these authors also invoke carbonaceous chondrites as a major asteroidal source for the water, which would indicate that many impactors were C-type asteroids \citep{Carry12}. Near-surface geological aspects of this ambiguity have been addressed in the literature in some cases \citep[e.g.,][for the Shackleton crater on the Moon]{Puga:etal16}, but the implications for the deep interior evolution have not received much attention so far.\par
In this paper, we consider different combinations of impactor characteristics that would all result in a crater of the same size (diameter) according to the scaling laws established by the theory of impact dynamics; we will call such events ``isocrater impacts''. Related scaling laws indicate that such isocrater impacts may still differ in their effects on the interior of the planet. By combining two-dimensional dynamical models of mantle convection with a parameterization of the principal effects of impacts based on scaling laws, we address the question whether the differences in interior dynamics effects of large isocrater impacts can be large enough to have significant consequences for the long-term evolution of the planet. The general theoretical considerations in the next section also allow us to exclude certain impactor types as causes of large impact basins for a given target body and thus also as important exogenic influences on interior dynamics. Furthermore, the models can also provide hints whether a population of now-extinct impactors might or must be considered to explain observed consequences of certain impacts.

\section{Theory}\label{sect:theory}
The final crater is the outcome of the collapse of the transient crater formed during the impact, and their diameters $D_\mathrm{f}$ and $D_\mathrm{tr}$ are related by the empirical relations
\begin{equation}
D_\mathrm{f}=
\begin{cases}
1.18D_\mathrm{tr}&\text{simple craters}\\
1.17D_\mathrm{tr}^{1.13}/D_\mathrm{s2c}^{0.13}&\text{complex craters}\\
\end{cases}\label{eq:Df-tr}
\end{equation}
\citep{JERichardson09,Melosh11}, where $D_\mathrm{s2c}$ is the diameter of transition from simple to complex crater shape. Following the practice of previous studies, we will apply the scaling laws for complex craters to the impact basins that result from giant impacts, because no generally applicable relations are available for these extreme cases at this time. However, although numerical studies give some indication that several scaling laws also hold for large basins \citep{Pott:etal15}, they also suggest that in particular the relations between the dimensions of the transient crater and the final structure are not as straightforwardly related as for complex craters and that other measures than the final crater diameter may be more useful \citep[e.g.,][]{Milj:etal16}. At any rate, although our focus in this study lies on basin-forming impacts because of their direct and substantial effect on mantle dynamics, we emphasize that the following considerations are at least as valid for smaller impacts.\par
The relation between the diameter of the transient crater and the characteristics of the impactor is derived by dimensional analysis:
\begin{equation}
D_\mathrm{tr}=1.16\left(\frac{\varrho_\mathrm{imp}}{\varrho}\right)^\frac{1}{3} D_\mathrm{imp}^{0.78}\frac{v_{\mathrm{imp}}^{0.44}}{g^{0.22}},\label{eq:Dtr-imp}
\end{equation}
where $D_\mathrm{imp}$ is the diameter of the impactor, $\varrho$ and $\varrho_\mathrm{imp}$ the densities of the target and the impactor, $v_{\mathrm{imp}}$ is the velocity of the impactor, and $g$ is gravity \citep[e.g.,][]{WeIv15}; following common practice \citep[e.g.,][]{CRChMcKi86}, we replace the velocity with its vertical component, thus introducing an implicit dependence on the angle $\theta$ with the horizontal in this and most of the following equations. The numerical values of the coefficient and exponents vary with certain target properties, especially porosity, and are chosen here to correspond to a frictionless, pore-free material because of our main focus on very large impacts; for porous targets with friction, the constants would be slightly different (see Supplement). Inserting Eq.~\ref{eq:Dtr-imp} into Eq.~\ref{eq:Df-tr} yields
\begin{subequations}\label{eq:Df-imp}
\begin{equation}
D_\mathrm{f}=1.3688\left(\frac{\varrho_\mathrm{imp}}{\varrho}\right)^\frac{1}{3} D_\mathrm{imp}^{0.78}\frac{v_{\mathrm{imp}}^{0.44}}{g^{0.22}}
\end{equation}
for simple craters and
\begin{equation}
D_\mathrm{f}=1.3836\left(\frac{\varrho_\mathrm{imp}}{\varrho}\right)^{0.377} \frac{D_\mathrm{imp}^{0.8814}}{D_\mathrm{s2c}^{0.13}}\frac{v_{\mathrm{imp}}^{0.4972}}{g^{0.2486}}
\end{equation}
\end{subequations}
for complex craters. For a given planet, $g$ and $\varrho$ are given and reasonably well known, and $D_\mathrm{s2c}$, which is inversely proportional to $g$, is also a constant characteristic for that planet deduced from crater geometry analysis. The remaining variables in eq.~\ref{eq:Dtr-imp}, $D_\mathrm{imp}$, $\varrho_\mathrm{imp}$, and $v_\mathrm{imp}$, however, are not known and may vary widely between different impactor types. Eqs.~\ref{eq:Df-imp} thus illustrate the dilemma of non-uniqueness of the impactor: we have only one observable, $D_\mathrm{f}$, but three unknown factors. In order to resolve the problem, two further independent conditions for these unknowns would be necessary. Unfortunately, various other geometrical or structural characteristics of a crater or basin that have been measured systematically are expressed in terms of $D_\mathrm{f}$, for instance the expression for the depth $h_\mathrm{f}$ of the final crater known as Schröter's rule, $h_\mathrm{f}=a_h D_\mathrm{f}^{b_h}$, where $a_h$ and $b_h$ are constants. However, characteristics of this form do not provide independent constraints. Furthermore, even if one could estimate the total mass of ejecta, this would not help distinguish different impactor types either, as can be shown using a relation between crater and impactor size and ejecta mass derived from dimensional analysis and experiments by \citet[eq.~18]{HoHo11}. They represent the total mass of ejecta launched from within a distance $x$ of the center of the impact as
\begin{equation}
M_\mathrm{ej}(x)=k_\mathrm{cr}\varrho\left[x^3-\left(\frac{\varkappa}{2} D_\mathrm{imp}\right)^3\right],
\end{equation}
where $k_\mathrm{cr}$ and $\varkappa\approx 1.2$ are constants; for the total mass of ejecta, $x=D_\mathrm{tr}/2$. At least for the target bodies considered here, however, $D_\mathrm{tr}\gg D_\mathrm{imp}$ so that one may neglect the term $\varkappa D_\mathrm{imp}/2$. By inserting the preceding relations then follows that for isocrater impacts, the resulting ejecta masses will be approximately equal as well, irrespective of the impactor type.\par
\begin{figure}
\includegraphics[viewport=10 0 420 210,clip,width=\textwidth]{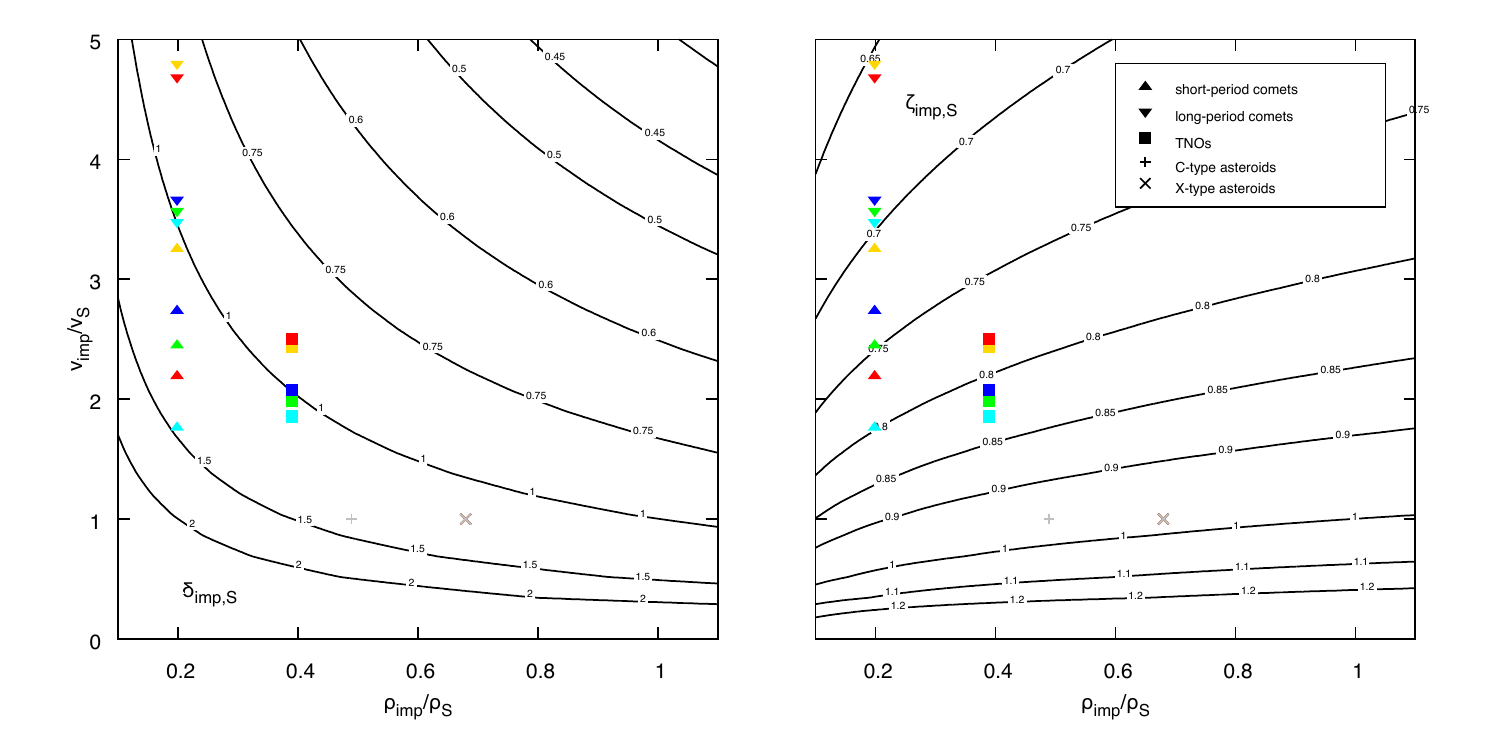}
\caption{Ratios of impactor diameters (left) and of depths to the center of the isobaric core (right) for isocrater impacts as functions of the ratios of densities and impactor velocities according to eqs.~\ref{eq:Dimprat} and \ref{eq:zicrat}, respectively. The reference bodies for calculating the ratios are S-type asteroids traveling at the velocities pertaining to the target planet (cyan: Mercury, green: Venus, blue: Earth, yellow: Moon, red: Mars, grey: all bodies). The data points in these and the following contour plots pertain to average values of the sizes of each impactor type and the impact velocities applicable to each target--impactor pair, both of which are generally subject to large variation and fairly poorly constrained.\label{fig:Dimpzic}}
\end{figure}

With no possibility to resolve the ambiguities on the basis of crater structure metrics, we now consider the condition for two impactors 1 and 2 to produce a final crater of the same diameter on the same target, which is given by $D_\mathrm{f1}=D_\mathrm{f2}$:
\begin{equation}
\frac{D_\mathrm{imp1}}{D_\mathrm{imp2}}=\left(\frac{\varrho_{\mathrm{imp}1}}{\varrho_{\mathrm{imp}2}}\right)^{-0.43}\left(\frac{v_{\mathrm{imp}1}}{v_{\mathrm{imp}2}}\right)^{-0.56}=\delta_{12},\label{eq:Dimprat}
\end{equation}
for both simple and complex craters; this defines a set of isocrater impacts. We have tacitly assumed that the target density $\varrho$ is the same for different impactors, although one may well argue that impactors of different sizes, which penetrate to different depths and affect different volumes as discussed below, ``see'' targets with different effective densities. However, the differences would be difficult to assess in detail and depend on the thermal evolution of the planet, and as they are expected to be minor, it seems legitimate to neglect them here. Figure~\ref{fig:Dimpzic} (left panel) shows the ratio of impactor diameters, $\delta_{12}$, for isocrater impacts according to Eq.~\ref{eq:Dimprat}, whereby impactor~2 is chosen as a common reference impactor, namely an S-type asteroid. The isolines thus show how strongly the size of impactor~1 must differ from that reference in order to produce a crater of the same size, for any combination of density and velocity, which are also normalized to the density and velocity, respectively, of the reference. An isocrater condition analogous to the one for the final diameter can also be formulated for the diameters of regions with thinned or thickened crust, $D_\mathrm{thin}$ and $D_\mathrm{thick}$, proposed by \citet{Milj:etal16} as characteristics for large impact basins. With their assumption that $\varrho_\mathrm{imp}=\varrho$, which we will refer to as a ``homogeneous impact'' in the following, this leads to a relation identical to Eq.~\ref{eq:Dimprat} (with a slightly different exponent for the velocity ratio due to their choice of exponent) for both diameters. As $D_\mathrm{thin}$ and $D_\mathrm{thick}$ are not independent, however, the impactor cannot be uniquely characterized even in a homogeneous basin-forming impact.\par
For the dynamics of the interior, it is the subsurface features of an impact rather than the crater that are of primary interest, but their geometry and properties are more difficult to study and less well described in terms of an analytical model. One measure of particular interest is the depth of the center of the isobaric core, $z_\mathrm{ic}$. All analytical models agree on a linear (or nearly linear) dependence on the impactor diameter but give different additional dependences on other variables, most significantly on the impactor velocity and on the ratio of impactor and target density. For instance, a numerical study by \citet{Pier:etal97} determined a power-law dependence on $v_{\mathrm{imp}}$ but did not account for the density ratio $\varrho_\mathrm{imp}/\varrho$ between impactor and target as both were assumed to consist of the same material in their models. In other studies that included this ratio, it usually enters the relations as a square or cube root. In view of this inconclusive situation, we decided to combine the square-root relation for the density with the \citet{Pier:etal97} formula, i.e.,
\begin{equation}
z_\mathrm{ic}=a_zD_\mathrm{imp} v_{\mathrm{imp}}^{b_z} \left(\frac{\varrho_\mathrm{imp}}{\varrho}\right)^{0.5}\label{eq:zic}
\end{equation}
with $a_z=0.1524$ and $b_z=0.361$. At least, this does not conflict with the original equation from \citet{Pier:etal97} for homogeneous impacts, but future numerical simulations of heterogeneous impacts should test whether empirical fits yield an exponent of the density ratio that is significantly different from 0.5. A brief summary of different expressions for $z_\mathrm{ic}$ from the literature is given in the Supplement.\nocite{Opik36a,Bruce62,Whipple63,DiWa70,Dehn86,OKeAh93}\par
The ratio of the depths of the isobaric cores for two isocrater impacts then follows by combination with eq.~\ref{eq:Dimprat}:
\begin{equation}
\frac{z_\mathrm{ic1}}{z_\mathrm{ic2}}=\left(\frac{\varrho_{\mathrm{imp}1}}{\varrho_{\mathrm{imp}2}}\right)^{0.07}\left(\frac{v_{\mathrm{imp}1}}{v_{\mathrm{imp}2}}\right)^{b_z-0.56}=\zeta_{12},\label{eq:zicrat}
\end{equation}
i.e., it is almost independent from the density ratio and changes approximately with the fifth root of the velocity ratio, if we use the value $b_z=0.361$ from \citet{Pier:etal97}. Figure~\ref{fig:Dimpzic} (right panel) shows how strongly the depth of penetration differs for isocrater events relative to the S-type asteroid reference case. Although this $b_z$ value was derived from a dataset without ice, the fit for ice yields a similar value, which indicates that ice lies well within the range of $b_z$ values for candidate materials; this point is important, because several types of potential impactors contain major ice fractions.\par
\begin{figure}
\includegraphics[viewport=10 35 410 630,clip,width=0.91\textwidth]{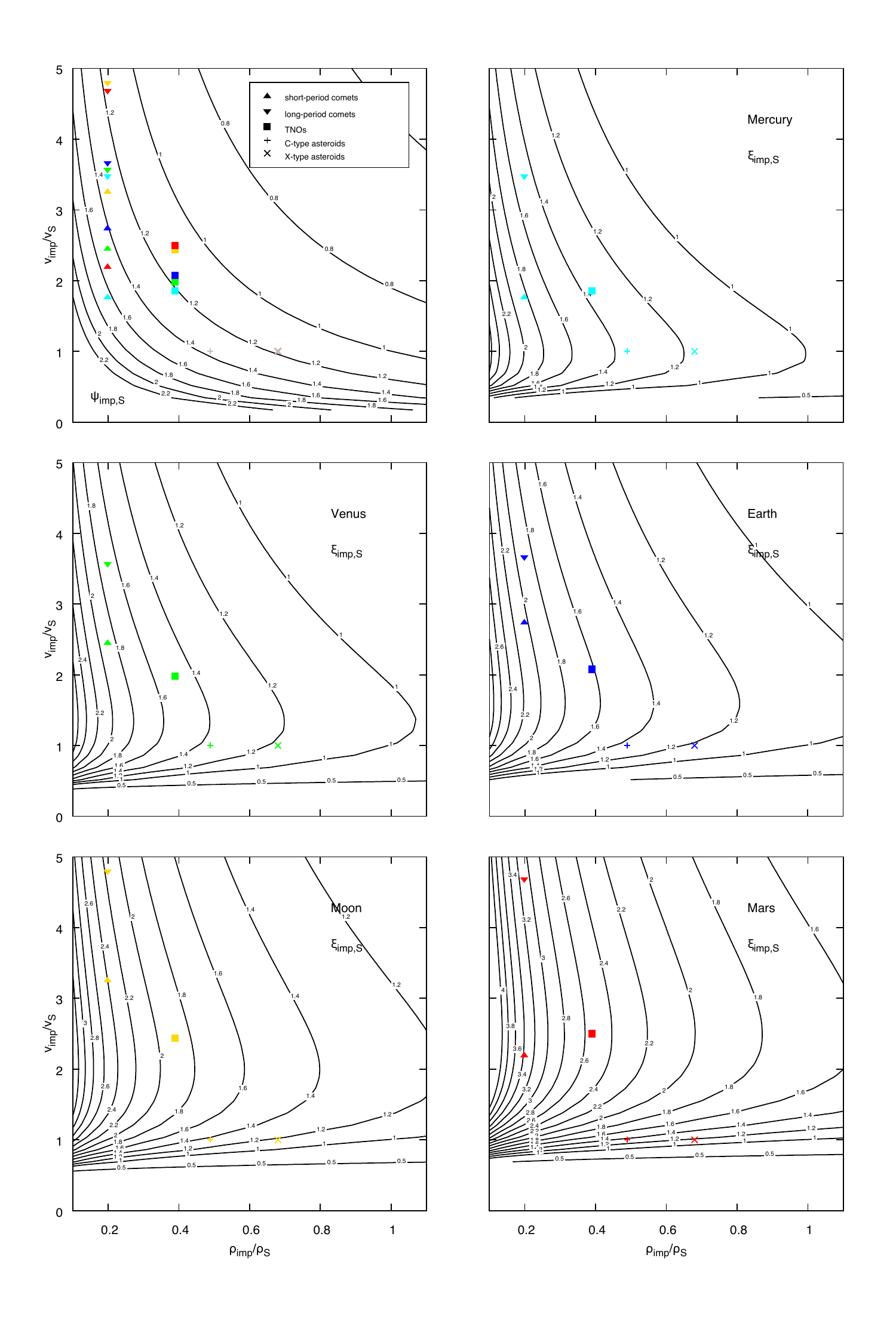}
\caption{Ratios of radii of the isobaric core for isocrater impacts as functions of the ratios of densities and impactor velocities according to Eq.~\ref{eq:ricrat} \citep[after][]{Pier:etal97} (top left) and the inflexion point formula Eq.~\ref{eq:rinflrat} \citep[after][]{Ruedas17a}, respectively. The reference bodies are the same as in Fig.~\ref{fig:Dimpzic}.\label{fig:rinflrat}}
\end{figure}

The other principal geometrical characteristic of an impact is the size of the shocked volume. As a first-order approximation, it has become somewhat common to subdivide a region affected by an impact into an isobaric core near the center of the impact, where the shock pressure shows relatively little variation, and some far-field outside that region, where it decays fairly quickly with distance. \citet{Pier:etal97} have also parameterized the radius of the roughly spherical (neglecting boundary effects from the surface) isobaric core using their homogeneous impacts and suggest the general formula
\begin{equation}
r_\mathrm{ic}=a_rD_\mathrm{imp} v_{\mathrm{imp}}^{b_r}\label{eq:ric}
\end{equation}
with $a_r=0.225$ and $b_r=0.211$ for various materials except ice; their $b_r$ for ice is smaller by a factor of 3.5, but as this seems to be mostly a consequence of a single point (at $v_{\mathrm{imp}}=10$\,km/s, the lowest velocity considered in their study) of their dataset, we suspect that even ice would be represented adequately by $b_r=0.211$. However, given the density dependence of $z_\mathrm{ic}$ and considering that the amplitude of the shock is determined by the thermoelastic properties of the projectile and the target, there is no obvious reason why the radius of the isobaric core, i.e., the region most severely affected by the shock, should be independent from these properties. In order to clarify this point, we take a different approach and start with one of the smooth global approximations for shock pressure decay proposed by \citet{Ruedas17a}, namely the ``inverse-$r$ approximation'', which reads
\begin{equation}
p(r)=p_\mathrm{IM}\frac{a_r}{b_r+\left(\frac{2r}{D_\mathrm{imp}}\right)^n}\label{eq:psr}
\end{equation}
in a form in which the shock pressure $p$ is normalized with the theoretical value from impedance matching, $p_\mathrm{IM}$, and the distance $r$ from the center of the impact is normalized with the impactor radius $D_\mathrm{imp}/2$; note that this formula does not require that $p(0)=p_\mathrm{IM}$. The parameters $a_r$, $b_r$, and $n$, which are functions of $v_{\mathrm{imp}}$, can be determined from fits to model results, whereby the model can be for a homogeneous impact. On the basis of the scarce available data, \citet{Ruedas17a} suggested that pressure decay for impacts with materials for which no data are available can be estimated using the parameters determined for other materials and the material properties that go into the impedance match solution in order to rescale $p$ with the appropriate $p_\mathrm{IM}$. He also suggested to replace the somewhat imprecise definition of the isobaric core from \citet{Pier:etal97} from which eq.~\ref{eq:ric} was derived with a definition that is directly based on the parameterization of the pressure decay, e.g., the distance at which a certain fraction $\varphi$ of the maximum pressure is reached,
\begin{equation}
r_\varphi=\frac{D_\mathrm{imp}}{2}\left(\frac{1-\varphi}{\varphi}b(v_{\mathrm{imp}})\right)^{\frac{1}{n(v_{\mathrm{imp}})}},
\end{equation}
or the position of the inflexion point,
\begin{equation}
r_\mathrm{infl}=\frac{D_\mathrm{imp}}{2}\left(\frac{n(v_{\mathrm{imp}})-1}{n(v_{\mathrm{imp}})+1}b(v_{\mathrm{imp}})\right)^{\frac{1}{n(v_{\mathrm{imp}})}}.\label{eq:rinfl}
\end{equation}
Such an analytic derivation from $p(r)$ offers, in principle, the possibility that the radius of the isobaric core thus redefined depends on both velocity and density, but as these equations show, the density dependence drops out at least for these choices; this would also be the case for the alternative pressure formula considered by \citet{Ruedas17a}. If one applies again the isocrater criterion eq.~\ref{eq:Dimprat}, the ratios of the isobaric cores of two isocrater impacts are
\begin{align}
\psi_{12}=\frac{r_{\mathrm{ic}1}}{r_{\mathrm{ic}2}}&=\left(\frac{\varrho_{\mathrm{imp}1}}{\varrho_{\mathrm{imp}2}}\right)^{-0.43}\left(\frac{v_{\mathrm{imp}1}}{v_{\mathrm{imp}2}}\right)^{b_r-0.56}\label{eq:ricrat}\\
\phi_{12}=\frac{r_{\varphi1}}{r_{\varphi2}}&=\left(\frac{\varrho_{\mathrm{imp}1}}{\varrho_{\mathrm{imp}2}}\right)^{-0.43}\left(\frac{v_{\mathrm{imp}1}}{v_{\mathrm{imp}2}}\right)^{-0.56} \frac{b_1^\frac{1}{n_1}}{b_2^\frac{1}{n_2}}\left(\frac{1-\varphi}{\varphi}\right)^{\frac{1}{n_1}-\frac{1}{n_2}}=\delta_{12}\frac{b_1^\frac{1}{n_1}}{b_2^\frac{1}{n_2}}\left(\frac{1-\varphi}{\varphi}\right)^{\frac{1}{n_1}-\frac{1}{n_2}}\label{eq:rphirat}\\
\xi_{12}=\frac{r_{\mathrm{infl}1}}{r_{\mathrm{infl}2}}&=\left(\frac{\varrho_{\mathrm{imp}1}}{\varrho_{\mathrm{imp}2}}\right)^{-0.43}\left(\frac{v_{\mathrm{imp}1}}{v_{\mathrm{imp}2}}\right)^{-0.56} \frac{\left(\frac{n_1-1}{n_1+1}b_1\right)^{\frac{1}{n_1}}}{\left(\frac{n_2-1}{n_2+1}b_2\right)^{\frac{1}{n_2}}}=\delta_{12}\frac{\left(\frac{n_1-1}{n_1+1}b_1\right)^{\frac{1}{n_1}}}{\left(\frac{n_2-1}{n_2+1}b_2\right)^{\frac{1}{n_2}}}\label{eq:rinflrat}
\end{align}
for the isobaric core definitions from \citet{Pier:etal97} and \citet{Ruedas17a}. For all three definitions, there would be the same dependence on approximately the square-root of the density ratio, but the velocity dependences are different, and they are complicated in the latter two formulas because the parameters $b$ and $n$ are, in fact, material-dependent functions of $v$ as well, and so the ratios become dependent on the target planet as well as on the impact angle. We use the non-linear forms preferred by \citet{Ruedas17a},
\begin{equation}
b(v_{\mathrm{imp}})=\alpha_b v_{\mathrm{imp}}^{\beta_b},\qquad
n(v_{\mathrm{imp}})=\alpha_n+\beta_n\lg v_{\mathrm{imp}},\label{eq:nb}
\end{equation}
with the coefficients $\alpha_b=0.266$, $\beta_b=1.161$, $\alpha_n=-0.203$, and $\beta_n=1.954$ for dunite, as better data are not available; strictly, coefficients for every material combination would be necessary, resulting in slightly different functions for every impactor type at every planet. Note that with the form eq.~\ref{eq:nb} and the chosen coefficients, $\phi_{12}$ has a pole at $v_{\mathrm{imp}}=1.27$\,km/s, and the highest $v_{\mathrm{imp}}$ at which $\xi_{12}$ has a pole is approximately 4.13\,km/s; this means that the formulae become unreliable at such low velocities and may need adjustment. The full set of ratios is shown in Figures~\ref{fig:rinflrat} and S1 for an impact angle of 45\textdegree.\par
Apart from the geometrical relations, there are also semi-empirical relations between impactor parameters and the amount of melt produced in an impact. While we will rather calculate melting in the framework of the thermal anomaly and the ensuing dynamics in the numerical models, applying the empirical scaling laws to the bulk amount of melt is a useful exercise for the purpose of a general comparison between different impactor classes. Starting from a scaling law after \citet[Eqs.~14,15]{BjHo87} and \citet{Abra:etal12}, we can write
\begin{equation}
V_\mathrm{m}=kD^3_\mathrm{imp}\left(\frac{\varrho_\mathrm{imp}}{\varrho}\right)^{\nu}\left(\frac{v_\mathrm{imp}^2}{E_\mathrm{melt}}\right)^\frac{3\mu}{2},\label{eq:Mmelt}
\end{equation}
where $V_\mathrm{m}$ is the melt volume, $E_\mathrm{melt}$ is a certain internal energy of the shocked target, and $k$, $\nu$, and $\mu$ are constants related to the target. Theoretically, the velocity exponent $\mu$ varies between 1/3 and 2/3 for momentum and energy scaling, respectively, but is experimentally found to lie between approximately 0.47 and 0.58 for commonly encountered materials, whereby the former is considered applicable to porous soils and the latter to non-porous materials such as metals \citep{HoSc87,BjHo87}. \citet[Eq.~11]{Abra:etal12} implies that the density exponent $\nu$ is 1, whereas \citet{BjHo87} suggested a value of 0.67 on the basis of crater growth scaling \citep[also see][]{HoSc87}. Combining this relation with eq.~\ref{eq:Dimprat} as before, we have
\begin{equation}
\lambda_{12}=\frac{V_\mathrm{m1}}{V_\mathrm{m2}}=\left(\frac{\varrho_{\mathrm{imp}1}}{\varrho_{\mathrm{imp}2}}\right)^{\nu-1.29}\left(\frac{v_{\mathrm{imp}1}}{v_{\mathrm{imp}2}}\right)^{3\mu-1.68}\label{eq:Vmrat}
\end{equation}
\begin{figure}
\includegraphics[viewport=10 104 420 327,clip,width=\textwidth]{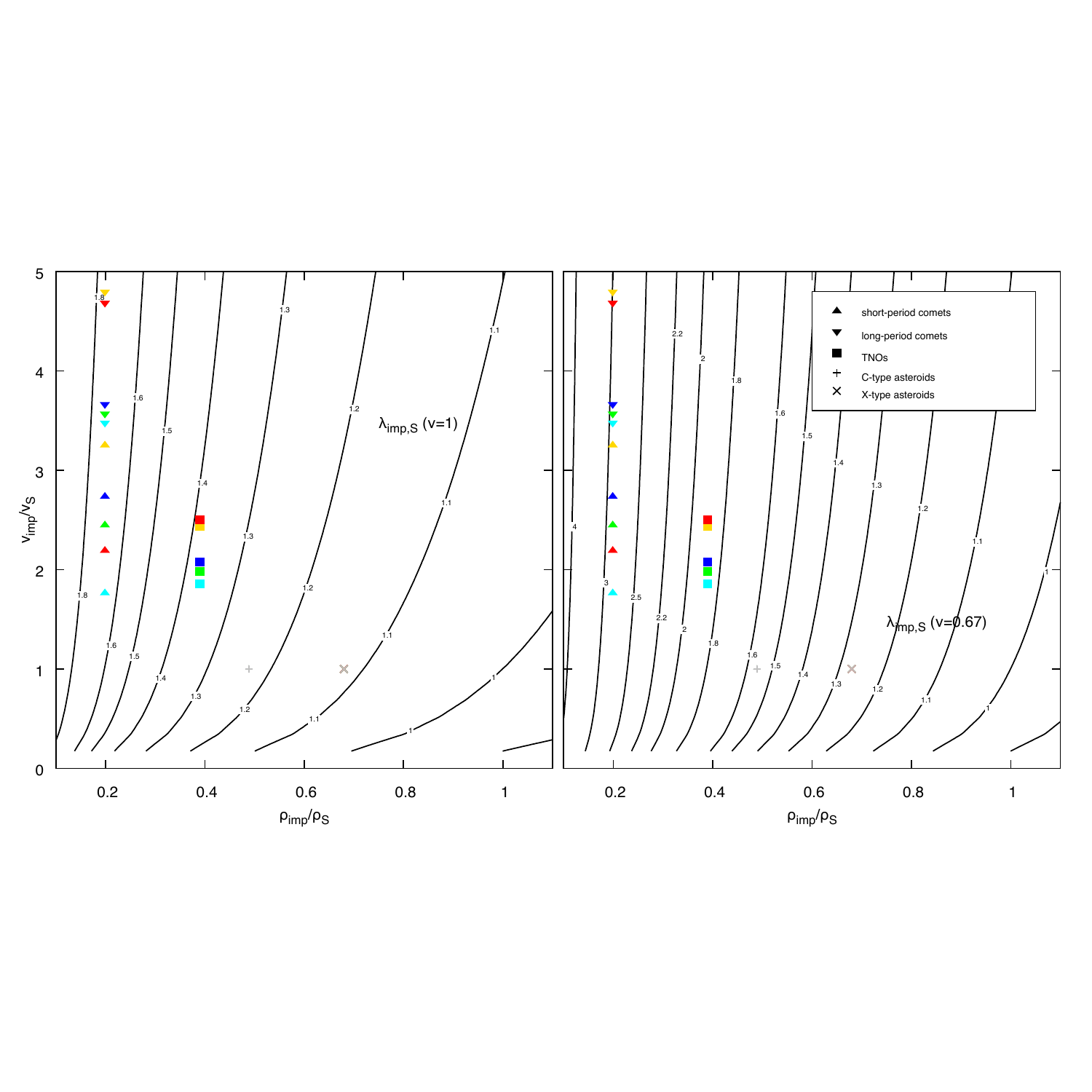}\\
\caption{Ratios of melt volumes for isocrater impacts as functions of the ratios of densities and impactor velocities according to eq.~\ref{eq:Vmrat}. $\mu=0.58$ as appropriate for metals was chosen to represent non-porous materials. The values $\nu=0.67$ and $\nu=1$ are suggested by \citet{BjHo87} and \citet{Abra:etal12}, respectively. The reference bodies for calculating the ratios are the same as in Fig.~\ref{fig:Dimpzic}.\label{fig:Vmelt}}
\end{figure}
for the ratio of impact melt volumes in isocrater impacts.\par
\begin{figure}
\includegraphics[width=\textwidth]{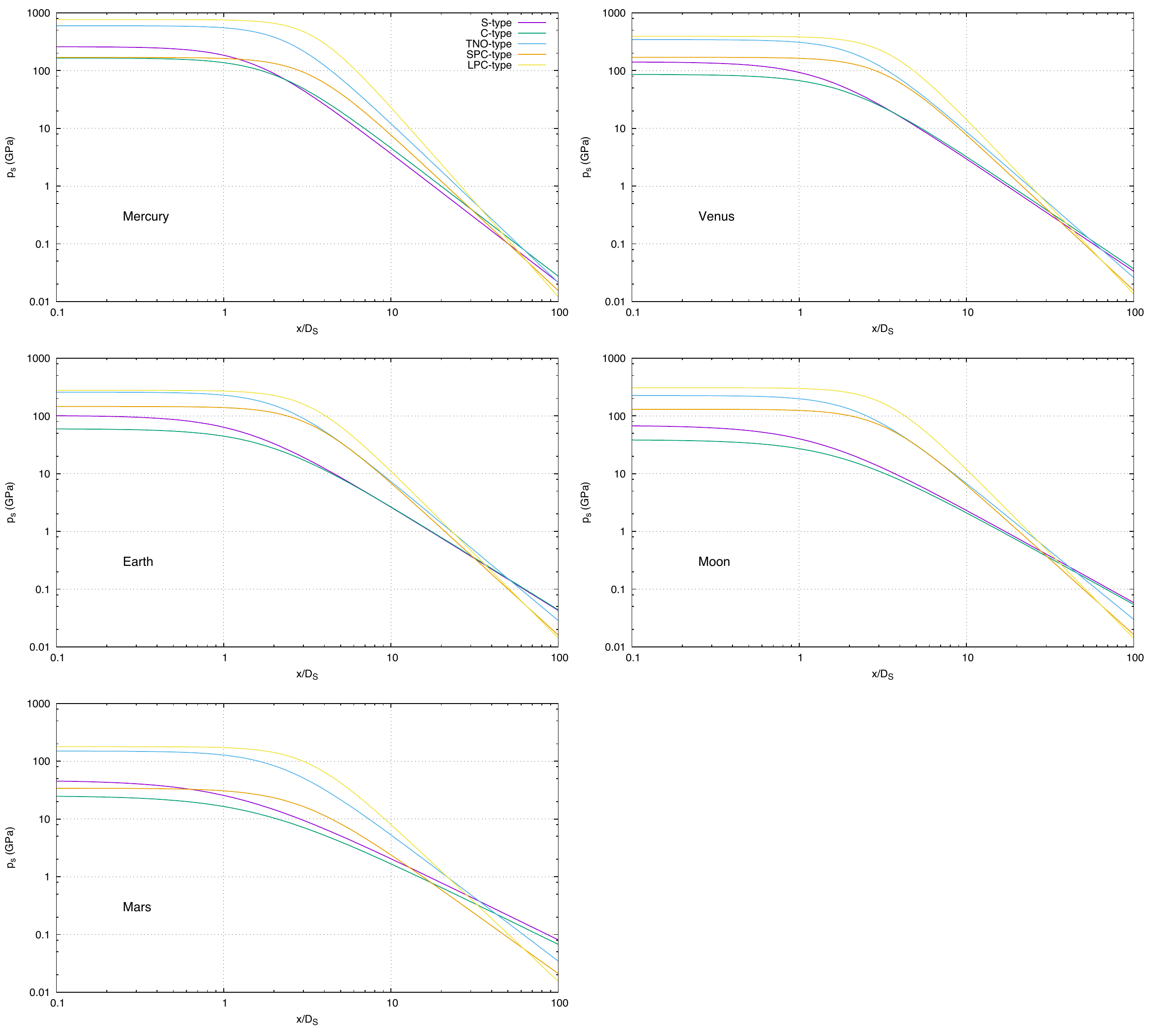}
\caption{Shock pressure as a function of distance from the impact point according to eq.~\ref{eq:psr} for isocrater impacts with different impactor types. The distance is normalized with the diameter of an S-type impactor for better comparison of the relative size of the shocked volume. For simplicity, constant densities of 3200--3300\,kg/m$^{3}$ have been assumed for each planet.\label{fig:psr}}
\end{figure}

While this approach may give an estimate of the total amount of melt produced, the calculation via the shock pressure and heating is more suitable for integration into a convection model. The downside is that the actual thermal anomalies, which can be calculated by combining the decay laws by \citet{Pier:etal97} or \citet{Ruedas17a} with the impedance-match solution for the shock of colliding infinite half-spaces, cannot be cast in a simple form anymore. The impedance match shock pressure is given by
\begin{equation}
p_\mathrm{IM}=\varrho(v_\mathrm{B}+S_\mathrm{h}u_\mathrm{IM})u_\mathrm{IM},\label{eq:pIM}
\end{equation}
with the density of the target, $\varrho$, in g/cm$^3$, the speed of sound in the target, $v_\mathrm{B}$, in km/s, and the Hugoniot slope $S_\mathrm{h}=(1+\gamma)/2$ being related to the Grüneisen parameter $\gamma$ \citep[App.~II]{Melosh89}. The particle velocity $u_\mathrm{IM} $ at the interface with the impactor is
\begin{equation}
u_\mathrm{IM}=\frac{-B_u+\sqrt{B_u^2-4A_uC_u}}{2A_u}\label{eq:uic}
\end{equation}
with
\addtocounter{equation}{-1}\begin{subequations}\begin{align}
A_u&=\varrho S_\mathrm{h}-\varrho_\mathrm{imp} S_\mathrm{h,imp}\\
B_u&=\varrho v_\mathrm{B}+\varrho_\mathrm{imp}(v_\mathrm{B,imp}+2 S_\mathrm{h,imp} v_{z,\mathrm{imp}})\\
C_u&=-\varrho_\mathrm{imp}v_{z,\mathrm{imp}}(v_\mathrm{B,imp}+S_\mathrm{h,imp} v_{z,\mathrm{imp}})
\end{align}
\end{subequations}
\citep[e.g.,][]{Melosh11}, whereby the properties used here are those of the unshocked material. The speed of sound of the target is derived from the thermoelastic properties as given by the mineralogical model for the planet via $v_\mathrm{B}=\sqrt{K_S/\varrho}$, where we use values for the adiabatic bulk modulus $K_S$ and for $\varrho$ averaged over the isobaric core volume in the convection models. In the impactor types that can be regarded as having a low porosity, the speed of sound in the impactor is simply determined using Birch's law, $v_\mathrm{B,imp}=2.36\varrho_\mathrm{imp}-1.75$ \citep[e.g.,][eq.~4.64]{Poirier00}; we treat the S- and C-type asteroids and the TNOs as belonging into this group, on the grounds that at the size of the impactors we consider the porosity tends towards low values \citep{Carry12}. In porous bodies, especially comets, the porosity causes not only the density, but also the bulk sound speed to drop to very low values. \citet{Davi:etal16} discuss various suggestions for the density, compressive strength, and porosity of comets from the literature and find porosities upwards of 60\% to be plausible; however, as we consider very large comets in this paper, we assume that some compaction has occurred and reduced the porosity to 50\%. Compressive strengths are found to be on the order of at most 1\,kPa for pristine material, which would give a speed of sound on the order of a few meters per second; this is also consistent with a Young's modulus of no more than 10\,kPa found in the Deep Impact experiment carried out on the comet 9P/Tempel~1 \citep{JERich:etal07}. Figure~\ref{fig:psr} shows the shock pressure decay with distance for isocrater impacts with different impactor types on the different targets; the temperature anomaly or waste heat production from the impact cannot be represented in such a general plot, because the dependence on the depth of penetration of the impactor enters via the pressure term and introduces an additional direct dependence on the impactor diameter. In almost all cases, the shock pressures reached in cometary and TNO-type impacts are substantially higher than those for asteroid impactors, and a much larger volume is exposed to them.\par
The impact velocities for most of the impactor types considered here are reasonably well known and constrained by observations of existing bodies. Therefore we will rely on data from the literature in connection with them. In the case of the hypothetical TNO-type impactors, however, the situation is different, because there are no known TNO-like objects in an orbit that reaches into the inner Solar System. For the purpose of including an interesting and extreme scenario, we posit their occasional appearance in the early history of the Solar System based on the assumption that gravitational interactions with the gas or ice giants, especially Neptune, may have perturbed the orbits of Kuiper belt objects (KBOs) in such a way that some of them have been deflected onto very eccentric paths that cross the orbits of the inner planets, somewhat similar to comets originating in that region. We use some results from orbital mechanics to derive values for the impact velocities at different planets that would be consistent with an origin somewhere in the Kuiper belt; the derivation is given in \ref{app:vimpTNO}.\par
Figures~\ref{fig:Dimpzic}--\ref{fig:Vmelt} give an impression of the variability of impactors that produce a crater of the same size, whereby one must keep in mind that the points that mark the impactor classes for different planets are only average values and that the values for individual impactors may differ substantially. The different impactor classes have properties quite different from the reference class, the S-type asteroids, but Fig.~\ref{fig:Dimpzic} (left) reveals that the differences in velocity can partly or completely compensate the differences in material effects. As a consequence, impactors of the same size but otherwise as different as S-type asteroids and long-period comets (LPCs) can create the same crater on Venus, Mercury, or the Earth; on Mars and the Moon, however, the comet would have to be almost one-fifth smaller according to Eq.~\ref{eq:Dimprat}, if we use the mean values for the densities and velocities from Table~\ref{tab:impactors}. In general, C-type asteroids are expected to show the largest size contrast with the reference for isocrater impacts.\par
The impactor diameter itself, however, is only of indirect significance with regard to the dynamical effects. The more directly relevant variables are the center depth and the radius of the isobaric core as well as the amount of impact melt produced, all of which are functions of $D_\mathrm{imp}$. Fig.~\ref{fig:Dimpzic} (right) shows that asteroidal impactors penetrate essentially to the same depths in isocrater impacts, irrespective of the class, as a consequence of the near-zero dependence on the density ratio, whereas comets and TNOs penetrate $\sim$20\% less deeply than a reference impactor in an isocrater impact and may therefore at first sight be expected to have a lesser effect on mantle dynamics. On the other hand, Eq.~\ref{eq:Vmrat} and Fig.~\ref{fig:Vmelt} suggest that these same impactor types should produce substantially more melt than the reference case as a consequence of the contrast in material properties, and Fig.~\ref{fig:psr} indicates that the shocked volumes are larger: These effects should more than compensate for the smaller $z_\mathrm{ic}$, and a stronger dynamical effect of cometary and TNO-type impactors is indeed confirmed by our numerical models, as will be shown below. The reason for the contrasting melt productivities lies partly in the size of the isobaric core, which for cometary and TNO-type impactors is always larger than for the reference type; this holds regardless whether the \citet{Pier:etal97} definition (top left panel in Figs.~\ref{fig:rinflrat} and S1) or the $\varphi$-value or inflexion point definitions from \citet{Ruedas17a} is used.\par
There are also differences between isocrater impactors in terms of atmospheric erosion, which are briefly outlined in the Supplement, as they are beyond the scope of this paper.\nocite{Shuvalov09,Shuv:etal14}

\section{Method and setup of numerical models}
We test the effects of different impactor types on the interiors of terrestrial planets with numerical models of mantle convection. The method has been described in detail elsewhere \citep[e.g.,][and references therein]{RuBr17c}, and so we will only give a brief outline of the approach and focus on the aspects of the model specific to this study. The convection algorithm is a modified version of the code StagYY \citep{Tackley96a,Tackley08} and solves the equations of conservation of mass, momentum, and energy in the compressible, anelastic approximation on a two-dimensional spherical annulus grid \citep{HeTa08} with free-slip, isothermal top and bottom boundaries. Melting and the transport of trace components such as water and radionuclides are modeled with tracer particles. In partially molten regions, we extract all melt that is less dense than the matrix, down to an extraction threshold of 0.7\% \citep[continuous or dynamic melting, cf.][]{DMShaw00} and add it instantaneously to the top to build the crust (see Supplement); in impacts, where complete melting may be achieved in a small volume, a hard upper bound of 60\% is imposed on melting. Incompatible components such as the internal heat sources are redistributed according to their partitioning coefficients. The convection model is coupled to a detailed model of the petrological and physical properties of the mantle and crust consisting of a parameterization of experimental phase diagrams of peridotite and basalt/eclogite as well as their thermoelastic properties such as the density, and the melting behavior. The viscosity is described by an extended Arrhenius-type flow law that includes a dependence on pressure, temperature, and water, iron, and melt content \citep[e.g.,][]{HiKo03}. The core is represented by a simplified one-dimensional model following \citet{Nimm:etal04} and \citet{JPWiNi04}, but we do not include the possibility of core crystallization, which is unimportant for the purpose of this study. The most important model parameters are given in Table~\ref{tab:models}, and some more technical details about the method are provided in the Supplement.\nocite{YHZhao:etal09,ZMJin:etal01,Mack:etal98,JiKa11,JiKa12,PHWaTa14,St-Ca:etal12}\par
We consider four target bodies, namely Mars, Moon, Venus, and Mercury; the Earth is left out in order to avoid possible complications with its plate tectonics and complex water cycle. Each of these targets has its peculiarities with regard to the response to impacts. There are also significant differences in Mg\# and alkali contents between the four targets under consideration, and so there are differences between the solidi of their mantles of up to 100\,K at any given pressure (cf. \ref{app:solidus}, Fig.~\ref{fig:solidi}), which have implications for the production of melt. As the models presented here are only a general survey meant to highlight some consequences of giant impacts specific for these four planets rather than a detailed study, we do not strive to reproduce specific observations, although we do account for certain important properties and consider the details of the Mars and Moon models in some more detail. The model sets are generally designed to include one impact-free model and two subsets with single impacts of either ``large'' or ``small'' size, designated by ``L-'' or ``S-'', and of one of the different impactor types (S, C, TNO, LPC).\par
Mars has about half the diameter of Venus and is twice as far from the Sun, which causes impacts of projectiles of a given size to be less energetic than on targets closer to the Sun. The martian core is assumed to be large enough to prevent the existence of a basal perovskite+ferropericlase layer, in agreement with recent structure models based on the moment of inertia \citep{KhCo08,Kono:etal11,Rivo:etal11}. Previous models \citep{RuBr17c} let us expect that on Mars, events much smaller than the Isidis-forming impact would not result in large-scale, long-lasting effects under the assumption of an S-type asteroid impactor, and so we take the event that formed the Hematite impact basin, which is a bit smaller than Isidis, as a lower bound. The corresponding basin has a diameter of 1065\,km \citep{Frey08,JHRobe:etal09}, implying a diameter of the assumed S-type asteroid impactor of 185\,km and an isobaric core diameter after eq.~\ref{eq:rinfl} of 102\,km (cf. Fig.~S4). As the upper limit we use the Utopia-forming event ($D_\mathrm{f}=3380$\,km), which is one of the largest basins in the Solar System and could have been caused by an S-type asteroid with $D_\mathrm{imp}=685$\,km. For better comparison, we will use these two S-type asteroid sizes as reference and derive from them isocrater impacts for Mars and Venus and, partly, for the Moon and Mercury; these are summarized in Table~\ref{tab:impactors}.\par
Venus, by contrast, is larger and closer to the Sun, which entails high average impact velocities for all sorts of impactors. The properties of the planet are assumed to correspond to those of the present, as far as they are known. The mantle is assumed to have a peridotitic petrology with $\textrm{Mg\#}=0.92$, corresponding to the chondritic model by \citet{MoAn80} \citep{Fegley14}, and the core is taken to be at the large end of the expected range from \citet{KoYo96}. The surface temperature is set to today's high value, which implies that the target material properties are somewhat different from those of other planets: the density and the bulk modulus, which control the speed of sound, are lower, the thermal expansivity is greater, the viscosity is smaller, and the mere fact that the temperature is closer to the solidus means that less energy has to be expended to bring the material to the melting point. The viscosity of the lower mantle has been tuned to approach the shape determined by \citet{Stei:etal10}. In order to avoid another parameter that would add more dynamical complexity to the model, we did not include plastic yielding of the lithosphere as other authors did \citep[e.g.,][]{ArTa12}. Giant impact basins are not known on Venus because of its geologically young surface, but have without doubt existed in its ancient past \citep{Kari:etal17}.\par
The Moon is the smallest target, but mostly thanks to its closer distance to the Sun, impact velocities are larger than on Mars (cf. Table~\ref{tab:impactors}). For its mantle, which accounts for most of the body, we assume an Earth-like mineralogy but with $\textrm{Mg\#}=0.83$ \citep{SRTaMcLe09}. As the Utopia-forming S-type impactor would have produced a crater substantially larger than the largest known basin (South Pole--Aitken basin, $D_\mathrm{f}\approx 2500$\,km) according to the scaling laws, whose applicability under such conditions would be questionable, we chose the latter as the upper limit for that body; it could have been produced by an S-type asteroid with $D_\mathrm{imp}=403$\,km.\par
Mercury is not much larger than the Moon, but has a very thin mantle with an estimated Mg\# in the range of approximately 0.94 to 0.97 \citep{MSRoTa01} and a surface gravity similar to Mars thanks to its unusually large core; on the other hand, its proximity to the Sun causes impact velocities to be much larger than at the other targets. The combination of high impact energies and a thin mantle is expected to result in a unique style of response to an impact, which will contrast with the other targets, whose mantles have much larger volumes with larger convective structures. Similar to the Moon, the Utopia-forming S-type impactor would have produced too large a crater, and the impactor would even have penetrated into the core. The diameter of the resulting basin would be about 25\% larger than the high-Mg anomaly ($D_\mathrm{f}\approx 2550$\,km) in the northwestern part of the planet, which is the largest hitherto identified potential (albeit not undisputed) impact site on that planet \citep[e.g.,][]{Weid:etal15,EAFran:etal17}. However, the center of the isobaric core for such an impactor would lie more than halfway to the core--mantle boundary (CMB) of Mercury, and the strongly shocked region is expected to intersect with it. \citet{Melosh84} has discussed in detail how the compressive shock wave turns into a tensile wave upon reflection at the free surface of a planet and how the interference of both results into partial annihilation of the shock in a certain area near the surface. For a deeply penetrating impact on a planet with a thin mantle and a liquid outer core like Mercury, we need to consider the potential effects of a lower strong reflector as well, because the shock-wave will partly be reflected at the CMB and will interfere with the direct wave. In order to estimate the character of this interference, we calculate the P-wave reflection coefficient for a P-wave reflected at the interface for two homogeneous half-spaces,
\begin{equation}
R_\mathrm{PP}=\frac{\varrho_\mathrm{c}v_{\mathrm{P,c}}\cos\varphi_\mathrm{m}-\varrho_\mathrm{m}v_{\mathrm{P,m}}\cos\varphi_\mathrm{c}}{\varrho_\mathrm{c}v_{\mathrm{P,c}}\cos\varphi_\mathrm{m}+\varrho_\mathrm{m}v_{\mathrm{P,m}}\cos\varphi_\mathrm{c}}
\end{equation}
\citep[e.g.,][]{GMuller07}, where $\varrho_\mathrm{m/c}$, $v_{\mathrm{P,m/c}}$, and $\varphi_\mathrm{m/c}$ are the density and the P-wave velocity in the mantle and core, and the angles of incidence and refraction of the seismic ray, respectively. For Mercury at a time close to 400\,My, our numerical convection models have $\varrho_\mathrm{m}=3266.5$\,kg/m\textsuperscript{3}, $v_{\mathrm{P,m}}\approx 8.15$\,km/s, $\varrho_\mathrm{c}=6692$\,kg/m\textsuperscript{3}, and $v_{\mathrm{P,c}}=5.03$\,km/s, which suggests that for small angles of incidence $\varphi_\mathrm{m}$, i.e., close to the center of the isobaric core, $R_\mathrm{PP}>0$. Interference would therefore be constructive in the most critical region and may result in an enhancement of the strongly heated region and melt production from the impact. Hence we conclude that the applicability of the simple impact model used here and in most other studies of this sort may be questionable in this case, and we decided to limit the Mercury models to a ``quasi-Hematite'' impact, which produces a basin of approximately the size of Caloris.\par
\begin{table}
\caption{Essential model parameters. Further parameters are listed in Table~\ref{tab:TNOtab} and in Table~S2.\label{tab:models}}
\centering
\begin{tabular}{lcccc}\toprule
&Mars&Moon&Venus&Mercury\\\midrule
$R_\mathrm{P}$ (km)&3389.5\textsuperscript{a}&1737.1513\textsuperscript{n}&6051.8\textsuperscript{a}&2439.36\textsuperscript{j}\\
$R_\mathrm{core}$ (km)&1730\textsuperscript{f,l}&350\textsuperscript{*}&3320\textsuperscript{g}&2020\textsuperscript{e}\\
$T_\mathrm{surf}$ (K)&218\textsuperscript{b}&250\textsuperscript{h}&735\textsuperscript{b}&440\textsuperscript{d}\\
$D_\mathrm{s2c}$ (km)&5.6\textsuperscript{m}&18.7\textsuperscript{k}&3.5\textsuperscript{\textdagger}&11.7\textsuperscript{p}\\
Initial $T_\mathrm{pot}$ (K)&1700&1550&1800&1700\\
Bulk silicate Mg\#&0.75\textsuperscript{q,r}&0.83\textsuperscript{q}&0.92\textsuperscript{c,i}&0.94\textsuperscript{i}\\
\bottomrule
\end{tabular}\\[1ex]
\begin{minipage}{\textwidth}
\textsuperscript{*}Estimated value\nocite{RCWebe:etal11,RFGarc:etal11}.\hspace{1em}
\textsuperscript{\textdagger}Calculated from empirical fit (see Supplement).\\
References: \textsuperscript{a}\citet{Arch:etal11}, \textsuperscript{b}\citet{Catling15}, \textsuperscript{c}\citet{Fegley14}, \textsuperscript{d}\citet{Grot:etal11a}, \textsuperscript{e}\citet{Hauc:etal13}, \textsuperscript{f}\citet{Kono:etal11}, \textsuperscript{g}\citet{KoYo96}, \textsuperscript{h}\citet{Lane:etal13}, \textsuperscript{i}\citet{MoAn80}, \textsuperscript{j}\citet{Perr:etal15}, \textsuperscript{k}\citet[Tab.~IX]{Pike89}, \textsuperscript{l}\citet{Rivo:etal11}, \textsuperscript{m}\citet[Tab.~3]{RoHy12b}, \textsuperscript{n}\citet{DESmit:etal17}, \textsuperscript{p}\citet{Suso:etal16}, \textsuperscript{q}\citet{SRTaMcLe09}, \textsuperscript{r}\citet{WaDr94}
\nocite{Chab:etal14,Clifford93,Han:etal14,Kief:etal12}
\end{minipage}
\end{table}
\begin{table}
\caption{Isocrater impact sets for Mars, Moon, Venus, and Mercury. The reference case is an S-type asteroid. All diameters and depths are given in kilometers, densities in kg/m\textsuperscript{3}, and absolute velocities in km/s. Impactor velocities for asteroids are from \citet[Mercury]{Grie:etal07}, \citet[Venus]{OKeAh94}, \citet[Moon]{Chyba91}, and \citet[Mars]{Ivanov01}, the long-period comet velocities are from \citet[Moon]{ShWo87} and \citet[Mars]{Steel98}, and the TNO velocities are from this study (cf. \ref{app:vimpTNO}).\label{tab:impactors}}
\centering
\begin{tabular}{lccccccc}\toprule
Planet&$D_\mathrm{f}$&Type&$\varrho_\mathrm{imp}$&$v_\mathrm{imp}$&$D_\mathrm{imp}$&$D_\mathrm{infl}$&$z_\mathrm{ic}$\\\midrule
Mars&1065 (S)&S&2720&9.6&185&102&81\\  
&&C&1330&9.6&251&138&77\\
&&LPC&540&45&154&312&61\\
&&TNO&1060&24&165&258&68\\
&3380 (L)&S&2720&9.6&685&378&301\\  
&&C&1330&9.6&930&513&286\\
&&TNO&1060&24&611&955&254\\
Moon&1259 (S)&S&2720&12&185&154&91\\
&&C&1330&12&251&209&86\\
&&LPC&540&58&152&330&68\\
&&TNO&1060&29.2&167&289&77\\
&2500 (L)&S&2720&12&403&336&197\\  
&&C&1330&12&547&456&187\\
&&TNO&1060&29.2&365&629&167\\
Venus&1257 (S)&S&2720&18&185&239&109\\
&&C&1330&18&251&324&103\\
&&TNO&1060&35.7&188&352&94\\
&3984 (L)&S&2720&18&685&885&403\\
&&C&1330&18&930&1201&383\\
&&TNO&1060&35.7&697&1304&348\\
Mercury&1579 (S)&S&2720&25&185&296&127\\  
&&C&1330&25&251&401&121\\
&&TNO&1060&46.4&195&399&111\\
\bottomrule
\end{tabular}
\end{table}
Inspection of Figures \ref{fig:Dimpzic}--\ref{fig:Vmelt} shows that certain combinations of impactor candidates are more likely to produce isocrater events with different dynamical effects in the deeper interior than others and that these combinations vary between planets. We shall now investigate some combinations in which those effects differ considerably, i.e., for which at least some of the ratios $\zeta_\mathrm{imp,S}$, $\lambda_\mathrm{imp,S}$, $\psi_\mathrm{imp,S}$, $\phi_\mathrm{imp,S}$, and $\xi_\mathrm{imp,S}$ differ significantly from 1. In general, the effects of X-type asteroids are most similar to those of the reference S-type impactors and will therefore not be considered further. C-type asteroids have a bit more different effects as a consequence of their lower density: they would be about one-third larger but penetrate to the same depth; the shocked volume and melt volume would be 2.5 times larger, however. TNO-type impactors are even less dense and twice as fast as the reference impactor, and although they would have to be of the same size, they penetrate some 20\% less deeply into the target. The radius of the isobaric core is nonetheless several times larger than in the reference case, especially for targets at greater distance from the Sun. Finally, comets have still lower densities than TNOs, but their velocities may be smaller or larger, depending on whether they are short-period or long-period comets. The former, whose periods are less than 200 years and whose orbits thus mostly lie within that of Neptune, would rather have to be a bit larger than the reference impactor, whereas the latter come from much further out and would tend to be a bit smaller. Both would penetrate between one-quarter and one-third less deeply but shock a much larger volume, even larger than TNOs. However, for very large impacts this implies comets of enormous size, which are not observed today and are rather unlikely to form; the largest cometary nucleus observed so far has a diameter of $\sim 96$\,km \citep{Kors:etal14}, and so we will assume a generous upper limit of $\sim 150$\,km for the size of comets, which is also in general agreement with the estimated maximum size of cometary impactors recently deduced by \citet{Rick:etal17}.

\section{Results}
In our brief survey of impact effects on the interiors, we will focus on the most salient features of the impact-generated temperature and composition anomalies and some essential dynamic characteristics. In the case of Mars and the Moon, which yield the richest model series, we also consider density and bulk sound velocity as potential observables of interest.
\subsection{Mars}
\begin{sidewaysfigure}
\centering
\includegraphics[width=\textwidth]{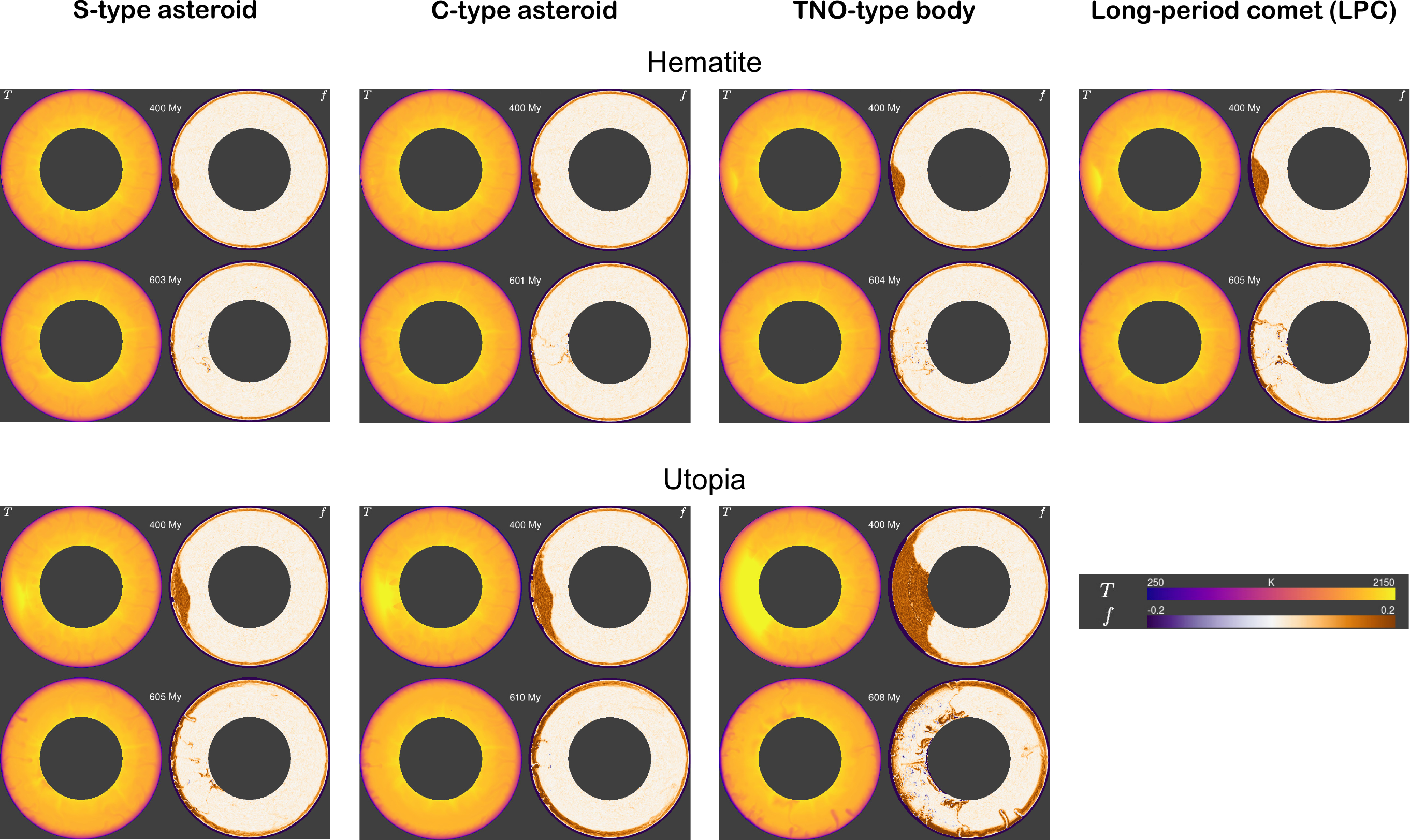}
\caption{Temperature ($T$, left part) and depletion ($f$, right part) of models with impacts on Mars by different impactor types and of different sizes. All impacts occur between the 8 and the 9 o'clock position and produce a basin with either the diameter of the Hematite basin ($D_\mathrm{f}=1065$\,km) or the diameter of the Utopia basin ($D_\mathrm{f}=3380$\,km) on Mars. The upper row of each impact size subset shows the state directly after the impact, the lower row the state 200\,Myr after the impact. The colorbars are clipped for clarity.\label{fig:mars-snap}}
\end{sidewaysfigure}

Fig.~\ref{fig:mars-snap} (upper part) shows snapshots of the temperature and the composition (depletion) field for the Hematite isocrater set at the time of impact and $\sim200$\,My later. This set consists of S- and C-type asteroid impactors as well as an LPC-type and a TNO-type impactor. The two asteroidal impacts are very similar in size and do not produce a major thermal anomaly; they dissipate quickly without influencing the plume distribution substantially, although the heat flux anomaly and a marked compositional anomaly remain at the site permanently. LPC- and TNO-type impactors both produce considerably larger thermal and compositional anomalies, which come close in size to the S-type Utopia-size event. They produce a thick crust, of which a part quickly delaminates, sinks to the CMB, and is then slowly mixed back into the mantle, whereas there is much lesser crustal thickening and hardly any delamination in the asteroidal impacts of the Hematite set.\par
The magnitude of the effects of the Utopia-sized impacts can clearly be put in the order S$<$C$<$TNO, whereby especially the TNO-type impact would shock-heat about one-eighth of the mantle and reach all the way down to the CMB (Fig.~\ref{fig:mars-snap}, lower part); considering that extrapolation of the scaling laws to such a large size may be problematic, this result should probably not be treated as quantitatively accurate. The effect on mantle convection is more diverse in the large impacts, but it is substantial for all three variants, and for the C-type asteroid and the TNO-type impactors, the compositional anomaly is expected to extend far beyond the final crater. The final compositional anomaly, which in this case resembles closely the pattern at $\sim 600$\,My, is fairly sharply bounded in the case of the small (Hematite-sized) impacts, but it transitions quite smoothly into the undisturbed mantle in the large (Utopia-sized) impacts. The reason is probably that in smaller events the depth of penetration of the impactor and the entire impact-induced anomaly are shallower and more closely coupled to the lithosphere, which would hamper its subsequent spreading. In a large event like Utopia, by contrast, the post-impact dynamics reach deeper into the mantle, and much of the ensuing anomaly develops in the sublithospheric mantle without encountering obstacles.\par
\begin{figure}
\includegraphics[width=\textwidth]{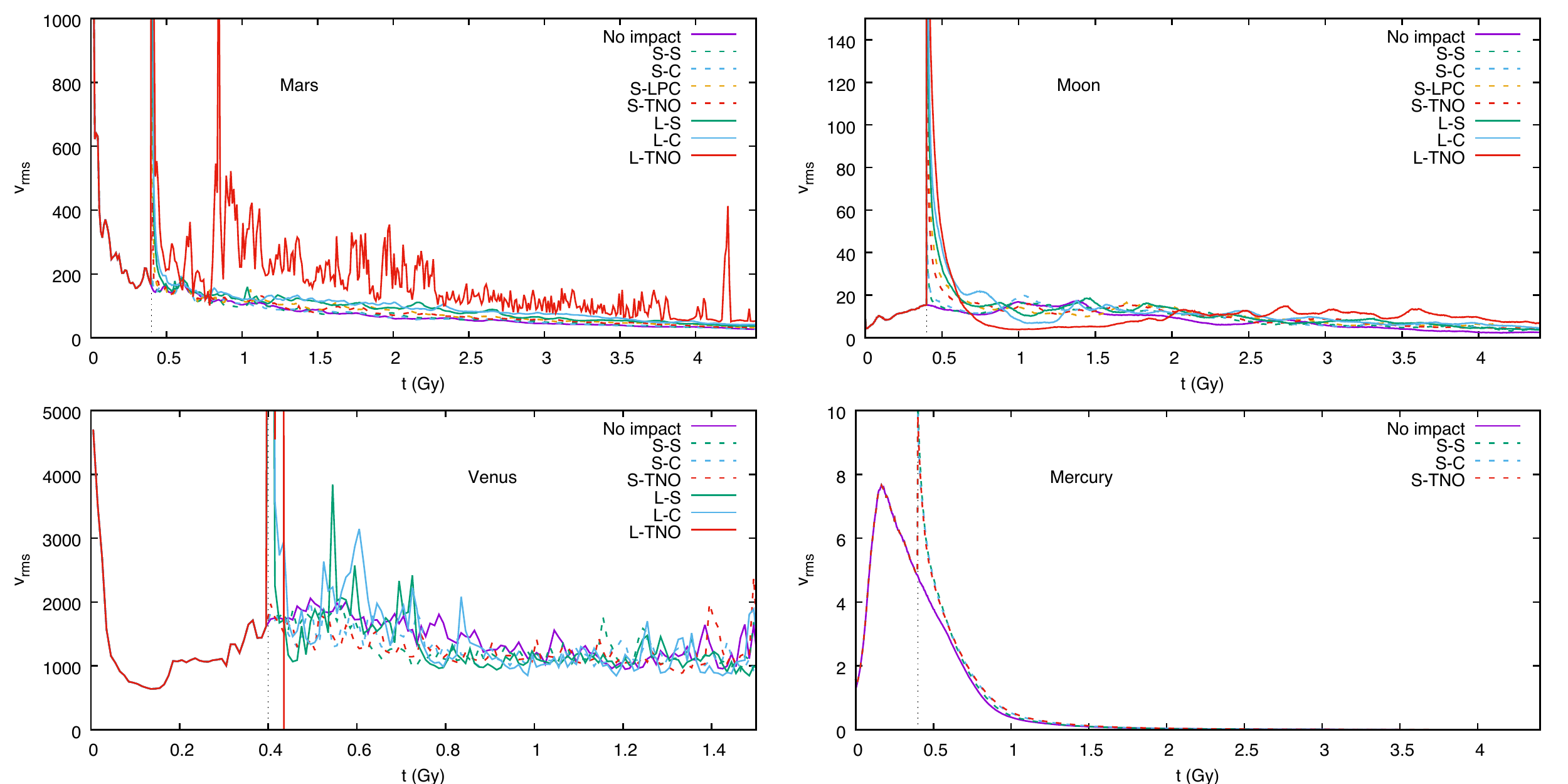}
\caption{Non-dimensional root-mean-square velocity of the Mars, Moon, Venus, and Mercury models as a function of time. For better readability, the vertical axes have been clipped, and $v_\mathrm{rms}$ was averaged over time intervals of 10\,My for the curves for Mars and Venus. For the same reason, the Venus evolution is plotted only for the first 1.5\,Gy, because at later stages, numerous velocity peaks related to lithospheric instabilities not relevant to the impact clutter the image.\label{fig:vrms}}
\end{figure}

\FloatBarrier
The root-mean-square velocity $v_\mathrm{rms}$ of the convective flow in the mantle, which serves us here as a measure of convective vigor, shows that there are only small differences between the impact-free model and the Hematite-size events. By contrast, the Utopia-size impactors cause substantial deviations in convective vigor that last for more than 100\,My or, in the extreme case of a TNO-type impactor, for several hundreds of millions of years (Fig.~\ref{fig:vrms}). The latter also features wide-spread appearances of lithospheric instability presumably caused by the massive production and partial eclogitization of cold crust overlying a strongly depleted sublithospheric layer from the impact, which are reflected in the numerous velocity maxima. The two cases with a large C- or TNO-type impactor also raise the mean temperature of the mantle more strongly relative to the impact-free model than all other cases.\par
\clearpage
\begin{figure}
\includegraphics[width=\textwidth]{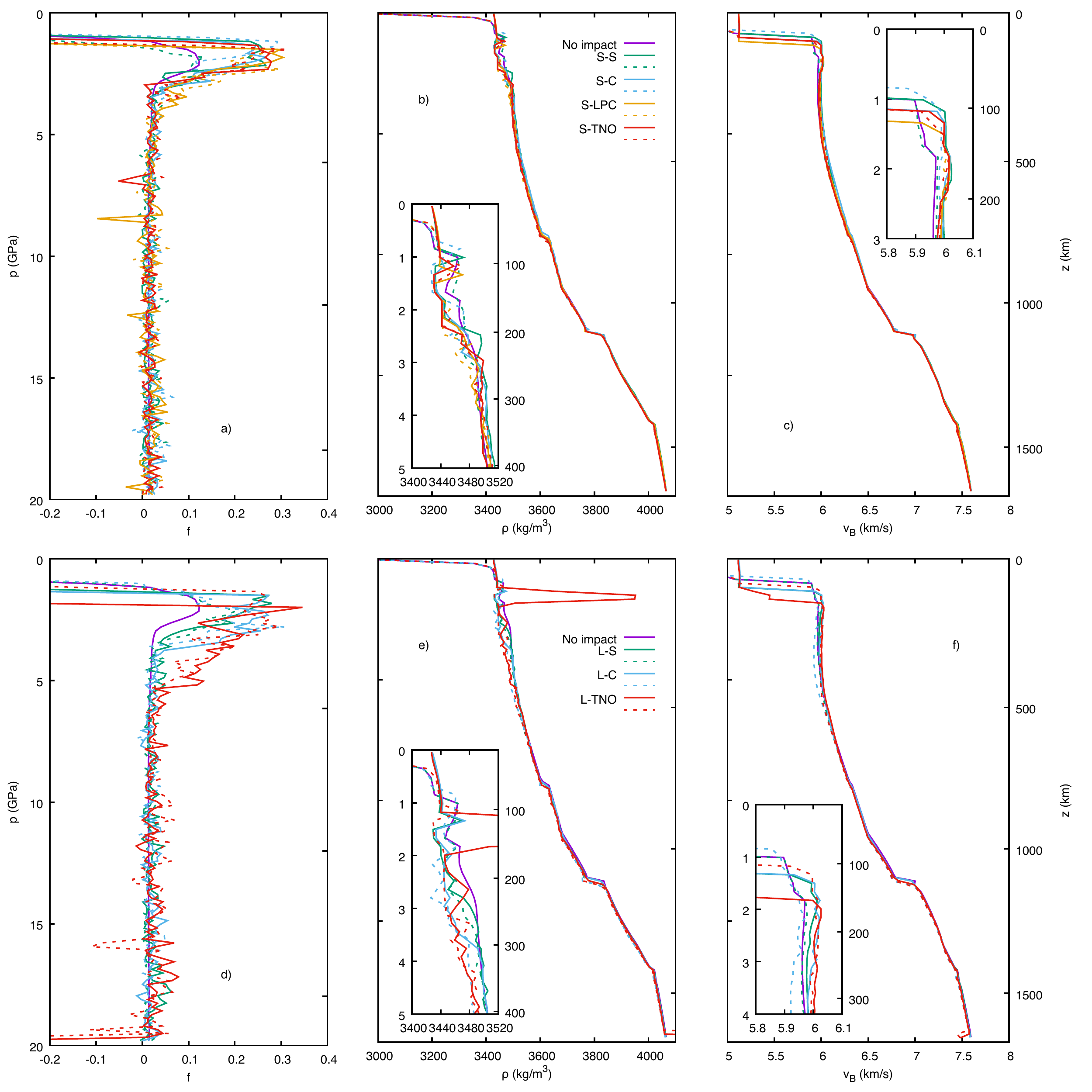}
\caption{Present-day compositional (a,d), density (b,e), and bulk sound speed (c,f) profiles for the small (Hematite, upper row) and large (Utopia, lower row) impact models on Mars. The ``no impact'' profiles are global lateral averages, the other profiles are local depth profiles taken at the impact site (solid lines) and at a more remote location in the region of the impact (dashed lines; Hematite: $\sim 10^\circ$, Utopia: $\sim 40^\circ$).\label{fig:mars_crhovB-z}}
\end{figure}
In order to test whether the different impactor types can be discerned on the basis of the present-day remnants of the impact-generated anomalies, we consider the radial profiles of the density and the bulk sound speed taken at the center of the impact basin and at a certain distance from it as derived from the final datasets of the model runs. Judging from the snapshots in Fig.~\ref{fig:mars-snap}, the anomalies in these two geophysical observables may be expected to differ in both magnitude and spatial extent for each impactor type, but the question is whether and under which conditions the differences would be detectable. The off-center profile is taken at a distance from the center that is just beyond the margin of the anomaly of the S-type impactor, which is always the smallest; this selection is meant to ensure that for the S-type case, there is no signal, whereas for the other ones there should still be a signature of the anomaly. For the small impact, this distance ($\sim 10^\circ$) is easily found due to the sharp boundary of the anomaly, but in the large-impact models, we chose instead a location at $\sim 40^\circ$ from the center of the crater, where the anomaly of the S-type impactor is already much weaker. The composition ($f$) profiles in Fig.~\ref{fig:mars_crhovB-z}a and d show that the anomaly is defined by an up to about three times stronger depletion of the mantle and that it is larger in the models with C-, LPC-, or TNO-type impactors. This stronger depletion translates into a lower density $\varrho$ of the anomaly, but the density deficit in the shallow sublithospheric mantle is similar in all cases, namely mostly on the order of 20 to 30\,kg/m$^3$ relative to the mean of the impact-free model (Fig.~\ref{fig:mars_crhovB-z}b,e). The negative density anomalies of the C-, LPC-, or TNO-type events, however, reach deeper, but only in the large impacts is the vertical extent of a distinct low-density volume larger by several tens or even more than a hundred kilometers such that it may be detectable in gravity data. A peculiar feature of the large TNO-type impact is the large eclogitized crustal root, which is visible as a high-density anomaly between ca.~120 and 140\,km and is expected to founder at some point, as well as eclogitic cumulates at the CMB. The bulk sound speed ($v_\mathrm{B}=\sqrt{K_S/\varrho}$) anomaly is also identifiable in all cases, but does not exceed 0.05\,km/s (Fig.~\ref{fig:mars_crhovB-z}c,f), and the different anomalies are not sufficiently distinct from each other to offer a realistic possibility of distinguishing them seismically. In the (admittedly extreme) case L-TNO, it must be noted that the anomaly ends up encompassing the entire planet, forming a more or less uniform layer whose irregularities are due to later secondary processes; such a layer would obviously not appear as an anomaly any more.
\subsection{Moon}
\afterpage{\clearpage}
\begin{figure}
\centering
\includegraphics[width=\textwidth]{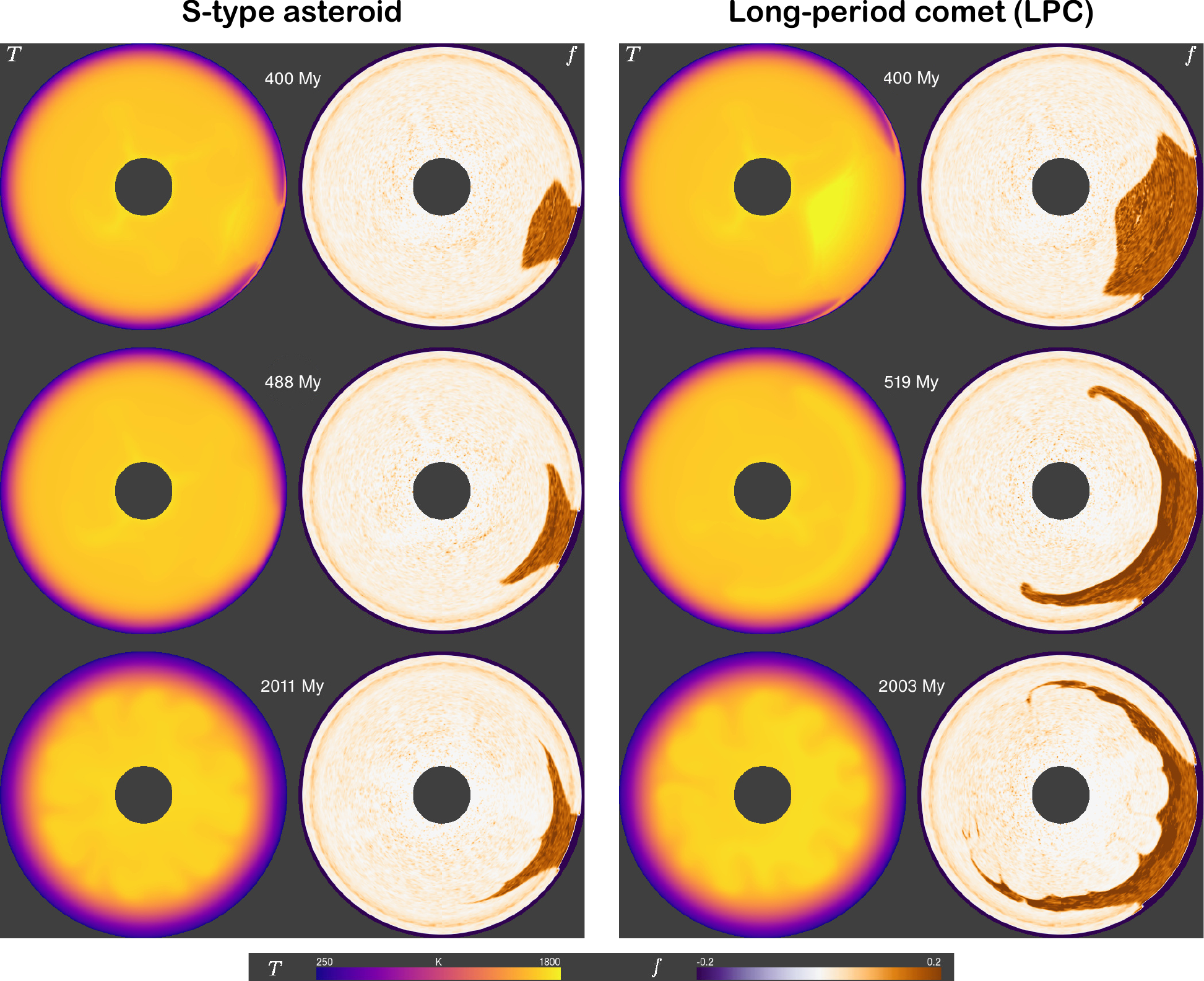}
\caption{Temperature ($T$) and depletion ($f$) of models with isocrater impacts on the Moon by an S-type asteroid and by a long-period comet. The impacts produce a basin with a diameter of $D_\mathrm{f}=1259$\,km, slightly larger than Mare Imbrium; the S-type asteroid had the same diameter as the one that produced the Hematite basin on Mars ($D_\mathrm{imp}=185$\,km), the comet had a diameter of $D_\mathrm{imp}=152$\,km. The top row shows the state directly after the impact, the bottom the state at 2\,Gyr, several hundreds of millions of years after the final geometry of the anomaly was essentially established. In the middle row, one can see how the anomaly rises after the impact and spreads beneath the lithosphere, attracting a preexisting plume (S-type) or triggering one at the CMB (LPC-type). The colorbars are clipped for clarity.\label{fig:imbrium}}
\end{figure}
\begin{figure}
\includegraphics[width=\textwidth]{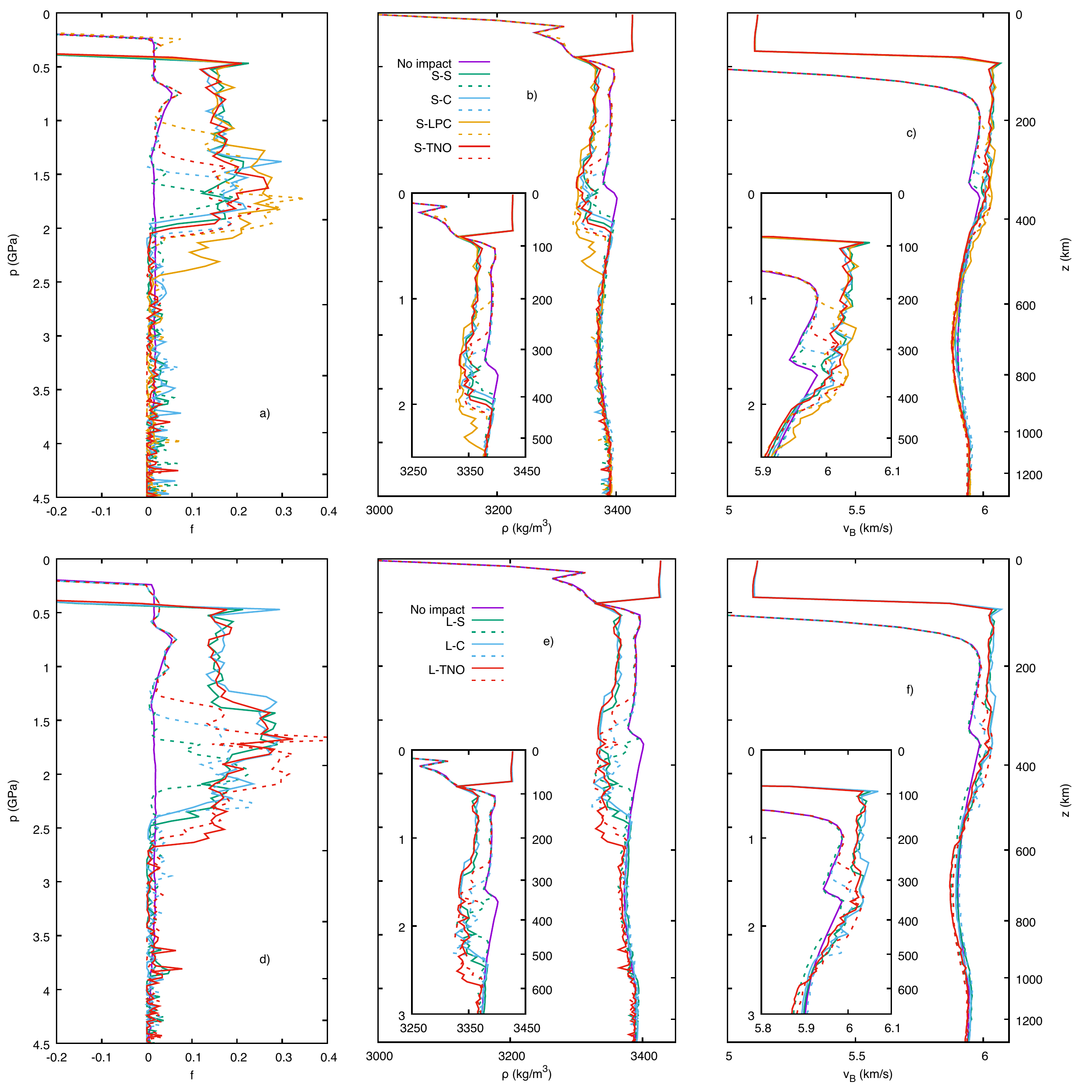}
\caption{Present-day compositional (a,d), density (b,e), and bulk sound speed (c,f) profiles for the small (Mare Imbrium, upper row) and large (South Pole--Aitken, lower row) impact models on the Moon. The ``no impact'' profiles are global lateral averages, the other profiles are local depth profiles taken at the impact site (solid lines) and at a more remote location in the region of the impact (dashed lines; Mare Imbrium: $\sim 40^\circ$, South Pole--Aitken: $\sim 90^\circ$).\label{fig:moon_crhovB-z}}
\end{figure}
The Moon is unusual for its very small core, whose surface does not provide space for a large number of mantle plumes to form. The relatively great depth of the mantle prevents even large impacts from affecting directly its entire depth range in spite of the Moon being a fairly small body. The plumes from the CMB rise sluggishly due to the small $g$ and largely dissipate before reaching the base of the lithosphere, which later becomes unstable and drives convection by cold downwellings; $v_\mathrm{rms}$ is several times smaller than in the Mars models (Fig.~\ref{fig:vrms}). The anomalies caused by large impacts trigger strong upwellings that attract plumes from the surroundings or draw new ones from the CMB behind them as they ascend (Fig.~\ref{fig:imbrium}); these plumes grow at the expense of other, more remote ones. The anomalies spread out over a much larger area than corresponds to the impact crater and introduce a significant degree-1 component into the global convection pattern. As one consequence of this pattern, the cold lithospheric downwellings do not start simultaneously everywhere but begin at the antipode of the impact site and then form successively closer to it. For the small isocrater set, which produces a basin approximately of the size of Mare Imbrium, the pattern is the same, albeit more locally confined. Convection in the lunar mantle is very sluggish, and although impacts do produce substantial thermal and compositional disturbances, they invigorate it only for a short time, i.e., usually on the order of tens of millions of years. After that, the resulting compositional anomalies are ``frozen in'' in a way similar as on Mars, but with much less long-term modification due to the smaller convective vigor in the lunar mantle (Fig.~\ref{fig:imbrium}). Contrary to Mars, we find that the general pattern of dynamical response is the same for all impacts of both sizes considered, i.e., the vigor of convection and the mean mantle temperature are increased by the impact.\par
\afterpage{\clearpage}
We can calculate the same set of synthetic observables as for Mars from these models and make a comparison based on vertical profiles through the center of the impact and at a certain distance. This time, we took the off-center profiles at locations at sites that would include the outer fringe of the anomaly even in the S-type case, specifically, at 40\textdegree\ from the center in the small-impact (Mare Imbrium-size) set and at 90\textdegree\ in the large-impact (South Pole--Aitken-size) set (Fig.~\ref{fig:moon_crhovB-z}). As in the Mars models, the anomalies have similar magnitudes but can, at least in principle, be distinguished by their spatial extent, as the low-density region spans different depth ranges in each case. However, the actual detection would amount to identifying a layer with a density only 20 to 40\,kg/m$^3$ lower than the background that thins to only a few tens of kilometers toward the outer fringes of the anomaly. As in the martian case, the seismic anomaly, although existent, has a small amplitude and will only be identifiable with a regional network.
\subsection{Venus}
\begin{figure}[!t]
\includegraphics[width=\textwidth]{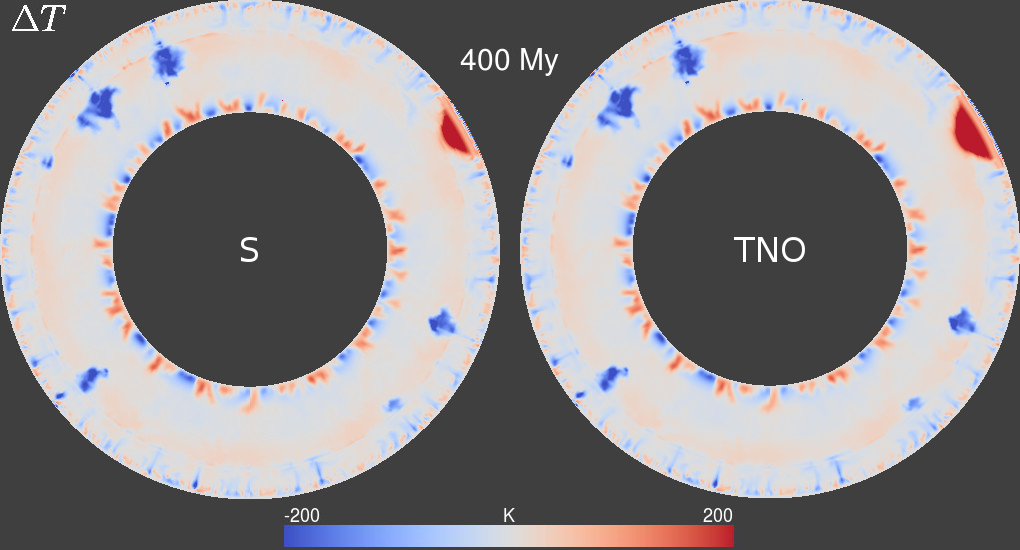}
\caption{Temperature anomaly $T(r,\vartheta)-T_\mathrm{ave}(r)$ of the small Venus isocrater models with an S-type impactor (left, $D_\mathrm{imp}=185$\,km) and a TNO-type impactor (right, $D_\mathrm{imp}=188$\,km) just after the impact. The final crater is 1257\,km in diameter, virtually the same size as the lunar one in Fig.~\ref{fig:imbrium} and 4.6 times larger than Mead, the largest crater on Venus in existence today. Note that the impactors have almost the same size; the larger anomaly of the TNO-type impactor is due to its higher velocity. Still, the anomalies of both impacts barely reach into the lower mantle. The cold downwellings (blue) consist of cold lithospheric material including eclogitized crust. The colorbar is clipped for clarity.\label{fig:venus}}
\end{figure}
In Venus, the mantle-to-core ratio is comparable to Mars, but contrary to the latter, Venus is much larger and thus has a massive perovskite+ferropericlase lower mantle, and the phase transition between the upper and the lower mantle acts as a partial barrier to mantle currents. For this reason, the primary effects even of large impacts on convection dynamics are mostly confined to the upper mantle, except for events in which the diameter of the impactor exceeds $\sim 300$\,km. The higher lithostatic pressure gradient narrows the depth range susceptible to melting, but this limitation is partly offset by the hot lithosphere, which is considerably closer to the melting point than in all other terrestrial bodies of our Solar System. Any impactor large enough to survive the encounter with the dense atmosphere, which is the case for all impactors considered here, therefore has to expend less of its energy on heating the target to the melting point and can, at least in principle, produce more melt. The steep pressure gradient also shifts phase boundaries to shallower depths than in smaller, less massive bodies, in particular the plagioclase--garnet transition of the basaltic crust. For this reason, crustal delamination is common even in the impact-free model and may in principle affect the upper crust in strong events, which would be in agreement with the observation that Venus's surface is geologically young and has been resurfaced on a global scale. In our models, the mantle of Venus experiences an initial heating phase similar to Mars due to the more intense radioactive heat production, but unlike the other planets, it does not cool subsequently because heat loss to the hot atmosphere is very inefficient; this particular feature of venusian thermal evolution is consistent with the findings of earlier studies \citep[e.g.,][]{ArTa12,Gill:etal16}.\par
As a consequence of the hot lithosphere, the shallow basalt--eclogite transition and the large size of the planet, the response to impacts on Venus is somewhat different from other, smaller planets. An impact does produce a regional upwelling, but has no immediate effect on the lower mantle in that it does not trigger a plume from the CMB. Quite to the contrary, the massive production of crust at the impact site soon lets the crust thicken into the stability field of eclogite, such that a major part of it delaminates. Delamination occurs on a much larger scale than on other planets and produces an avalanche-like flow of relatively cool, eclogite-rich material into the lower mantle; such downwellings appear in the temperature anomaly image Fig.~\ref{fig:venus} as dark blue regions. This mass flux induces a return flow of hot, pristine material into the upper mantle, often in the form of plume-like pulses in regions adjacent to the impact site. The vigorous convection including delamination of the crust, both in general and regionally in the aftermath of an impact, has the same effect on the mantle as on the surface: it quickly obliterates coherent, geophysically identifiable traces of impact-generated anomalies, leaving no hope to detect vestiges of ancient basin-forming impacts in the modern lithosphere of Venus. The enhanced delamination of relatively cool lithospheric material causes models with impacts to have mean mantle temperatures that generally lie a few tens of kelvins lower than that of the impact-free model, except for a short period of a few tens of millions of years directly after the impact. This thermal effect and the fact that the impacts affect a smaller fraction of the mantle are probably the cause for a reduction of $v_\mathrm{rms}$ in the aftermath of an impact (Fig.~\ref{fig:vrms}).
\subsection{Mercury}
\begin{figure}
‚\includegraphics[width=\textwidth]{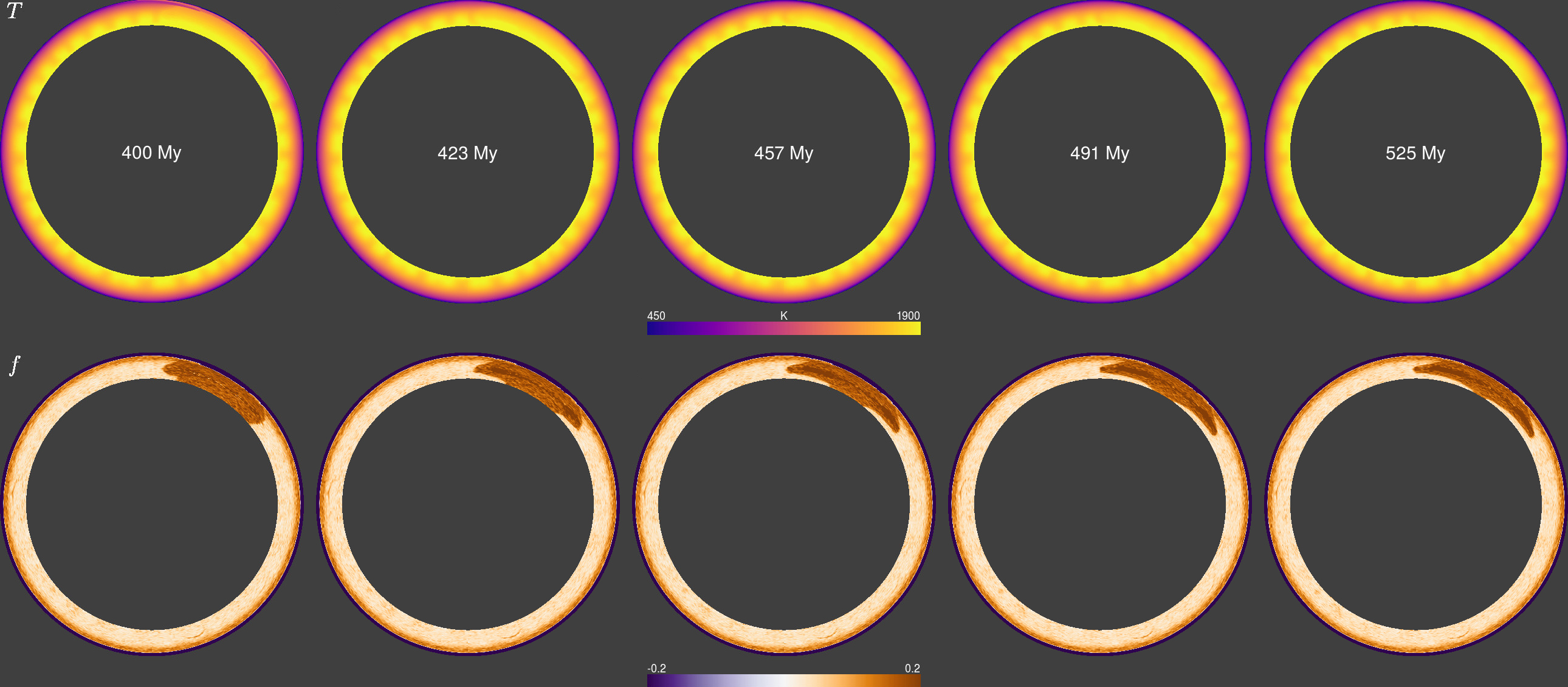}
\caption{Evolution of the temperature (top) and the compositional field (melting degree, bottom) of Mercury after the impact of an S-type asteroid that forms a basin of approximately the size of Caloris (model S-S). The impact occurred at 400\,My at the one o'clock position of the cross-section. The associated anomaly does not spread much further than shown in the last panel and stalls permanently at around 850\,My.}\label{fig:mercury}
\end{figure}

Structurally, Mercury is in some sense the opposite of the Moon, as it has a very large core and a thin mantle. As a consequence, large impacts shock a major fraction of, or even the entire, depth range of the mantle and induce melting in a large part of it. Melt production to great depth is further facilitated by the relatively flat lithostatic gradient in the mantle, although this is partly offset by the unusually high solidus of the Fe-poor peridotite (see Fig.~\ref{fig:solidi}). The small depth of the mantle strongly diminishes its Rayleigh number, causes $v_\mathrm{rms}$ to be more than an order of magnitude lower than that of the Moon (Fig.~\ref{fig:vrms}), and suppresses the formation of convection cells with large lateral extent in favor of a small-scale pattern with many cells. As a consequence, any heterogeneities that may form in the mantle are prone to be trapped within the realm of one or a few of those numerous small cells that hinder lateral transport over long distances, because the effective path traversed by any volume is much longer and the flow velocities are low. Thus, even the massive anomalies produced by basin-forming impacts are not transported much beyond the region where they were created: in our two-dimensional models, the pre-impact mantle has almost three dozen convection cells, but the impacts affect only between four and six of them directly and at most one or two more indirectly in their aftermath (Fig.~\ref{fig:mercury}). The density structure is relatively homogeneous throughout the affected region, which is partly due to the fact that the post-impact temperature in its supersolidus part is set to a value just above the solidus to account for the consumption of latent heat by melting. This region with little lateral density variation spans essentially the entire height of the convecting layer and inhibits long-lived vigorous post-impact convection in the impact zone, because cooler mantle material from the surroundings cannot penetrate beneath it easily in large amounts. As can be seen especially in the $f$ field in Fig.~\ref{fig:mercury}, low-$f$ material from the surroundings of the impact site, which is marked by a massive high-$f$ region touching the CMB, takes almost 150\,My to separate the anomaly fully from the core. The effects of the impacts we considered on global dynamical characteristics are therefore very minor and vanish within a few hundred million years, but the anomalies in the mantle are preserved permanently and with little modification, as the small-scale structure of convection generally inhibits their lateral spreading. Although impact-generated compositional anomalies do spread to some extent, the differences between different impactor types are small and will be hard to detect.

\section{Discussion}
The scaling laws summarized in Sect.~\ref{sect:theory} lead to the conclusion that very large impact basins, i.e., significantly larger than the martian Hematite basin, cannot have been produced by comets, as such comets would exceed the 150\,km in diameter we have set as an upper limit of probable comet sizes. Up to that size, however, very large long-period cometary impactors would in principle be conceivable on Mars and on the Moon, but on Venus and Mercury, the required size of LPC impactors would lie above the limit. TNO-type impactors, however, would be possible on all targets, as several of the known TNOs in the Kuiper belt have the required diameter on the order of a few hundred kilometers, and although their penetration depth would not be very different from that of S-type impactors, the isobaric cores would be substantially larger. The effects caused by C-type asteroids are in many respects similar to those of S-type asteroids, in spite of their lower density. In brief, there is a fairly clear distinction between asteroidal and non-asteroidal impactors in terms of their effects on the interior, which are substantially stronger in the latter case. It is worth recalling that while the TNO-type impactors considered here are an extreme and hypothetical example, there are real objects in the Solar System like centaurs that could take an intermediate place between these subsets and produce effects in the interior that are larger than those of asteroidal impactors. At any rate, the size--frequency distributions of comets and other non-asteroidal impactors, as far as they are known at all, suggest that bodies of the size required even for the small basins such as Hematite are extremely rare or nonexistent, making impacts of such bodies relatively unlikely. This conclusion is in agreement with independent arguments for a predominantly asteroidal origin of the lunar craters and extends, at least in principle, to other bodies in the inner Solar System. It should hold at least qualitatively even if one admits that the scaling laws used here to derive the impactor size and other impact characteristics may not yield a very accurate representation of the characteristics of a basin-forming impact and its ensuing crater \citep{Bier:etal11}. Concerns regarding the scaling laws will be stronger the larger the impact and the smaller the target is, for instance, because curvature effects begin to become important. At least for the Moon, \citet{Pott:etal15} found that target curvature must be accounted for in numerical impact models if $D_\mathrm{tr}/(2R_\mathrm{P})\gtrsim 0.3$; this would apply to both lunar impacts considered here, and if the criterion also holds for other planets, it would also apply to the Utopia event on Mars. The conclusion about the low probability of non-asteroidal impacts does not imply, however, that smaller craters, whose impact effects are restricted to the lithosphere, could not have been caused by cometary impactors, and we suggest that the geophysical signature of such smaller craters should be considered with this possibility in mind.\par
\citet{RuBr17c} found that giant impacts produce compositional anomalies in the martian mantle that are self-preserving due to their own buoyancy, which anchors them within the growing lithosphere, and that such anomalies should leave their signature, for instance, in the gravity field. The results from this study confirm this result for bodies of the size of Mars or smaller, which can form thick, stable crusts. It is reassuring that for all differences in model definition and setup, the models of very large basin-forming impacts on Mars, the Moon, and Mercury by \citet{Ar-HaGh11} indicate a trend of self-stabilization and ``freezing'' of the impact-generated anomalous mantle structure similar to ours, although they do not cover the long-term evolution to the extent we do. On Venus, however, the extensive delamination of crust in our models largely disrupts and obliterates such anomalies. The stability of such anomalies would then turn out to depend on the lithostatic gradient and the melt production of the planet, i.e., ultimately on its mass and size. On those planets whose dynamical evolution preserves impact-generated anomalies, gravimetry offers probably the most promising approach for detecting them and putting constraints on the type of impactor, although the non-uniqueness of gravity field signatures will make additional evidence necessary. Furthermore, the small amplitude of the mantle density anomalies and the decreasing thickness of the layer at great distances from the point of impact, which are both crucial to estimating its spatial extent, will require a good crustal thickness model and make detection difficult.\par
Moreover, the set of small S-type asteroid impactors, which are all of the same size, highlights the fact that impactor size alone is not the only, and not even necessarily the most important parameter that controls the effects of an impact. For a given impactor type, the effective velocity of the projectile, its size relative to the target, and the global structure of the target, specifically the mantle-to-core ratio, are at least as important. For example, very large impacts on Mars have been found to trigger long-lived plumes at the impact site or attract nearby plumes to it under certain circumstances \citep[e.g.,][]{Rees:etal02,Rees:etal04,RuBr17c}. On Venus, by contrast, we did not observe a similar tendency, although the impacts did focus upwellings and volcanism for some time towards the impact site. The more vigorous convection, the delamination of the crust, and also the larger separation between the shocked mantle volume and the CMB seem to prevent such a stabilization of mantle flow patterns. This difference was already noted by \citet{Gill:etal16} in a series of models of Venus that did not couple the rheology to the water content but implemented a sophisticated coupling of the interior with the changing atmospheric conditions; thus it seems to be a fairly robust feature. The importance of the proximity of the shocked region to the CMB can be qualitatively assessed by comparing our Mercury models with other models from the literature. The models of the Caloris impact by \citet{JHRoBa12} and \citet{Pado:etal17} show a dynamical behavior beneath the impact that is more similar to that of a martian mantle in that a normal plume forms and rises from the CMB. The various differences in model setup and impact parameters and geometry result in a smaller size of their heated volume and a greater distance of the isobaric core from the CMB that allows the thermal anomaly to be transported and obliterated more easily than in our models.\par
There are of course some caveats in the interpretation of the results of our models. One of the most important ones is the problem of melt production by the impact itself and the fate of that melt. We have assumed that all of the melt that is produced in the impact and is, in principle, extractable is indeed extracted and forms crust. This assumption is responsible for substantial impact-related crust production and represents an endmember case that puts an upper bound on both the post-impact crustal thickness and the magnitude of the interior compositional anomaly. Fitting impact melt volumes from all models to a power law of the form of Eq.~\ref{eq:Mmelt} with $\mu=0.58$ fixed, we also find that a decent fit can be obtained with $D_\mathrm{imp}^{2.48}(\varrho_\mathrm{imp}/\varrho)^{0.16}$. However, one must keep in mind that the melting model is not calibrated for impact shock melting and that this fit was derived from two-dimensional models assuming axial symmetry about the radial axis of the anomaly. The relation should therefore be considered a tentative observation and not be used for quantitative estimates, even though independent work based on impact simulations by \citet{Mari:etal11} and \citet{MaWu17} points to a comparable reduction of the exponent 3 of $D_\mathrm{imp}$ in the traditional scaling law Eq.~\ref{eq:Mmelt}. At any rate, a close agreement with Eq.~\ref{eq:Mmelt} should probably not be expected given that the equation was derived for homogeneous materials. The other extreme would be to assume that no impact-generated melt is extracted, i.e., that the melt produced by the impact shock itself recrystallizes in the interior, possibly in situ, and that all post-impact crust formation is due to longer-term, secondary volcanism. An approach of that sort, which has been taken by \citet{Pado:etal17} in models of Mercury, leads to a thinner post-impact crust and would result in a much smaller compositional anomaly, which would also be even harder to observe. The fate of the impact melt is an unresolved issue and will need further work in the future.\par
Another aspect are the uncertainties in the impact parameters. The velocities and densities of impactors are not tightly constrained but span a large range. Furthermore, as mentioned above, we assume an impact angle of 45\textdegree, but in reality, any other angle is also possible. In order to avoid the addition of $\theta$ as another free parameter to the analysis, one may therefore choose to read the equations and figures as applicable to events with the same $\theta$ or, more appropriately, as comparing the vertical components of $v_\mathrm{imp}$. The latter interpretation expands the range of possible velocities even further, but it has the advantage of being more general, as it allows, in principle, the comparison of impacts with different $\theta$. Moreover, one can also use velocity--density ratio contour plots such as Figs.~\ref{fig:Dimpzic}--\ref{fig:Vmelt} for comparing impacts of S-type asteroids with different $\theta$, as these would lie on the vertical line through $\varrho_\mathrm{imp}/\varrho_\mathrm{S}=1$. Most of the equations and figures would remain unchanged if a different reference angle than 45\textdegree\ had been chosen, because the $\sin\theta$ terms that are now implicit in the impactor velocity cancel out in the ratios; however, care must be taken in those formulae in which velocities do not appear only as ratios, i.e., Eqs.~\ref{eq:rphirat} and \ref{eq:rinflrat}, where $b$ and $n$ are functions of the velocity. The most important aspect of oblique impacts that cannot be captured at all by this approach is the asymmetry of the deformation and pressure pattern. Especially for $\theta\lesssim 30^\circ$, the asymmetry may often be visible enough in the shape of the crater to facilitate some constraints on the vertical component to adjust the parameters in a model. The asymmetry in the subsurface post-impact temperature distribution, however, is not included in the numerical models. We expect it to have some influence on local post-impact convective and melt production patterns, but they are probably minor and are not well resolved in global-scale models. Future improvements in the representation of the impact shock pressure distribution may remove this shortcoming. When reading Figs.~\ref{fig:Dimpzic}--\ref{fig:Vmelt}, it must therefore always be kept in mind that the colored dots that mark impactor--target combinations are averages and come with large and overlapping errorbars.

\section{Conclusions}
Impactors of different types and with different impact parameters may produce a final crater of the same diameter on any given target if certain conditions are met. On the basis of scaling laws, we derived the condition that defines such ``isocrater impacts'' as well as relations that describe how the different types of impactors of such a set differ in terms of their effects on the interior of the target planet. Most importantly, non-asteroidal impactors produce substantially greater anomalies than asteroidal impactors in any set of isocrater impacts. These differences may in principle help to identify the properties of the projectile that formed a given crater and thus constrain the character and abundance of different possible impactor populations relevant for the early inner Solar System. If extremely large geophysical anomalies associated with large impact basins as expected from non-asteroidal impactors turn out to be absent, that would imply that cometary or similar impactors of the required size are extremely rare or do not exist.\par
The specific response to a given impactor depends not only on the parameters of the projectile, however, but just as much on those of the target. The very different dynamics of the four bodies considered show that the anomalies produced by the impact evolve very differently depending on the vigor of convection and the relative size of core and mantle: in Mars, anomalies tend to spread at the base of the lithosphere and are eventually integrated into it; in the Moon, anomalies stall much more quickly and tend to be preserved as structures with a substantial depth extent, as opposed to the thin anomaly remnants in Mars; in Venus, anomalies are quickly obliterated by the generally vigorous convection and in particular by crustal delamination; in Mercury, almost the entire depth range of the mantle may be affected, but the small scale of convective structures and efficient cooling inhibit large-scale lateral spread and prevent the development of very marked differences in the signatures of different impactor types. Given the destructive effect of crustal recycling on lithospheric anomalies in Venus, one should expect that comparable anomalies in the Earth's lithosphere have suffered a similar fate due to subduction or rifting, unless they formed in stable continental lithosphere.

\section*{Acknowledgments}
We thank Roland Wagner and Ekkehard Kührt for helpful discussions and advice. The constructive comments by Gregor Golabek and an anonymous referee are very much appreciated. TR was supported by DFG grant Ru 1839/1-1, with additional funding from the Helmholtz Alliance project ``Planetary evolution and life'' and the DFG programme SFB-TRR~170. DB was supported by SFB-TRR~170. This is TRR~170 publication no.~35. The numerical calculations were carried out on the computational resource ForHLR~II at the Steinbuch Centre for Computing, Karlsruhe Institute of Technology, funded by the Ministry of Science, Research and the Arts Baden-Württemberg and DFG. This research has made use of data and/or services provided by the International Astronomical Union's Minor Planet Center.

\appendix
\setcounter{table}{0}
\setcounter{figure}{0}
\section{TNO-type impactor velocities}\label{app:vimpTNO}
First of all, we shall emphasize that this is a hypothetical scenario designed to produce an extreme impact event with a relatively straightforward orbital dynamics model. We have therefore deliberately chosen not to model impacts of objects like centaurs or similar bodies whose origins lie similarly far away from the inner Solar System and whose complicated orbital histories have brought them closer and may in some cases turn them into potential impactors; such objects have been shown to have a non-zero collision probability with terrestrial planets \citep[e.g.,][]{IpMa04} and have been linked to mass extinctions on Earth \citep[e.g.,][]{Napier15}.\par
For simplicity, we assume that the perturbing giant planet that nudges the TNO into its highly eccentric orbit does not accelerate it significantly so that the orbit of the TNO can still be treated like an elliptical cometary Keplerian orbit with semimajor axis $a_\mathrm{c}$, numerical eccentricity $e$, and inclination $i$ relative to the orbit of its target. Many of the known present-day TNOs have inclinations of only a few degrees \citep[e.g.,][]{dePaLi10} (also see \url{http://www.minorplanetcenter.net/iau/lists/TNOs.html}), and orbital evolution models of Kuiper belt objects suggest that the gravitational effects acting on these objects preferentially remove low-inclination objects from that population, at least from the inner Kuiper belt \citep{Kuch:etal02}. Therefore we assume that our hypothetical TNO-type impactor was transferred by such processes onto its final high-eccentricity orbit that brought it into the inner Solar System and that this body had a low inclination. A condition for a collision to be possible at all is that $e$ exceeds a certain minimum value. In polar coordinates $r,\varphi$ with the origin at the center of the Sun, the target planet's orbit, which we assume to be circular, is defined by its radius $a_\mathrm{p}$, and the elliptical orbit of the impactor with the Sun in one focal point is
\begin{equation}
r_\mathrm{c}=a_\mathrm{c}\frac{1-e^2}{1+e\cos\varphi}.
\end{equation}
Equating both curves, i.e., $r_\mathrm{c}=a_\mathrm{p}$, gives the coordinate $\varphi$ of the intersections:
\begin{equation}
\varphi=\arccos\left\lbrace\frac{1}{e}\left[\frac{a_\mathrm{c}}{a_\mathrm{p}}(1-e^2)-1\right]\right\rbrace.
\end{equation}
The argument of the arc cosine must lie in the interval $[-1,1]$, for given $a_\mathrm{c}$ and $a_\mathrm{p}$. Furthermore, it is extremely improbable that a projectile entering the high-eccentricity elliptic orbit hits its target during its first approach, although it is in principle possible; therefore we permit it to pass its orbit several times and thus have to prevent it from approaching the Sun closer than a certain distance, which for convenience we express here in terms of the semilatus rectum being the $c$-fold of the Sun's radius $R_\mathrm{S}$ (with some suitably defined constant $c$), and being destroyed. These conditions constrain the eccentricity of candidate impactors with aphelion distance $d_\mathrm{ap,c}$ as
\begin{equation}
\begin{aligned}
1-\frac{a_\mathrm{p}}{a_\mathrm{c}}&\leq e\leq \sqrt{1-\frac{cR_\mathrm{s}}{a_\mathrm{c}}}\\
\Leftrightarrow \frac{1-\frac{a_\mathrm{p}}{d_\mathrm{ap,c}}}{1+\frac{a_\mathrm{p}}{d_\mathrm{ap,c}}}&\leq e\leq 1-\frac{cR_\mathrm{s}}{d_\mathrm{ap,c}}
\end{aligned}
\end{equation}
Hence, for the putative TNO-type impactors, for which $a_\mathrm{c}\gg a_\mathrm{p}$, there would be only a quite narrow range of eccentricities that would allow them to hit a target in the inner Solar System. We choose $c=10$ and assume that $d_\mathrm{ap,c}=40$\,AU, which puts the lower limit on $e$ at 0.98, 0.9645, 0.951, and 0.9266 for Mercury, Venus, the Earth--Moon system, and Mars, respectively, and the upper limit almost at 1.\par
The velocity of the impactor in a stationary reference frame is composed of two contributions, namely, its movement in the gravitational potential of the central star of mass $M_\mathrm{s}$ and its acceleration by the gravitational attraction of the target body with mass $M_\mathrm{P}$ and radius $R_\mathrm{P}$. The first contribution is governed by the vis viva equation, by which the velocity of the impactor at a distance $r$ from the central star is
\begin{equation}
v_\mathrm{c}=\sqrt{G_0 M_\mathrm{s}\left(\frac{2}{r}-\frac{1}{a_\mathrm{c}}\right)},
\end{equation}
and its period according to Kepler's Third Law is $P=2\pi\sqrt{a_\mathrm{c}^3/(G_0 M_\mathrm{s})}$ \citep[e.g.,][]{dePaLi10}; $G_0=\dpow{6.67408}{-11}$\,m$^3$/(kg\,s$^2$) is Newton's constant of gravitation \citep{CODATA2014}. An analogous expression holds for the orbital velocity of the target planet. The relevant velocity with respect to the impact, however, is the relative velocity between impactor and target at the point of intersection of their orbits. \citet[Eq.~21]{Wetherill67} has generalized the derivation by \citet{Opik51} for the collision of two bodies on elliptical orbits, and his solution would be the most accurate. However, as we are considering a hypothetical object and concentrate on Solar System target bodies with nearly circular orbits, the simpler solution from \citet{Opik51} or \citet{HuWi00},
\begin{equation}
v_\mathrm{cp}=\sqrt{G_0 M_\mathrm{s}\left(\frac{3}{a_\mathrm{p}}-2\cos i \sqrt{\frac{a_\mathrm{c}(1-e^2)}{a_\mathrm{p}^3}}-\frac{1}{a_\mathrm{c}}\right)}=
\sqrt{G_0 M_\mathrm{s}\left(\frac{3}{a_\mathrm{p}}-2\cos i \sqrt{\frac{d_\mathrm{ap,c}(1-e)}{a_\mathrm{p}^3}}-\frac{1+e}{d_\mathrm{ap,c}}\right)},
\end{equation}
is sufficient. As we assumed that the impactor has a small inclination, we set $i=0$ in the following, i.e., the target and impactor orbits are coplanar. The second contribution is in fact the escape velocity of the target planet, $v_\mathrm{esc}=\sqrt{2G_0 M_\mathrm{P}/R_\mathrm{P}}$, and the mean impact velocity is the geometric mean of both velocities:
\begin{equation}
v_\mathrm{imp}=\sqrt{v_\mathrm{cp}^2+v_\mathrm{esc}^2}
\end{equation}
\citep[e.g.,][]{HuWi00}. The parameters for the inner Solar System objects are listed in Table~\ref{tab:TNOtab}, and the velocities for the target bodies using the mean $e$ in the permissible range of each are 46.38, 35.67, 31.16, 29.23, and 24.01\,km/s for Mercury, Venus, Earth, Moon, and Mars, respectively; for Earth we used the total mass of the Earth--Moon system and the radius of the Earth, for the Moon we estimated $v_\mathrm{esc}$ from the lunar mass and radius combined with the Earth's mass at a distance of 182\,000\,km as an approximate value for the radius of the lunar orbit in the first few hundred millions of years of the Solar System \citep{BiRa99}.
\begin{table}
\caption{Mass, mean radius, and mean distance from the Sun for the major bodies in the inner Solar System. $a_\mathrm{p}$ are the values for the epoch J2000 as provided at \texttt{http://ssd.jpl.nasa.gov/txt/p\_elem\_t2.txt}; for the Earth and the Moon, we use the value of their barycenter.\label{tab:TNOtab}}
\centering
\begin{tabular}{lccc}\toprule
&$M_\mathrm{s}$, $M_\mathrm{P}$ (kg)&$R_\mathrm{s}$, $R_\mathrm{P}$ (km)&$a_\mathrm{p}$ (AU)\\\midrule
Sun&\dpow{1.98847541595}{30} \textsuperscript{c}&696342\textsuperscript{b}&--\\
Mercury&\dpow{3.30111}{23} \textsuperscript{g}&2439.36\textsuperscript{h}&0.38709843\\
Venus&\dpow{4.867}{24} \textsuperscript{f}&6051.8\textsuperscript{a}&0.72332102\\
Earth&\dpow{5.972365}{24} \textsuperscript{j}&6371\textsuperscript{d}&1.00000018\\
Moon&\dpow{7.34603154}{22} \textsuperscript{i}&1737.1513\textsuperscript{k}&1.00000018\\
Mars&\dpow{6.41712}{23} \textsuperscript{f,e}&3389.5\textsuperscript{a}&1.52371243\\\bottomrule
\multicolumn{4}{c}{\begin{minipage}{0.6\textwidth}
{\small \textsuperscript{a}\citet{Arch:etal11}, \textsuperscript{b}\citet{Emil:etal12}, \textsuperscript{c} IAU 2012 and $G_0$ from \citet{CODATA2014}, \textsuperscript{d}\citet{IERS2010}, \textsuperscript{e}\citet{Jacobson10}, \textsuperscript{f}\citet{Kono:etal99}, \textsuperscript{g}\citet{Maza:etal14}, \textsuperscript{h}\citet{Perr:etal15}, \textsuperscript{i}\citet{PiSt09}, \textsuperscript{j}\citet{Ries:etal92}, \textsuperscript{k}\citet{DESmit:etal17}}
\end{minipage}}
\end{tabular}
\end{table}

\setcounter{table}{0}
\setcounter{figure}{0}
\section{Mantle solidi}\label{app:solidus}
A well-defined reference is the terrestrial peridotite solidus \citep{MMHirs:etal09}, whose section above 23.5\,GPa we have updated with additional newer data \citep{Andr:etal11,Fiqu:etal10,Nomu:etal14} and refitted with a modified Simon--Glatzel equation. Applying the downward shift suggested by \citet{Katz:etal03} with respect to the earlier version of the low-$p$ solidus from \citet{Hirschmann00}, which had not been used by \citet{MMHirs:etal09}, the terrestrial solidus then is
\begin{equation}
T_\mathrm{s,E}(p)=
\begin{cases}
1358.81061+132.899012p-5.1404654p^2&p\leq 10\,\mathrm{GPa}\\
2173.15+32.39(p-10)-1.092(p-10)^2&10\,\mathrm{GPa}<p\leq 23.5\,\mathrm{GPa}\\
786.966(1+4.312p)^{0.2497}&\text{else},
\end{cases}\label{eq:Ts-earth}
\end{equation}
with $p$ in GPa and $T_\mathrm{s}$ in K. A martian solidus has been constructed by \citet{RuBr17c} from experimental data:
\begin{equation}
T_\mathrm{s,M}(p)=
\begin{cases}
0.118912p^3-6.37695p^2+130.33p+1340.38&\text{for $p<23$\,GPa}\\
62.5p+975&\mathrm{else.}
\end{cases}\label{eq:Ts-mars}
\end{equation}
For the other targets, we do not know of specific data, and so we attempt to construct suitable solidi by interpolation. The Moon has a Mg\# between that of the Earth and Mars, and so we interpolate between eqs.~\ref{eq:Ts-earth} and \ref{eq:Ts-mars}, whereby the latter is corrected for the effect of alkalis as described below. The interpolated value is then again adjusted for the alkali content of the bulk silicate Moon. Venus and Mercury have higher Mg\# than the Earth, and therefore we interpolate between eq.~\ref{eq:Ts-earth} and a solidus for the iron- and alkali-free CMAS system,
\begin{equation}
T_\mathrm{s,CMAS}(p)=1477.54+139.391p-7.7545p^2+0.160258p^3,\label{eq:Ts-CMAS}
\end{equation}
which we constructed from experimental data \citep{Asah:etal98,AsOh01,GuPr96,GuPr00,Herz:etal90,KlStCONe00,LiOh02,XLiONe04a,MiPr98,
Pres:etal79}. This solidus is only constrained for pressures up to 21.5\,GPa, as no data are available for lower-mantle pressures, and eq.~\ref{eq:Ts-CMAS} should not be used for extrapolation to much higher pressures; in order to fill the need in that pressure range for Venus models, we set the solidus to $T_\mathrm{s,CMAS}(23.5)-T_\mathrm{s,E}(23.5)\approx50$\,K above eq.~\ref{eq:Ts-earth} at $p>23.5$\,GPa. Eq.~\ref{eq:Ts-CMAS} may yield slightly too high values between 21.5 and 23.5\,GPa, but this is of little practical relevance here.\par
\begin{figure}
\includegraphics[width=\textwidth]{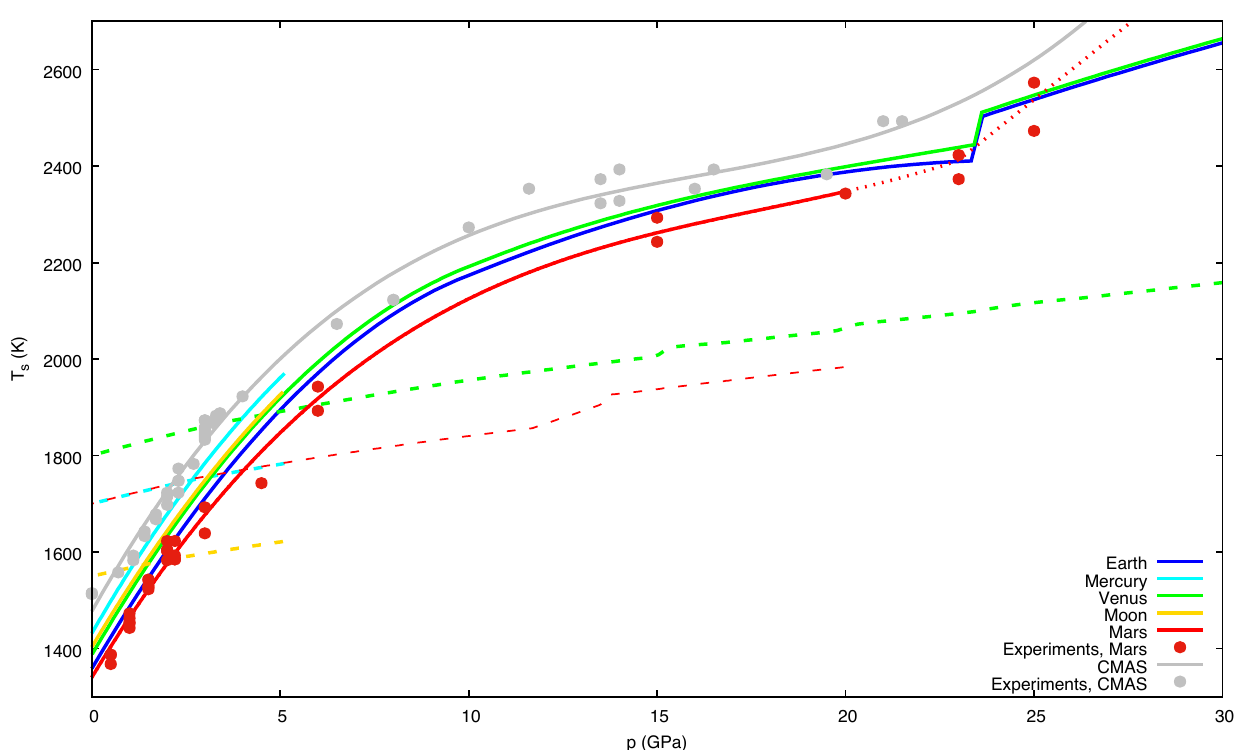}
\caption{The upper 30\,GPa of the water-free solidi of the terrestrial bodies and the synthetic CMAS system as used in this study. Also shown are the initial adiabats (dashed lines) for the four bodies considered.\label{fig:solidi}}
\end{figure}

For the correction for the effect of alkalis we considered the plots of $T_\mathrm{s}$ as a function of the combined mass fraction $X_\mathrm{NK}$ of K$_2$O and Na$_2$O plotted in Figures~4b and 5b from \citet{Hirschmann00}. Fitting each of these datasets with a linear function, we find that the slope $\mathrm{d}T_\mathrm{s}/\mathrm{d}X_\mathrm{NK}$ does not change significantly between 1.5 and 3\,GPa, and so we use the mean value of $-14.952$\,K/wt.\% (K$_2$O+Na$_2$O) for the correction at all pressures before the appearance of a perovskite+ferropericlase assemblage. For the lower mantle, no correction is applied \citep[cf.][]{WWaTa00}. The calculated solidi and experimental data used for fitting are shown in Figure~\ref{fig:solidi}.\par
The different solidi control the extent of melting in a planet at a given temperature. Models that are aimed at constraining the amount of melt and crust produced via the temperature of the interior should therefore do so in the context of the solidus appropriate for the composition if a compositional model can be reasonably constructed for a planet, rather than simply assume the solidus of a terrestrial peridotite. For instance, using the initial martian adiabat from our models together with the terrestrial solidus would let us underestimate the maximum pressure of regular melting by 0.5\,GPa, which corresponds to about 40\,km. By contrast, in Mercury, for which we used an almost identical adiabat, use of the terrestrial solidus would result in overestimating that maximum pressure by 0.9\,GPa, which translates into more than 70\,km even for the relatively low Mg\# of 0.94 we assumed. In the Moon, which has a very flat pressure gradient, the pressure overestimate of some 0.4\,GPa for our adiabat would even lead to a depth overestimate of about 90\,km.

\section*{Supplementary Material}
Supplementary material associated with this article can be found, in the online version, at \url{doi: http://dx.doi.org/10.1016/j.icarus.2018.02.005}.



\begin{thebibliography}{140}
\expandafter\ifx\csname natexlab\endcsname\relax\def\natexlab#1{#1}\fi
\providecommand{\url}[1]{\texttt{#1}}
\providecommand{\href}[2]{#2}
\providecommand{\path}[1]{#1}
\providecommand{\DOIprefix}{doi:}
\providecommand{\ArXivprefix}{arXiv:}
\providecommand{\URLprefix}{URL: }
\providecommand{\Pubmedprefix}{pmid:}
\providecommand{\doi}[1]{\href{http://dx.doi.org/#1}{\path{#1}}}
\providecommand{\Pubmed}[1]{\href{pmid:#1}{\path{#1}}}
\providecommand{\bibinfo}[2]{#2}
\ifx\xfnm\relax \def\xfnm[#1]{\unskip,\space#1}\fi
\bibitem[{Abramov et~al.(2012)Abramov, Wong and Kring}]{Abra:etal12}
\bibinfo{author}{Abramov, O.}, \bibinfo{author}{Wong, S.M.},
  \bibinfo{author}{Kring, D.A.}, \bibinfo{year}{2012}.
\newblock \bibinfo{title}{Differential melt scaling for oblique impacts on
  terrestrial planets}.
\newblock \bibinfo{journal}{Icarus} \bibinfo{volume}{218},
  \bibinfo{pages}{906--916}.
\newblock \DOIprefix\doi{10.1016/j.icarus.2011.12.022}.
\bibitem[{Andrault et~al.(2011)Andrault, Bolfan-Casanova, Lo~Nigro, Bouhifd,
  Garbarino and Mezouar}]{Andr:etal11}
\bibinfo{author}{Andrault, D.}, \bibinfo{author}{Bolfan-Casanova, N.},
  \bibinfo{author}{Lo~Nigro, G.}, \bibinfo{author}{Bouhifd, M.A.},
  \bibinfo{author}{Garbarino, G.}, \bibinfo{author}{Mezouar, M.},
  \bibinfo{year}{2011}.
\newblock \bibinfo{title}{{Solidus and liquidus profiles of chondritic mantle:
  Implication for melting of the Earth across its history}}.
\newblock \bibinfo{journal}{Earth Planet. Sci. Lett.} \bibinfo{volume}{304},
  \bibinfo{pages}{251--259}.
\newblock \DOIprefix\doi{10.1016/j.epsl.2011.02.006}.
\bibitem[{Archinal et~al.(2011)Archinal, A'Hearn, Bowell, Conrad, Consolmagno,
  Courtin, Fukushima, Hestroffer, Hilton, Krasinsky, Neumann, Oberst,
  Seidelmann, Stooke, Tholen, Thomas and Williams}]{Arch:etal11}
\bibinfo{author}{Archinal, B.A.}, \bibinfo{author}{A'Hearn, M.F.A.},
  \bibinfo{author}{Bowell, E.}, \bibinfo{author}{Conrad, A.},
  \bibinfo{author}{Consolmagno, G.J.}, \bibinfo{author}{Courtin, R.},
  \bibinfo{author}{Fukushima, T.}, \bibinfo{author}{Hestroffer, D.},
  \bibinfo{author}{Hilton, J.L.}, \bibinfo{author}{Krasinsky, G.A.},
  \bibinfo{author}{Neumann, G.}, \bibinfo{author}{Oberst, J.},
  \bibinfo{author}{Seidelmann, P.K.}, \bibinfo{author}{Stooke, P.},
  \bibinfo{author}{Tholen, D.J.}, \bibinfo{author}{Thomas, P.C.},
  \bibinfo{author}{Williams, I.P.}, \bibinfo{year}{2011}.
\newblock \bibinfo{title}{{Report of the IAU Working Group on Cartographic
  Coordinates and Rotational Elements: 2009}}.
\newblock \bibinfo{journal}{Celest. Mech. Dyn. Astr.} \bibinfo{volume}{109},
  \bibinfo{pages}{101--135}.
\newblock \DOIprefix\doi{10.1007/s10569-010-9320-4}. \bibinfo{note}{erratum in
  110, 401--403}.
\bibitem[{Arkani-Hamed and Ghods(2011)}]{Ar-HaGh11}
\bibinfo{author}{Arkani-Hamed, J.}, \bibinfo{author}{Ghods, A.},
  \bibinfo{year}{2011}.
\newblock \bibinfo{title}{Could giant impacts cripple core dynamos of small
  terrestrial planets?}
\newblock \bibinfo{journal}{Icarus} \bibinfo{volume}{212},
  \bibinfo{pages}{920--934}.
\newblock \DOIprefix\doi{10.1016/j.icarus.2011.01.020}.
\bibitem[{Armann and Tackley(2012)}]{ArTa12}
\bibinfo{author}{Armann, M.}, \bibinfo{author}{Tackley, P.J.},
  \bibinfo{year}{2012}.
\newblock \bibinfo{title}{{Simulating the thermo-chemical magmatic and tectonic
  evolution of Venus' mantle and lithosphere 1. Two-dimensional models}}.
\newblock \bibinfo{journal}{J. Geophys. Res.} \bibinfo{volume}{117}.
\newblock \DOIprefix\doi{10.1029/2012JE004231}.
\bibitem[{Asahara and Ohtani(2001)}]{AsOh01}
\bibinfo{author}{Asahara, Y.}, \bibinfo{author}{Ohtani, E.},
  \bibinfo{year}{2001}.
\newblock \bibinfo{title}{{Melting relations of the hydrous primitive mantle in
  the CMAS--\chem{H}{2}O system at high pressures and temperatures, and
  implications for generation of komatiites}}.
\newblock \bibinfo{journal}{Phys. Earth Planet. Inter.} \bibinfo{volume}{125},
  \bibinfo{pages}{31--44}.
\newblock \DOIprefix\doi{10.1016/S0031-9201(01)00208-4}.
\bibitem[{Asahara et~al.(1998)Asahara, Ohtani and Suzuki}]{Asah:etal98}
\bibinfo{author}{Asahara, Y.}, \bibinfo{author}{Ohtani, E.},
  \bibinfo{author}{Suzuki, A.}, \bibinfo{year}{1998}.
\newblock \bibinfo{title}{Melting relations of hydrous and dry mantle
  compositions and the genesis of komatiites}.
\newblock \bibinfo{journal}{Geophys. Res. Lett.} \bibinfo{volume}{25},
  \bibinfo{pages}{2201--2204}.
\newblock \DOIprefix\doi{10.1029/98GL01527}.
\bibitem[{Barnes et~al.(2016)Barnes, Kring, Tart{\`e}se, Franchi, Anand and
  Russell}]{JJBarn:etal16}
\bibinfo{author}{Barnes, J.J.}, \bibinfo{author}{Kring, D.A.},
  \bibinfo{author}{Tart{\`e}se, R.}, \bibinfo{author}{Franchi, I.A.},
  \bibinfo{author}{Anand, M.}, \bibinfo{author}{Russell, S.S.},
  \bibinfo{year}{2016}.
\newblock \bibinfo{title}{{An asteroidal origin for water in the Moon}}.
\newblock \bibinfo{journal}{Nat. Comm.} \bibinfo{volume}{7}.
\newblock \DOIprefix\doi{10.1038/ncomms11684}.
\bibitem[{Bierhaus et~al.(2011)Bierhaus, W{\"u}nnemann, Elbeshausen and
  Collins}]{Bier:etal11}
\bibinfo{author}{Bierhaus, M.}, \bibinfo{author}{W{\"u}nnemann, K.},
  \bibinfo{author}{Elbeshausen, D.}, \bibinfo{author}{Collins, G.S.},
  \bibinfo{year}{2011}.
\newblock \bibinfo{title}{{Numerical modeling of basin-forming impacts:
  Implications for the heat budget of planetary interiors}}.
\newblock \bibinfo{journal}{Lunar Planet. Sci.} \bibinfo{volume}{XLII}.
\newblock \URLprefix
  \url{http://www.lpi.usra.edu/meetings/lpsc2011/pdf/2128.pdf}.
\bibitem[{Bills and Ray(1999)}]{BiRa99}
\bibinfo{author}{Bills, B.G.}, \bibinfo{author}{Ray, R.D.},
  \bibinfo{year}{1999}.
\newblock \bibinfo{title}{{Lunar orbital evolution: A synthesis of recent
  results}}.
\newblock \bibinfo{journal}{Geophys. Res. Lett.} \bibinfo{volume}{26},
  \bibinfo{pages}{3045--3048}.
\newblock \DOIprefix\doi{10.1029/1999GL008348}.
\bibitem[{Bjorkman and Holsapple(1987)}]{BjHo87}
\bibinfo{author}{Bjorkman, M.D.}, \bibinfo{author}{Holsapple, K.A.},
  \bibinfo{year}{1987}.
\newblock \bibinfo{title}{Velocity scaling impact melt volume}.
\newblock \bibinfo{journal}{Int. J. Impact Engng.} \bibinfo{volume}{5},
  \bibinfo{pages}{155--163}.
\newblock \DOIprefix\doi{10.1016/0734-743X(87)90035-2}.
\bibitem[{Bottke et~al.(2017)Bottke, Nesvorn{\'y}, Roig, Marchi and
  Vokrouhlicky}]{Bott:etal17}
\bibinfo{author}{Bottke, W.F.}, \bibinfo{author}{Nesvorn{\'y}, D.},
  \bibinfo{author}{Roig, F.}, \bibinfo{author}{Marchi, S.},
  \bibinfo{author}{Vokrouhlicky, D.}, \bibinfo{year}{2017}.
\newblock \bibinfo{title}{{Evidence for two impacting populations in the early
  bombardment of Mars and the Moon}}, in: \bibinfo{booktitle}{Lunar Planet.
  Sci.}
\newblock \URLprefix
  \url{http://www.hou.usra.edu/meetings/lpsc2017/pdf/2572.pdf}.
\bibitem[{Bruce(1962)}]{Bruce62}
\bibinfo{author}{Bruce, E.P.}, \bibinfo{year}{1962}.
\newblock \bibinfo{title}{Review and analysis of high velocity impact data},
  in: \bibinfo{booktitle}{Proceedings of the Fifth Symposium on Hypervelocity
  Impact}, \bibinfo{address}{Denver}. pp. \bibinfo{pages}{439--474}.
\newblock \URLprefix
  \url{http://www.dtic.mil/cgi-bin/GetTRDoc?AD=AD0284280#page=67}.
\bibitem[{Carry(2012)}]{Carry12}
\bibinfo{author}{Carry, B.}, \bibinfo{year}{2012}.
\newblock \bibinfo{title}{Density of asteroids}.
\newblock \bibinfo{journal}{Planet. Space Sci.} \bibinfo{volume}{73},
  \bibinfo{pages}{98--118}.
\newblock \DOIprefix\doi{10.1016/j.pss.2012.03.009}.
\bibitem[{Catling(2015)}]{Catling15}
\bibinfo{author}{Catling, D.C.}, \bibinfo{year}{2015}.
\newblock \bibinfo{title}{Planetary atmospheres}, in: \bibinfo{editor}{Spohn,
  T.} (Ed.), \bibinfo{booktitle}{Physics of Terrestrial Planets and Moons}.
  \bibinfo{edition}{2nd} ed.. \bibinfo{publisher}{Elsevier}.
  volume~\bibinfo{volume}{10} of \textit{\bibinfo{series}{Treatise on
  Geophysics}}. chapter \bibinfo{chapter}{10.13}, pp.
  \bibinfo{pages}{429--472}.
\newblock \DOIprefix\doi{10.1016/B978-0-444-53802-4.00185-8}.
\bibitem[{Chabot et~al.(2014)Chabot, Wollack, Klima and Minitti}]{Chab:etal14}
\bibinfo{author}{Chabot, N.L.}, \bibinfo{author}{Wollack, E.A.},
  \bibinfo{author}{Klima, R.L.}, \bibinfo{author}{Minitti, M.E.},
  \bibinfo{year}{2014}.
\newblock \bibinfo{title}{{Experimental constraints on Mercury's core
  composition}}.
\newblock \bibinfo{journal}{Earth Planet. Sci. Lett.} \bibinfo{volume}{390},
  \bibinfo{pages}{199--208}.
\newblock \DOIprefix\doi{10.1016/j.epsl.2014.01.004}.
\bibitem[{Chapman and McKinnon(1986)}]{CRChMcKi86}
\bibinfo{author}{Chapman, C.R.}, \bibinfo{author}{McKinnon, W.B.},
  \bibinfo{year}{1986}.
\newblock \bibinfo{title}{Cratering of planetary satellites}, in:
  \bibinfo{editor}{Burns, J.A.}, \bibinfo{editor}{Matthews, M.S.} (Eds.),
  \bibinfo{booktitle}{Satellites}. \bibinfo{publisher}{University of Arizona
  Press}, \bibinfo{address}{Tucson}. chapter~\bibinfo{chapter}{11}, pp.
  \bibinfo{pages}{492--580}.
\bibitem[{Chyba(1987)}]{Chyba87}
\bibinfo{author}{Chyba, C.F.}, \bibinfo{year}{1987}.
\newblock \bibinfo{title}{{The cometary contribution to the oceans of primitive
  Earth}}.
\newblock \bibinfo{journal}{Nature} \bibinfo{volume}{330},
  \bibinfo{pages}{632--635}.
\newblock \DOIprefix\doi{10.1038/330632a0}.
\bibitem[{Chyba(1991)}]{Chyba91}
\bibinfo{author}{Chyba, C.F.}, \bibinfo{year}{1991}.
\newblock \bibinfo{title}{Terrestrial mantle siderophiles and the lunar impact
  record}.
\newblock \bibinfo{journal}{Icarus} \bibinfo{volume}{92},
  \bibinfo{pages}{217--233}.
\newblock \DOIprefix\doi{10.1016/0019-1035(91)90047-W}.
\bibitem[{Clifford(1993)}]{Clifford93}
\bibinfo{author}{Clifford, S.M.}, \bibinfo{year}{1993}.
\newblock \bibinfo{title}{{A model for the hydrologic and climatic behavior of
  water on Mars}}.
\newblock \bibinfo{journal}{J. Geophys. Res.} \bibinfo{volume}{98},
  \bibinfo{pages}{10973--11016}.
\newblock \DOIprefix\doi{10.1029/93JE00225}.
\bibitem[{Davidsson et~al.(2016)Davidsson, Sierks, G{\"u}ttler, Marzari,
  Pajola, Rickman, A'Hearn, Auger, El-Maarry, Fornasier, Guti{\'e}rrez, Keller,
  Massironi, Snodgrass, Vincent, Barbieri, Lamy, Rodrigo, Koschny, Barucci,
  Bertaux, Bertini, Cremonese, Da~Deppo, Debei, De~Cecco, Feller, Fulle,
  Groussin, Hviid, H{\"o}fner, Ip, Jorda, Knollenberg, Kov{\'a}cs, Kramm,
  K{\"u}hrt, K{\"u}ppers, La~Forgia, Lara, Lazzarin, L{\'o}pez~Moreno,
  Moissl-Fraund, Mottola, Naletto, Oklay, Thomas and Tubiana}]{Davi:etal16}
\bibinfo{author}{Davidsson, B.J.R.}, \bibinfo{author}{Sierks, H.},
  \bibinfo{author}{G{\"u}ttler, C.}, \bibinfo{author}{Marzari, F.},
  \bibinfo{author}{Pajola, M.}, \bibinfo{author}{Rickman, H.},
  \bibinfo{author}{A'Hearn, M.F.}, \bibinfo{author}{Auger, A.T.},
  \bibinfo{author}{El-Maarry, M.R.}, \bibinfo{author}{Fornasier, S.},
  \bibinfo{author}{Guti{\'e}rrez, P.J.}, \bibinfo{author}{Keller, H.U.},
  \bibinfo{author}{Massironi, M.}, \bibinfo{author}{Snodgrass, C.},
  \bibinfo{author}{Vincent, J.B.}, \bibinfo{author}{Barbieri, C.},
  \bibinfo{author}{Lamy, P.L.}, \bibinfo{author}{Rodrigo, R.},
  \bibinfo{author}{Koschny, D.}, \bibinfo{author}{Barucci, M.A.},
  \bibinfo{author}{Bertaux, J.L.}, \bibinfo{author}{Bertini, I.},
  \bibinfo{author}{Cremonese, G.}, \bibinfo{author}{Da~Deppo, V.},
  \bibinfo{author}{Debei, S.}, \bibinfo{author}{De~Cecco, M.},
  \bibinfo{author}{Feller, C.}, \bibinfo{author}{Fulle, M.},
  \bibinfo{author}{Groussin, O.}, \bibinfo{author}{Hviid, S.F.},
  \bibinfo{author}{H{\"o}fner, S.}, \bibinfo{author}{Ip, W.H.},
  \bibinfo{author}{Jorda, L.}, \bibinfo{author}{Knollenberg, J.},
  \bibinfo{author}{Kov{\'a}cs, G.}, \bibinfo{author}{Kramm, J.R.},
  \bibinfo{author}{K{\"u}hrt, E.}, \bibinfo{author}{K{\"u}ppers, M.},
  \bibinfo{author}{La~Forgia, F.}, \bibinfo{author}{Lara, L.M.},
  \bibinfo{author}{Lazzarin, M.}, \bibinfo{author}{L{\'o}pez~Moreno, J.J.},
  \bibinfo{author}{Moissl-Fraund, R.}, \bibinfo{author}{Mottola, S.},
  \bibinfo{author}{Naletto, G.}, \bibinfo{author}{Oklay, N.},
  \bibinfo{author}{Thomas, N.}, \bibinfo{author}{Tubiana, C.},
  \bibinfo{year}{2016}.
\newblock \bibinfo{title}{{The primordial nucleus of comet
  67P/Churyumov--Gerasimenko}}.
\newblock \bibinfo{journal}{Astron. Astrophys.} \bibinfo{volume}{592}.
\newblock \DOIprefix\doi{10.1051/0004-6361/201526968}.
\bibitem[{Dehn(1986)}]{Dehn86}
\bibinfo{author}{Dehn, J.T.}, \bibinfo{year}{1986}.
\newblock \bibinfo{title}{A unified theory of penetration}.
\newblock \bibinfo{type}{Technical Report} \bibinfo{number}{BRL-TR-2770}.
  Ballistic Research Laboratory, U.S. Army. \bibinfo{address}{Aberdeen, MD}.
\newblock \URLprefix \url{http://www.dtic.mil/cgi-bin/GetTRDoc?AD=ADA176249}.
\bibitem[{DeMeo and Carry(2013)}]{DeMeCa13}
\bibinfo{author}{DeMeo, F.E.}, \bibinfo{author}{Carry, B.},
  \bibinfo{year}{2013}.
\newblock \bibinfo{title}{The taxonomic distribution of asteroids from
  multi-filter all-sky photometric surveys}.
\newblock \bibinfo{journal}{Icarus} \bibinfo{volume}{226},
  \bibinfo{pages}{723--741}.
\newblock \DOIprefix\doi{10.1016/j.icarus.2013.06.027}.
\bibitem[{Dienes and Walsh(1970)}]{DiWa70}
\bibinfo{author}{Dienes, J.K.}, \bibinfo{author}{Walsh, J.M.},
  \bibinfo{year}{1970}.
\newblock \bibinfo{title}{Theory of impact: Some general principles and the
  method of eulerian codes}, in: \bibinfo{editor}{Kinslow, R.} (Ed.),
  \bibinfo{booktitle}{High-Velocity Impact Phenomena}.
  \bibinfo{publisher}{Academic Press}. chapter~\bibinfo{chapter}{3}, pp.
  \bibinfo{pages}{45--104}.
\newblock \DOIprefix\doi{10.1016/B978-0-12-408950-1.50008-2}.
\bibitem[{Emilio et~al.(2012)Emilio, Kuhn, Bush and Scholl}]{Emil:etal12}
\bibinfo{author}{Emilio, M.}, \bibinfo{author}{Kuhn, J.R.},
  \bibinfo{author}{Bush, R.I.}, \bibinfo{author}{Scholl, I.F.},
  \bibinfo{year}{2012}.
\newblock \bibinfo{title}{{Measuring the solar radius from space during the
  2003 and 2006 Mercury transits}}.
\newblock \bibinfo{journal}{Astrophys. J.} \bibinfo{volume}{750}.
\newblock \DOIprefix\doi{10.1088/0004-637X/750/2/135}.
\bibitem[{Fegley(2014)}]{Fegley14}
\bibinfo{author}{Fegley, Jr., B.}, \bibinfo{year}{2014}.
\newblock \bibinfo{title}{Venus}, in: \bibinfo{editor}{Davis, A.M.} (Ed.),
  \bibinfo{booktitle}{Planets, Asteroids, Comets and The Solar System}.
  \bibinfo{edition}{2nd} ed. \bibinfo{publisher}{Elsevier}.
  volume~\bibinfo{volume}{2} of \textit{\bibinfo{series}{Treatise on
  Geochemistry}}. chapter \bibinfo{chapter}{2.7}, pp.
  \bibinfo{pages}{127--148}.
\newblock \DOIprefix\doi{10.1016/B978-0-08-095975-7.00122-4}.
\bibitem[{Fiquet et~al.(2010)Fiquet, Auzende, Siebert, Corgne, Bureau, Ozawa
  and Garbarino}]{Fiqu:etal10}
\bibinfo{author}{Fiquet, G.}, \bibinfo{author}{Auzende, A.L.},
  \bibinfo{author}{Siebert, J.}, \bibinfo{author}{Corgne, A.},
  \bibinfo{author}{Bureau, H.}, \bibinfo{author}{Ozawa, H.},
  \bibinfo{author}{Garbarino, G.}, \bibinfo{year}{2010}.
\newblock \bibinfo{title}{Melting of peridotite to 140 gigapascals}.
\newblock \bibinfo{journal}{Science} \bibinfo{volume}{329},
  \bibinfo{pages}{1516--1518}.
\newblock \DOIprefix\doi{10.1126/science.1192448}.
\bibitem[{Frank et~al.(2017)Frank, Potter, Abramov, James, Klima, Mojzsis and
  Nittler}]{EAFran:etal17}
\bibinfo{author}{Frank, E.A.}, \bibinfo{author}{Potter, R.W.K.},
  \bibinfo{author}{Abramov, O.}, \bibinfo{author}{James, P.B.},
  \bibinfo{author}{Klima, R.L.}, \bibinfo{author}{Mojzsis, S.J.},
  \bibinfo{author}{Nittler, L.R.}, \bibinfo{year}{2017}.
\newblock \bibinfo{title}{{Evaluating an impact origin for Mercury's
  high-magnesium region}}.
\newblock \bibinfo{journal}{J. Geophys. Res.} \bibinfo{volume}{122},
  \bibinfo{pages}{614--632}.
\newblock \DOIprefix\doi{10.1002/2016JE005244}.
\bibitem[{Frey(2008)}]{Frey08}
\bibinfo{author}{Frey, H.}, \bibinfo{year}{2008}.
\newblock \bibinfo{title}{{Ages of very large impact basins on Mars:
  Implications for the late heavy bombardment in the inner solar system}}.
\newblock \bibinfo{journal}{Geophys. Res. Lett.} \bibinfo{volume}{35}.
\newblock \DOIprefix\doi{10.1029/2008GL033515}.
\bibitem[{Garcia et~al.(2011)Garcia, Gagnepain-Beyneix, Chevrot and
  Lognonn{\'e}}]{RFGarc:etal11}
\bibinfo{author}{Garcia, R.F.}, \bibinfo{author}{Gagnepain-Beyneix, J.},
  \bibinfo{author}{Chevrot, S.}, \bibinfo{author}{Lognonn{\'e}, P.},
  \bibinfo{year}{2011}.
\newblock \bibinfo{title}{{Very preliminary reference Moon model}}.
\newblock \bibinfo{journal}{Phys. Earth Planet. Inter.} \bibinfo{volume}{188},
  \bibinfo{pages}{96--113}.
\newblock \DOIprefix\doi{10.1016/j.pepi.2011.06.015}. \bibinfo{note}{correction
  in vol.~202--203, 89--91 (2012)}.
\bibitem[{Gillmann et~al.(2016)Gillmann, Golabek and Tackley}]{Gill:etal16}
\bibinfo{author}{Gillmann, C.}, \bibinfo{author}{Golabek, G.J.},
  \bibinfo{author}{Tackley, P.J.}, \bibinfo{year}{2016}.
\newblock \bibinfo{title}{{Effect of a single large impact on the coupled
  atmosphere-interior evolution of Venus}}.
\newblock \bibinfo{journal}{Icarus} \bibinfo{volume}{268},
  \bibinfo{pages}{295--312}.
\newblock \DOIprefix\doi{10.1016/j.icarus.2015.12.024}.
\bibitem[{Gomes et~al.(2005)Gomes, Levison, Tsiganis and
  Morbidelli}]{Gome:etal05}
\bibinfo{author}{Gomes, R.}, \bibinfo{author}{Levison, H.F.},
  \bibinfo{author}{Tsiganis, K.}, \bibinfo{author}{Morbidelli, A.},
  \bibinfo{year}{2005}.
\newblock \bibinfo{title}{{Origin of the cataclysmic Late Heavy Bombardment
  period of the terrestrial planets}}.
\newblock \bibinfo{journal}{Nature} \bibinfo{volume}{435},
  \bibinfo{pages}{466--469}.
\newblock \DOIprefix\doi{10.1038/nature03676}.
\bibitem[{Grieve et~al.(2007)Grieve, Cintala and Tagle}]{Grie:etal07}
\bibinfo{author}{Grieve, R.A.F.}, \bibinfo{author}{Cintala, M.J.},
  \bibinfo{author}{Tagle, R.}, \bibinfo{year}{2007}.
\newblock \bibinfo{title}{Planetary impacts}, in: \bibinfo{editor}{McFadden,
  L.A.}, \bibinfo{editor}{Weissman, P.R.}, \bibinfo{editor}{Johnson, T.V.}
  (Eds.), \bibinfo{booktitle}{Encyclopedia of the Solar System}.
  \bibinfo{edition}{2nd} ed.. \bibinfo{publisher}{Academic Press}.
  chapter~\bibinfo{chapter}{43}, pp. \bibinfo{pages}{813--828}.
\bibitem[{Grott et~al.(2011)Grott, Breuer and Laneuville}]{Grot:etal11a}
\bibinfo{author}{Grott, M.}, \bibinfo{author}{Breuer, D.},
  \bibinfo{author}{Laneuville, M.}, \bibinfo{year}{2011}.
\newblock \bibinfo{title}{{Thermo-chemical evolution and global contraction of
  Mercury}}.
\newblock \bibinfo{journal}{Earth Planet. Sci. Lett.} \bibinfo{volume}{307},
  \bibinfo{pages}{135--146}.
\newblock \DOIprefix\doi{10.1016/j.epsl.2011.04.040}.
\bibitem[{Gudfinnsson and Presnall(1996)}]{GuPr96}
\bibinfo{author}{Gudfinnsson, G.H.}, \bibinfo{author}{Presnall, D.C.},
  \bibinfo{year}{1996}.
\newblock \bibinfo{title}{{Melting relations of model lherzolite in the system
  CaO-MgO-\chem{Al}{2}\chem{O}{3}-Si\chem{O}{2} at 2.4--3.4\,GPa and the
  generation of komatiites}}.
\newblock \bibinfo{journal}{J. Geophys. Res.} \bibinfo{volume}{101},
  \bibinfo{pages}{27701--27710}.
\newblock \DOIprefix\doi{10.1029/96JB02462}.
\bibitem[{Gudfinnsson and Presnall(2000)}]{GuPr00}
\bibinfo{author}{Gudfinnsson, G.H.}, \bibinfo{author}{Presnall, D.C.},
  \bibinfo{year}{2000}.
\newblock \bibinfo{title}{{Melting behaviour of model lherzolite in the system
  CaO--MgO--\chem{Al}{2}\chem{O}{3}--Si\chem{O}{2}--FeO at 0.7--2.8\,GPa}}.
\newblock \bibinfo{journal}{J. Petrol.} \bibinfo{volume}{41},
  \bibinfo{pages}{1241--1269}.
\newblock \DOIprefix\doi{10.1093/petrology/41.8.1241}.
\bibitem[{Han et~al.(2014)Han, Schmerr, Neumann and Holmes}]{Han:etal14}
\bibinfo{author}{Han, S.C.}, \bibinfo{author}{Schmerr, N.},
  \bibinfo{author}{Neumann, G.}, \bibinfo{author}{Holmes, S.},
  \bibinfo{year}{2014}.
\newblock \bibinfo{title}{{Global characteristics of porosity and density
  stratification within the lunar crust from GRAIL gravity and LOLA topography
  data}}.
\newblock \bibinfo{journal}{Geophys. Res. Lett.}
  \DOIprefix\doi{10.1002/2014GL059378}.
\bibitem[{Hauck et~al.(2013)Hauck, Margot, Solomon, Phillips, Johnson, Lemoine,
  Mazarico, McCoy, Padovan, Peale, Perry, Smith and Zuber}]{Hauc:etal13}
\bibinfo{author}{Hauck, II, S.A.}, \bibinfo{author}{Margot, J.L.},
  \bibinfo{author}{Solomon, S.C.}, \bibinfo{author}{Phillips, R.J.},
  \bibinfo{author}{Johnson, C.L.}, \bibinfo{author}{Lemoine, F.G.},
  \bibinfo{author}{Mazarico, E.}, \bibinfo{author}{McCoy, T.J.},
  \bibinfo{author}{Padovan, S.}, \bibinfo{author}{Peale, S.J.},
  \bibinfo{author}{Perry, M.E.}, \bibinfo{author}{Smith, D.E.},
  \bibinfo{author}{Zuber, M.T.}, \bibinfo{year}{2013}.
\newblock \bibinfo{title}{{The curious case of Mercury's internal structure}}.
\newblock \bibinfo{journal}{J. Geophys. Res.} \bibinfo{volume}{118},
  \bibinfo{pages}{1204--1220}.
\newblock \DOIprefix\doi{10.1002/jgre.20091}.
\bibitem[{Hernlund and Tackley(2008)}]{HeTa08}
\bibinfo{author}{Hernlund, J.W.}, \bibinfo{author}{Tackley, P.J.},
  \bibinfo{year}{2008}.
\newblock \bibinfo{title}{Modeling mantle convection in the spherical annulus}.
\newblock \bibinfo{journal}{Phys. Earth Planet. Inter.} \bibinfo{volume}{171},
  \bibinfo{pages}{48--54}.
\newblock \DOIprefix\doi{10.1016/j.pepi.2008.07.037}.
\bibitem[{Herrick and Hynek(2017)}]{HeHy17}
\bibinfo{author}{Herrick, R.R.}, \bibinfo{author}{Hynek, B.M.},
  \bibinfo{year}{2017}.
\newblock \bibinfo{title}{Investigating target versus impactor influences on
  martian crater morphology at the simple-complex transition}.
\newblock \bibinfo{journal}{Meteorit. Planet. Sci.} \bibinfo{volume}{52},
  \bibinfo{pages}{1722--1743}.
\newblock \DOIprefix\doi{10.1111/maps.12884}.
\bibitem[{Herzberg et~al.(1990)Herzberg, Gasparik and Sawamoto}]{Herz:etal90}
\bibinfo{author}{Herzberg, C.}, \bibinfo{author}{Gasparik, T.},
  \bibinfo{author}{Sawamoto, H.}, \bibinfo{year}{1990}.
\newblock \bibinfo{title}{Origin of mantle peridotite: Constraints from melting
  experiments to {16.5\,GPa}}.
\newblock \bibinfo{journal}{J. Geophys. Res.} \bibinfo{volume}{95},
  \bibinfo{pages}{15779--15803}.
\newblock \DOIprefix\doi{10.1029/JB095iB10p15779}.
\bibitem[{Hirschmann(2000)}]{Hirschmann00}
\bibinfo{author}{Hirschmann, M.M.}, \bibinfo{year}{2000}.
\newblock \bibinfo{title}{{Mantle solidus: Experimental constraints and the
  effects of peridotite composition}}.
\newblock \bibinfo{journal}{Geochem. Geophys. Geosyst.} \bibinfo{volume}{1}.
\newblock \DOIprefix\doi{10.1029/2000GC000070}.
\bibitem[{Hirschmann et~al.(2009)Hirschmann, Tenner, Aubaud and
  Withers}]{MMHirs:etal09}
\bibinfo{author}{Hirschmann, M.M.}, \bibinfo{author}{Tenner, T.},
  \bibinfo{author}{Aubaud, C.}, \bibinfo{author}{Withers, A.C.},
  \bibinfo{year}{2009}.
\newblock \bibinfo{title}{{Dehydration melting of nominally anhydrous mantle:
  The primacy of partitioning}}.
\newblock \bibinfo{journal}{Phys. Earth Planet. Inter.} \bibinfo{volume}{176},
  \bibinfo{pages}{54--68}.
\newblock \DOIprefix\doi{10.1016/j.pepi.2009.04.001}.
\bibitem[{Hirth and Kohlstedt(2003)}]{HiKo03}
\bibinfo{author}{Hirth, G.}, \bibinfo{author}{Kohlstedt, D.L.},
  \bibinfo{year}{2003}.
\newblock \bibinfo{title}{{Rheology of the upper mantle and the mantle wedge: A
  view from the experimentalists}}, in: \bibinfo{editor}{Eiler, J.} (Ed.),
  \bibinfo{booktitle}{Inside the Subduction Factory}.
  \bibinfo{publisher}{American Geophysical Union},
  \bibinfo{address}{Washington, D.C.}. volume \bibinfo{volume}{138} of
  \textit{\bibinfo{series}{AGU Geophysical Monograph}}, pp.
  \bibinfo{pages}{83--105}.
\bibitem[{Holsapple and Schmidt(1987)}]{HoSc87}
\bibinfo{author}{Holsapple, K.A.}, \bibinfo{author}{Schmidt, R.M.},
  \bibinfo{year}{1987}.
\newblock \bibinfo{title}{Point source solutions and coupling parameters in
  cratering mechanics}.
\newblock \bibinfo{journal}{J. Geophys. Res.} \bibinfo{volume}{92},
  \bibinfo{pages}{6350--6376}.
\newblock \DOIprefix\doi{10.1029/JB092iB07p06350}.
\bibitem[{Housen and Holsapple(2011)}]{HoHo11}
\bibinfo{author}{Housen, K.R.}, \bibinfo{author}{Holsapple, K.A.},
  \bibinfo{year}{2011}.
\newblock \bibinfo{title}{Ejecta from impact craters}.
\newblock \bibinfo{journal}{Icarus} \bibinfo{volume}{211},
  \bibinfo{pages}{856--875}.
\newblock \DOIprefix\doi{10.1016/j.icarus.2010.09.017}.
\bibitem[{Hughes and Williams(2000)}]{HuWi00}
\bibinfo{author}{Hughes, D.W.}, \bibinfo{author}{Williams, I.P.},
  \bibinfo{year}{2000}.
\newblock \bibinfo{title}{The velocity distributions of periodic comets and
  stream meteoroids}.
\newblock \bibinfo{journal}{Mon. Not. R. Astron. Soc.} \bibinfo{volume}{315},
  \bibinfo{pages}{629--634}.
\newblock \DOIprefix\doi{10.1046/j.1365-8711.2000.03435.x}.
\bibitem[{IERS2010(2010)}]{IERS2010}
\bibinfo{author}{IERS2010}, \bibinfo{year}{2010}.
\newblock \bibinfo{title}{IERS Conventions (2010)}.
\newblock \bibinfo{type}{Technical Note} \bibinfo{number}{36}. International
  Earth Rotation and Reference Systems Service (IERS).
  \bibinfo{address}{Frankfurt am Main}.
\newblock \URLprefix
  \url{https://www.iers.org/IERS/EN/Publications/TechnicalNotes/tn36.html}.
\bibitem[{Ipatov and Mather(2004)}]{IpMa04}
\bibinfo{author}{Ipatov, S.I.}, \bibinfo{author}{Mather, J.C.},
  \bibinfo{year}{2004}.
\newblock \bibinfo{title}{Comet and asteroid hazard to the terrestrial
  planets}.
\newblock \bibinfo{journal}{Adv. Space Res.} \bibinfo{volume}{33},
  \bibinfo{pages}{1524--1533}.
\newblock \DOIprefix\doi{10.1016/S0273-1177(03)00451-4}.
\bibitem[{Ivanov(2001)}]{Ivanov01}
\bibinfo{author}{Ivanov, B.A.}, \bibinfo{year}{2001}.
\newblock \bibinfo{title}{{Mars/Moon cratering rate ratio estimates}}.
\newblock \bibinfo{journal}{Space Sci. Rev.} \bibinfo{volume}{96},
  \bibinfo{pages}{87--104}.
\newblock \DOIprefix\doi{10.1023/A:1011941121102}.
\bibitem[{Ivanov et~al.(2002)Ivanov, Neukum, Bottke and
  Hartmann}]{BAIvan:etal02}
\bibinfo{author}{Ivanov, B.A.}, \bibinfo{author}{Neukum, G.},
  \bibinfo{author}{Bottke, Jr., W.F.}, \bibinfo{author}{Hartmann, W.K.},
  \bibinfo{year}{2002}.
\newblock \bibinfo{title}{The comparison of size-frequency distributions of
  impact craters and asteroids and the planetary cratering rate}, in:
  \bibinfo{editor}{Bottke, Jr., W.F.}, \bibinfo{editor}{Cellino, A.},
  \bibinfo{editor}{Paolicchi, P.}, \bibinfo{editor}{Binzel, R.P.} (Eds.),
  \bibinfo{booktitle}{Asteroids III}. \bibinfo{publisher}{University of Arizona
  Press}, \bibinfo{address}{Tucson, Arizona}, pp. \bibinfo{pages}{89--101}.
\bibitem[{Jacobson(2010)}]{Jacobson10}
\bibinfo{author}{Jacobson, R.A.}, \bibinfo{year}{2010}.
\newblock \bibinfo{title}{{The orbits and masses of the martian satellites and
  the libration of Phobos}}.
\newblock \bibinfo{journal}{Astron. J.} \bibinfo{volume}{139},
  \bibinfo{pages}{668--679}.
\newblock \DOIprefix\doi{10.1088/0004-6256/139/2/668}.
\bibitem[{Jin et~al.(2001)Jin, Zhang, Green and Jin}]{ZMJin:etal01}
\bibinfo{author}{Jin, Z.M.}, \bibinfo{author}{Zhang, J.},
  \bibinfo{author}{Green, II, H.W.}, \bibinfo{author}{Jin, S.},
  \bibinfo{year}{2001}.
\newblock \bibinfo{title}{{Eclogite rheology: Implications for subducted
  lithosphere}}.
\newblock \bibinfo{journal}{Geology} \bibinfo{volume}{29},
  \bibinfo{pages}{667--670}.
\bibitem[{Jing and Karato(2011)}]{JiKa11}
\bibinfo{author}{Jing, Z.}, \bibinfo{author}{Karato, S.i.},
  \bibinfo{year}{2011}.
\newblock \bibinfo{title}{{A new approach to the equation of state of silicate
  melts: An application of the theory of hard sphere mixtures}}.
\newblock \bibinfo{journal}{Geochim. Cosmochim. Acta} \bibinfo{volume}{75},
  \bibinfo{pages}{6780--6802}.
\newblock \DOIprefix\doi{10.1016/j.gca.2011.09.004}.
\bibitem[{Jing and Karato(2012)}]{JiKa12}
\bibinfo{author}{Jing, Z.}, \bibinfo{author}{Karato, S.i.},
  \bibinfo{year}{2012}.
\newblock \bibinfo{title}{{Effect of \chem{H}{2}O on the density of silicate
  melts at high pressures: Static experiments and the application of a modified
  hard-sphere model of equation of state}}.
\newblock \bibinfo{journal}{Geochim. Cosmochim. Acta} \bibinfo{volume}{85},
  \bibinfo{pages}{357--372}.
\newblock \DOIprefix\doi{10.1016/j.gca.2012.03.001}.
\bibitem[{Karimi et~al.(2017)Karimi, Ojha and Lewis}]{Kari:etal17}
\bibinfo{author}{Karimi, S.}, \bibinfo{author}{Ojha, L.},
  \bibinfo{author}{Lewis, K.}, \bibinfo{year}{2017}.
\newblock \bibinfo{title}{{Searching for large buried craters on Venus}}, in:
  \bibinfo{booktitle}{Lunar Planet. Sci.}
\newblock \URLprefix
  \url{http://www.hou.usra.edu/meetings/lpsc2017/pdf/2831.pdf}.
\bibitem[{Katz et~al.(2003)Katz, Spiegelman and Langmuir}]{Katz:etal03}
\bibinfo{author}{Katz, R.F.}, \bibinfo{author}{Spiegelman, M.},
  \bibinfo{author}{Langmuir, C.}, \bibinfo{year}{2003}.
\newblock \bibinfo{title}{A new parameterization of hydrous mantle melting}.
\newblock \bibinfo{journal}{Geochem. Geophys. Geosyst.} \bibinfo{volume}{4}.
\newblock \DOIprefix\doi{10.1029/2002GC000433}.
\bibitem[{Khan and Connolly(2008)}]{KhCo08}
\bibinfo{author}{Khan, A.}, \bibinfo{author}{Connolly, J.A.D.},
  \bibinfo{year}{2008}.
\newblock \bibinfo{title}{{Constraining the composition and thermal state of
  Mars from inversion of geophysical data}}.
\newblock \bibinfo{journal}{J. Geophys. Res.} \bibinfo{volume}{113}.
\newblock \DOIprefix\doi{10.1029/2007JE002996}.
\bibitem[{Kiefer et~al.(2012)Kiefer, Macke, Britt, Irving and
  Consolmagno}]{Kief:etal12}
\bibinfo{author}{Kiefer, W.S.}, \bibinfo{author}{Macke, R.J.},
  \bibinfo{author}{Britt, D.T.}, \bibinfo{author}{Irving, A.J.},
  \bibinfo{author}{Consolmagno, G.J.}, \bibinfo{year}{2012}.
\newblock \bibinfo{title}{The density and porosity of lunar rocks}.
\newblock \bibinfo{journal}{Geophys. Res. Lett.} \bibinfo{volume}{39}.
\newblock \DOIprefix\doi{10.1029/2012GL051319}.
\bibitem[{Klemme and O'Neill(2000)}]{KlStCONe00}
\bibinfo{author}{Klemme, S.}, \bibinfo{author}{O'Neill, H.S.C.},
  \bibinfo{year}{2000}.
\newblock \bibinfo{title}{The near-solidus transition from garnet lherzolite to
  spinel lherzolite}.
\newblock \bibinfo{journal}{Contrib. Mineral. Petrol.} \bibinfo{volume}{138},
  \bibinfo{pages}{237--248}.
\newblock \DOIprefix\doi{10.1007/s004100050560}.
\bibitem[{Konopliv et~al.(2011)Konopliv, Asmar, Folkner, Karatekin, Nunes,
  Smrekar, Yoder and Zuber}]{Kono:etal11}
\bibinfo{author}{Konopliv, A.S.}, \bibinfo{author}{Asmar, S.W.},
  \bibinfo{author}{Folkner, W.M.}, \bibinfo{author}{Karatekin, {\"O}.},
  \bibinfo{author}{Nunes, D.C.}, \bibinfo{author}{Smrekar, S.E.},
  \bibinfo{author}{Yoder, C.F.}, \bibinfo{author}{Zuber, M.T.},
  \bibinfo{year}{2011}.
\newblock \bibinfo{title}{{Mars high resolution gravity fields from MRO, Mars
  seasonal gravity, and other dynamical parameters}}.
\newblock \bibinfo{journal}{Icarus} \bibinfo{volume}{211},
  \bibinfo{pages}{401--428}.
\newblock \DOIprefix\doi{10.1016/j.icarus.2010.10.004}.
\bibitem[{Konopliv et~al.(1999)Konopliv, Banerdt and Sjogren}]{Kono:etal99}
\bibinfo{author}{Konopliv, A.S.}, \bibinfo{author}{Banerdt, W.B.},
  \bibinfo{author}{Sjogren, W.L.}, \bibinfo{year}{1999}.
\newblock \bibinfo{title}{Venus gravity: 180th degree and order model}.
\newblock \bibinfo{journal}{Icarus} \bibinfo{volume}{139},
  \bibinfo{pages}{3--18}.
\newblock \DOIprefix\doi{10.1006/icar.1999.6086}.
\bibitem[{Konopliv and Yoder(1996)}]{KoYo96}
\bibinfo{author}{Konopliv, A.S.}, \bibinfo{author}{Yoder, C.F.},
  \bibinfo{year}{1996}.
\newblock \bibinfo{title}{{Venusian $k_2$ tidal Love number from Magellan and
  PVO tracking data}}.
\newblock \bibinfo{journal}{Geophys. Res. Lett.} \bibinfo{volume}{23},
  \bibinfo{pages}{1857--1860}.
\bibitem[{Korsun et~al.(2014)Korsun, Rousselot, Kulyk, Afanasiev and
  Ivanova}]{Kors:etal14}
\bibinfo{author}{Korsun, P.P.}, \bibinfo{author}{Rousselot, P.},
  \bibinfo{author}{Kulyk, I.V.}, \bibinfo{author}{Afanasiev, V.L.},
  \bibinfo{author}{Ivanova, O.V.}, \bibinfo{year}{2014}.
\newblock \bibinfo{title}{{Distant activity of Comet C/2002 VQ94 (LINEAR):
  Optical spectrophotometric monitoring between 8.4 and 16.8\,au from the
  Sun}}.
\newblock \bibinfo{journal}{Icarus} \bibinfo{volume}{232},
  \bibinfo{pages}{88--96}.
\newblock \DOIprefix\doi{10.1016/j.icarus.2014.01.006}.
\bibitem[{Kuchner et~al.(2002)Kuchner, Brown and Holman}]{Kuch:etal02}
\bibinfo{author}{Kuchner, M.J.}, \bibinfo{author}{Brown, M.E.},
  \bibinfo{author}{Holman, M.}, \bibinfo{year}{2002}.
\newblock \bibinfo{title}{{Long-term dynamics and the orbital inclinations of
  the classical Kuiper belt objects}}.
\newblock \bibinfo{journal}{Astrophys. J.} \bibinfo{volume}{124},
  \bibinfo{pages}{1221--1230}.
\bibitem[{Laneuville et~al.(2013)Laneuville, Wieczorek, Breuer and
  Tosi}]{Lane:etal13}
\bibinfo{author}{Laneuville, M.}, \bibinfo{author}{Wieczorek, M.A.},
  \bibinfo{author}{Breuer, D.}, \bibinfo{author}{Tosi, N.},
  \bibinfo{year}{2013}.
\newblock \bibinfo{title}{{Asymmetric thermal evolution of the Moon}}.
\newblock \bibinfo{journal}{J. Geophys. Res.} \bibinfo{volume}{118},
  \bibinfo{pages}{1435--1452}.
\newblock \DOIprefix\doi{10.1002/jgre.20103}.
\bibitem[{Litasov and Ohtani(2002)}]{LiOh02}
\bibinfo{author}{Litasov, K.}, \bibinfo{author}{Ohtani, E.},
  \bibinfo{year}{2002}.
\newblock \bibinfo{title}{{Phase relations and melt compositions in
  CMAS--pyrolite--\chem{H}{2}O system up to 25\,GPa}}.
\newblock \bibinfo{journal}{Phys. Earth Planet. Inter.} \bibinfo{volume}{134},
  \bibinfo{pages}{105--127}.
\newblock \DOIprefix\doi{10.1016/S0031-9201(02)00152-8}.
\bibitem[{Liu and O'Neill(2004)}]{XLiONe04a}
\bibinfo{author}{Liu, X.}, \bibinfo{author}{O'Neill, H.S.C.},
  \bibinfo{year}{2004}.
\newblock \bibinfo{title}{{Partial melting of spinel lherzolite in the system
  CaO--MgO--\chem{Al}{2}\chem{O}{3}--\chem{SiO}{2}\textpm\chem{K}{2}O at
  1.1\,GPa}}.
\newblock \bibinfo{journal}{J. Petrol.} \bibinfo{volume}{45},
  \bibinfo{pages}{1339--1368}.
\newblock \DOIprefix\doi{10.1093/petrology/egh021}.
\bibitem[{Mackwell et~al.(1998)Mackwell, Zimmerman and Kohlstedt}]{Mack:etal98}
\bibinfo{author}{Mackwell, S.J.}, \bibinfo{author}{Zimmerman, M.E.},
  \bibinfo{author}{Kohlstedt, D.L.}, \bibinfo{year}{1998}.
\newblock \bibinfo{title}{{High-temperature deformation of dry diabase with
  application to tectonics on Venus}}.
\newblock \bibinfo{journal}{J. Geophys. Res.} \bibinfo{volume}{103},
  \bibinfo{pages}{975--984}.
\bibitem[{Manske and W{\"u}nnemann(2017)}]{MaWu17}
\bibinfo{author}{Manske, L.}, \bibinfo{author}{W{\"u}nnemann, K.},
  \bibinfo{year}{2017}.
\newblock \bibinfo{title}{Impact-induced melting by collision events --
  implications for the formation of magma oceans on terrestrial planets}, in:
  \bibinfo{booktitle}{Accretion and Early Differentiation of the Earth and
  Terrestrial Planets}, \bibinfo{address}{Nice}.
\newblock \URLprefix \url{https://www-n.oca.eu/morby/abstracts.tar}.
\bibitem[{Marinova et~al.(2011)Marinova, Aharonson and Asphaug}]{Mari:etal11}
\bibinfo{author}{Marinova, M.M.}, \bibinfo{author}{Aharonson, O.},
  \bibinfo{author}{Asphaug, E.}, \bibinfo{year}{2011}.
\newblock \bibinfo{title}{{Geophysical consequences of planetary-scale impacts
  into a Mars-like planet}}.
\newblock \bibinfo{journal}{Icarus} \bibinfo{volume}{211},
  \bibinfo{pages}{960--985}.
\newblock \DOIprefix\doi{10.1016/j.icarus.2010.10.032}.
\bibitem[{Mazarico et~al.(2014)Mazarico, Genova, Goossens, Lemoine, Neumann,
  Zuber, Smith and Solomon}]{Maza:etal14}
\bibinfo{author}{Mazarico, E.}, \bibinfo{author}{Genova, A.},
  \bibinfo{author}{Goossens, S.}, \bibinfo{author}{Lemoine, F.G.},
  \bibinfo{author}{Neumann, G.A.}, \bibinfo{author}{Zuber, M.T.},
  \bibinfo{author}{Smith, D.E.}, \bibinfo{author}{Solomon, S.C.},
  \bibinfo{year}{2014}.
\newblock \bibinfo{title}{{The gravity field, orientation, and ephemeris of
  Mercury from MESSENGER observations after three years in orbit}}.
\newblock \bibinfo{journal}{J. Geophys. Res.} \bibinfo{volume}{119},
  \bibinfo{pages}{2417--2436}.
\newblock \DOIprefix\doi{10.1002/2014JE004675}.
\bibitem[{Melosh(1984)}]{Melosh84}
\bibinfo{author}{Melosh, H.J.}, \bibinfo{year}{1984}.
\newblock \bibinfo{title}{Impact ejection, spallation, and the origin of
  meteorites}.
\newblock \bibinfo{journal}{Icarus} \bibinfo{volume}{59},
  \bibinfo{pages}{234--260}.
\newblock \DOIprefix\doi{10.1016/0019-1035(84)90026-5}.
\bibitem[{Melosh(1989)}]{Melosh89}
\bibinfo{author}{Melosh, H.J.}, \bibinfo{year}{1989}.
\newblock \bibinfo{title}{Impact cratering: a geologic process}.
\newblock Number~\bibinfo{number}{11} in \bibinfo{series}{Oxford Monographs on
  Geology and Geophysics}, \bibinfo{publisher}{Oxford University Press}.
\bibitem[{Melosh(2011)}]{Melosh11}
\bibinfo{author}{Melosh, H.J.}, \bibinfo{year}{2011}.
\newblock \bibinfo{title}{Planetary Surface Processes}.
\newblock Number~\bibinfo{number}{13} in \bibinfo{series}{Cambridge Planetary
  Science}, \bibinfo{publisher}{Cambridge University Press}.
\bibitem[{Milholland and Presnall(1998)}]{MiPr98}
\bibinfo{author}{Milholland, C.S.}, \bibinfo{author}{Presnall, D.C.},
  \bibinfo{year}{1998}.
\newblock \bibinfo{title}{{Liquidus phase relations in the
  CaO--MgO--\chem{Al}{2}\chem{O}{3}--\chem{SiO}{2} system at 3.0\,GPa: the
  aluminous pyroxene thermal divide and high-pressure fractionation of picritic
  and komatiitic magmas}}.
\newblock \bibinfo{journal}{J. Petrol.} \bibinfo{volume}{39},
  \bibinfo{pages}{3--27}.
\newblock \DOIprefix\doi{10.1093/petroj/39.1.3}.
\bibitem[{Miljkovi{\'c} et~al.(2016)Miljkovi{\'c}, Collins, Wieczorek, Johnson,
  Soderlund, Neumann and Zuber}]{Milj:etal16}
\bibinfo{author}{Miljkovi{\'c}, K.}, \bibinfo{author}{Collins, G.S.},
  \bibinfo{author}{Wieczorek, M.A.}, \bibinfo{author}{Johnson, B.C.},
  \bibinfo{author}{Soderlund, J.M.}, \bibinfo{author}{Neumann, G.A.},
  \bibinfo{author}{Zuber, M.T.}, \bibinfo{year}{2016}.
\newblock \bibinfo{title}{Subsurface morphology and scaling of lunar impact
  basins}.
\newblock \bibinfo{journal}{J. Geophys. Res.} \bibinfo{volume}{121},
  \bibinfo{pages}{1695--1712}.
\newblock \DOIprefix\doi{10.1002/2016JE005038}.
\bibitem[{Mohr et~al.(2016)Mohr, Newell and Taylor}]{CODATA2014}
\bibinfo{author}{Mohr, P.J.}, \bibinfo{author}{Newell, D.B.},
  \bibinfo{author}{Taylor, B.N.}, \bibinfo{year}{2016}.
\newblock \bibinfo{title}{Codata recommended values of the fundamental physical
  constants: 2014}.
\newblock \bibinfo{journal}{J. Phys. Chem. Ref. Data} \bibinfo{volume}{45}.
\newblock \DOIprefix\doi{10.1063/1.4954402}.
\bibitem[{Morgan and Anders(1980)}]{MoAn80}
\bibinfo{author}{Morgan, J.W.}, \bibinfo{author}{Anders, E.},
  \bibinfo{year}{1980}.
\newblock \bibinfo{title}{{Chemical composition of Earth, Venus, and Mercury}}.
\newblock \bibinfo{journal}{Proc. Nat. Acad. Sci.} \bibinfo{volume}{77},
  \bibinfo{pages}{6973--6977}.
\bibitem[{M{\"u}ller(2007)}]{GMuller07}
\bibinfo{author}{M{\"u}ller, G.}, \bibinfo{year}{2007}.
\newblock \bibinfo{title}{Theory of elastic waves}.
\newblock \bibinfo{type}{Scientific Technical Report STR}
  \bibinfo{number}{07/03}. GeoForschungsZentrum GFZ.
  \bibinfo{address}{Potsdam}.
\newblock \DOIprefix\doi{10.2312/GFZ.b103-07037}.
\bibitem[{Napier(2015)}]{Napier15}
\bibinfo{author}{Napier, W.M.}, \bibinfo{year}{2015}.
\newblock \bibinfo{title}{Giant comets and mass extinctions of life}.
\newblock \bibinfo{journal}{Mon. Not. R. Astron. Soc.} \bibinfo{volume}{448},
  \bibinfo{pages}{27--36}.
\newblock \DOIprefix\doi{10.1093/mnras/stu2681}.
\bibitem[{Nimmo et~al.(2004)Nimmo, Price, Brodholt and Gubbins}]{Nimm:etal04}
\bibinfo{author}{Nimmo, F.}, \bibinfo{author}{Price, G.D.},
  \bibinfo{author}{Brodholt, J.}, \bibinfo{author}{Gubbins, D.},
  \bibinfo{year}{2004}.
\newblock \bibinfo{title}{The influence of potassium on core and geodynamo
  evolution}.
\newblock \bibinfo{journal}{Geophys. J. Int.} \bibinfo{volume}{156},
  \bibinfo{pages}{363--376}.
\bibitem[{Nomura et~al.(2014)Nomura, Hirose, Uesugi, Ohishi, Tsuchiyama, Miyake
  and Ueno}]{Nomu:etal14}
\bibinfo{author}{Nomura, R.}, \bibinfo{author}{Hirose, K.},
  \bibinfo{author}{Uesugi, K.}, \bibinfo{author}{Ohishi, Y.},
  \bibinfo{author}{Tsuchiyama, A.}, \bibinfo{author}{Miyake, A.},
  \bibinfo{author}{Ueno, Y.}, \bibinfo{year}{2014}.
\newblock \bibinfo{title}{Low core--mantle boundary temperature inferred from
  the solidus of pyrolite}.
\newblock \bibinfo{journal}{Science} \bibinfo{volume}{343},
  \bibinfo{pages}{522--525}.
\newblock \DOIprefix\doi{10.1126/science.1248186}.
\bibitem[{O'Keefe and Ahrens(1982)}]{OKeAh82}
\bibinfo{author}{O'Keefe, J.D.}, \bibinfo{author}{Ahrens, T.J.},
  \bibinfo{year}{1982}.
\newblock \bibinfo{title}{Cometary and meteorite swarm impact on planetary
  surfaces}.
\newblock \bibinfo{journal}{J. Geophys. Res.} \bibinfo{volume}{87},
  \bibinfo{pages}{6668--6680}.
\newblock \DOIprefix\doi{10.1029/JB087iB08p06668}.
\bibitem[{O'Keefe and Ahrens(1993)}]{OKeAh93}
\bibinfo{author}{O'Keefe, J.D.}, \bibinfo{author}{Ahrens, T.J.},
  \bibinfo{year}{1993}.
\newblock \bibinfo{title}{Planetary cratering mechanics}.
\newblock \bibinfo{journal}{J. Geophys. Res.} \bibinfo{volume}{98},
  \bibinfo{pages}{17011--17028}.
\newblock \DOIprefix\doi{10.1029/93JE01330}.
\bibitem[{O'Keefe and Ahrens(1994)}]{OKeAh94}
\bibinfo{author}{O'Keefe, J.D.}, \bibinfo{author}{Ahrens, T.J.},
  \bibinfo{year}{1994}.
\newblock \bibinfo{title}{Impact-induced melting of planetary surfaces}, in:
  \bibinfo{editor}{Dressler, B.O.}, \bibinfo{editor}{Grieve, R.A.F.},
  \bibinfo{editor}{Sharpton, V.L.} (Eds.), \bibinfo{booktitle}{Large Meteorite
  Impacts and Planetary Evolution}. \bibinfo{publisher}{Geological Society of
  America}, \bibinfo{address}{Boulder, Colorado}. number \bibinfo{number}{293}
  in \bibinfo{series}{Special Papers}, pp. \bibinfo{pages}{103--109}.
\bibitem[{Olsson-Steel(1987)}]{Olsson-Steel87}
\bibinfo{author}{Olsson-Steel, D.}, \bibinfo{year}{1987}.
\newblock \bibinfo{title}{{Collisions in the Solar System -- IV. Cometary
  impacts upon the planets}}.
\newblock \bibinfo{journal}{Mon. Not. R. Astron. Soc.} \bibinfo{volume}{227},
  \bibinfo{pages}{501--524}.
\newblock \DOIprefix\doi{10.1093/mnras/227.2.501}.
\bibitem[{{\"O}pik(1936)}]{Opik36a}
\bibinfo{author}{{\"O}pik, E.}, \bibinfo{year}{1936}.
\newblock \bibinfo{title}{{Researches on the physical theory of meteor
  phenomena. I. Theory of the formation of meteor craters}}.
\newblock \bibinfo{journal}{Acta Comm. Univ. Tartu.} \bibinfo{volume}{A30},
  \bibinfo{pages}{3--12}.
\bibitem[{{\"O}pik(1951)}]{Opik51}
\bibinfo{author}{{\"O}pik, E.J.}, \bibinfo{year}{1951}.
\newblock \bibinfo{title}{Collision probabilities with the planets and the
  distribution of interplanetary matter}.
\newblock \bibinfo{journal}{Proc. R. Irish Acad.} \bibinfo{volume}{A54},
  \bibinfo{pages}{165--199}.
\newblock \URLprefix \url{www.jstor.org/stable/20488532}.
\bibitem[{Padovan et~al.(2017)Padovan, Tosi, Plesa and Ruedas}]{Pado:etal17}
\bibinfo{author}{Padovan, S.}, \bibinfo{author}{Tosi, N.},
  \bibinfo{author}{Plesa, A.C.}, \bibinfo{author}{Ruedas, T.},
  \bibinfo{year}{2017}.
\newblock \bibinfo{title}{{Impact-induced changes in source depth and volume of
  magmatism on Mercury and their observational signatures}}.
\newblock \bibinfo{journal}{Nat. Comm.} \bibinfo{volume}{8}.
\newblock \DOIprefix\doi{10.1038/s41467-017-01692-0}.
\bibitem[{de~Pater and Lissauer(2010)}]{dePaLi10}
\bibinfo{author}{de~Pater, I.}, \bibinfo{author}{Lissauer, J.J.},
  \bibinfo{year}{2010}.
\newblock \bibinfo{title}{Planetary Sciences}.
\newblock \bibinfo{edition}{2nd} ed., \bibinfo{publisher}{Cambridge University
  Press}.
\bibitem[{Perry et~al.(2015)Perry, Neumann, Phillips, Barnouin, Ernst, Kahan,
  Solomon, Zuber, Smith, Hauck, Peale, Margot, Mazarico, Johnson, Gaskell,
  Roberts, McNutt and Oberst}]{Perr:etal15}
\bibinfo{author}{Perry, M.E.}, \bibinfo{author}{Neumann, G.A.},
  \bibinfo{author}{Phillips, R.J.}, \bibinfo{author}{Barnouin, O.S.},
  \bibinfo{author}{Ernst, C.M.}, \bibinfo{author}{Kahan, D.S.},
  \bibinfo{author}{Solomon, S.C.}, \bibinfo{author}{Zuber, M.T.},
  \bibinfo{author}{Smith, D.E.}, \bibinfo{author}{Hauck, II, S.A.},
  \bibinfo{author}{Peale, S.J.}, \bibinfo{author}{Margot, J.L.},
  \bibinfo{author}{Mazarico, E.}, \bibinfo{author}{Johnson, C.L.},
  \bibinfo{author}{Gaskell, R.W.}, \bibinfo{author}{Roberts, J.H.},
  \bibinfo{author}{McNutt, Jr., R.L.}, \bibinfo{author}{Oberst, J.},
  \bibinfo{year}{2015}.
\newblock \bibinfo{title}{{The low-degree shape of Mercury}}.
\newblock \bibinfo{journal}{Geophys. Res. Lett.} \bibinfo{volume}{42},
  \bibinfo{pages}{6951--6958}.
\newblock \DOIprefix\doi{10.1002/2015GL065101}.
\bibitem[{Pierazzo et~al.(1997)Pierazzo, Vickery and Melosh}]{Pier:etal97}
\bibinfo{author}{Pierazzo, E.}, \bibinfo{author}{Vickery, A.M.},
  \bibinfo{author}{Melosh, H.J.}, \bibinfo{year}{1997}.
\newblock \bibinfo{title}{A reevaluation of impact melt production}.
\newblock \bibinfo{journal}{Icarus} \bibinfo{volume}{127},
  \bibinfo{pages}{408--423}.
\newblock \DOIprefix\doi{10.1006/icar.1997.5713}.
\bibitem[{Pike(1989)}]{Pike89}
\bibinfo{author}{Pike, R.J.}, \bibinfo{year}{1989}.
\newblock \bibinfo{title}{Geomorphology of impact craters on mercury}, in:
  \bibinfo{editor}{Vilas, F.}, \bibinfo{editor}{Chapman, C.R.},
  \bibinfo{editor}{Matthews, M.S.} (Eds.), \bibinfo{booktitle}{Mercury}.
  \bibinfo{publisher}{University of Arizona Press}, \bibinfo{address}{Tucson},
  pp. \bibinfo{pages}{165--273}.
\bibitem[{Pitjeva and Standish(2009)}]{PiSt09}
\bibinfo{author}{Pitjeva, E.V.}, \bibinfo{author}{Standish, E.M.},
  \bibinfo{year}{2009}.
\newblock \bibinfo{title}{{Proposals for the masses of the three largest
  asteroids, the Moon-Earth mass ratio and the Astronomical Unit}}.
\newblock \bibinfo{journal}{Celest. Mech. Dyn. Astr.} \bibinfo{volume}{103},
  \bibinfo{pages}{365}.
\newblock \DOIprefix\doi{10.1007/s10569-009-9203-8}.
\bibitem[{Poirier(2000)}]{Poirier00}
\bibinfo{author}{Poirier, J.P.}, \bibinfo{year}{2000}.
\newblock \bibinfo{title}{Introduction to the Physics of the Earth's Interior}.
\newblock \bibinfo{edition}{2} ed., \bibinfo{publisher}{Cambridge University
  Press}.
\bibitem[{Potter et~al.(2015)Potter, Kring and Collins}]{Pott:etal15}
\bibinfo{author}{Potter, R.W.K.}, \bibinfo{author}{Kring, D.A.},
  \bibinfo{author}{Collins, G.S.}, \bibinfo{year}{2015}.
\newblock \bibinfo{title}{Scaling of basin-sized impacts and the influence of
  target temperature}, in: \bibinfo{editor}{Osinski, G.R.},
  \bibinfo{editor}{Kring, D.A.} (Eds.), \bibinfo{booktitle}{Large Meteorite
  Impacts and Planetary Evolution V}. \bibinfo{publisher}{Geological Society of
  America}. number \bibinfo{number}{518} in \bibinfo{series}{Special Papers},
  pp. \bibinfo{pages}{99--113}.
\newblock \DOIprefix\doi{10.1130/2015.2518(06)}.
\bibitem[{Presnall et~al.(1979)Presnall, Dixon, O'Donnell and
  Dixon}]{Pres:etal79}
\bibinfo{author}{Presnall, D.C.}, \bibinfo{author}{Dixon, J.R.},
  \bibinfo{author}{O'Donnell, T.H.}, \bibinfo{author}{Dixon, S.A.},
  \bibinfo{year}{1979}.
\newblock \bibinfo{title}{Generation of mid-ocean ridge tholeiites}.
\newblock \bibinfo{journal}{J. Petrol.} \bibinfo{volume}{20},
  \bibinfo{pages}{3--35}.
\newblock \DOIprefix\doi{10.1093/petrology/20.1.3}.
\bibitem[{Pugacheva et~al.(2016)Pugacheva, Feoktistova and
  Shevchenko}]{Puga:etal16}
\bibinfo{author}{Pugacheva, S.G.}, \bibinfo{author}{Feoktistova, E.A.},
  \bibinfo{author}{Shevchenko, V.V.}, \bibinfo{year}{2016}.
\newblock \bibinfo{title}{{On the nature of the impactor that formed the
  Shackleton crater on the Moon}}.
\newblock \bibinfo{journal}{Earth Moon Planets} \bibinfo{volume}{118},
  \bibinfo{pages}{27--50}.
\newblock \DOIprefix\doi{10.1007/s11038-016-9489-y}.
\bibitem[{Reese et~al.(2002)Reese, Solomatov and Baumgardner}]{Rees:etal02}
\bibinfo{author}{Reese, C.C.}, \bibinfo{author}{Solomatov, V.S.},
  \bibinfo{author}{Baumgardner, J.R.}, \bibinfo{year}{2002}.
\newblock \bibinfo{title}{Survival of impact-induced thermal anomalies in the
  martian mantle}.
\newblock \bibinfo{journal}{J. Geophys. Res.} \bibinfo{volume}{107}.
\newblock \DOIprefix\doi{10.1029/2000JE001474}.
\bibitem[{Reese et~al.(2004)Reese, Solomatov, Baumgardner, Stegman and
  Vezolainen}]{Rees:etal04}
\bibinfo{author}{Reese, C.C.}, \bibinfo{author}{Solomatov, V.S.},
  \bibinfo{author}{Baumgardner, J.R.}, \bibinfo{author}{Stegman, D.R.},
  \bibinfo{author}{Vezolainen, A.V.}, \bibinfo{year}{2004}.
\newblock \bibinfo{title}{{Magmatic evolution of impact-induced Martian mantle
  plumes and the origin of Tharsis}}.
\newblock \bibinfo{journal}{J. Geophys. Res.} \bibinfo{volume}{109}.
\newblock \DOIprefix\doi{10.1029/2003JE002222}.
\bibitem[{Richardson(2009)}]{JERichardson09}
\bibinfo{author}{Richardson, J.E.}, \bibinfo{year}{2009}.
\newblock \bibinfo{title}{{Cratering saturation and equilibrium: A new model
  looks at an old problem}}.
\newblock \bibinfo{journal}{Icarus} \bibinfo{volume}{204},
  \bibinfo{pages}{697--715}.
\newblock \DOIprefix\doi{10.1016/j.icarus.2009.07.029}.
\bibitem[{Richardson et~al.(2007)Richardson, Melosh, Lisse and
  Carcich}]{JERich:etal07}
\bibinfo{author}{Richardson, J.E.}, \bibinfo{author}{Melosh, H.J.},
  \bibinfo{author}{Lisse, C.M.}, \bibinfo{author}{Carcich, B.},
  \bibinfo{year}{2007}.
\newblock \bibinfo{title}{{A ballistics analysis of the Deep Impact ejecta
  plume: Determining Comet Tempel 1's gravity, mass, and density}}.
\newblock \bibinfo{journal}{Icarus} \bibinfo{volume}{190},
  \bibinfo{pages}{357--390}.
\newblock \DOIprefix\doi{10.1016/j.icarus.2007.08.001}.
\bibitem[{Rickman et~al.(2017)Rickman, Wi{\'s}niowski, Gabryszewski, Wajer,
  W{\'o}jcikowski, Szutowicz, Valsecchi and Morbidelli}]{Rick:etal17}
\bibinfo{author}{Rickman, H.}, \bibinfo{author}{Wi{\'s}niowski, T.},
  \bibinfo{author}{Gabryszewski, R.}, \bibinfo{author}{Wajer, P.},
  \bibinfo{author}{W{\'o}jcikowski, K.}, \bibinfo{author}{Szutowicz, S.},
  \bibinfo{author}{Valsecchi, G.B.}, \bibinfo{author}{Morbidelli, A.},
  \bibinfo{year}{2017}.
\newblock \bibinfo{title}{{Cometary impact rates on the Moon and planets during
  the late heavy bombardment}}.
\newblock \bibinfo{journal}{Astron. Astrophys.} \bibinfo{volume}{598}.
\newblock \DOIprefix\doi{10.1051/0004-6361/201629376}.
\bibitem[{Ries et~al.(1992)Ries, Eanes, Shum and Watkins}]{Ries:etal92}
\bibinfo{author}{Ries, J.C.}, \bibinfo{author}{Eanes, R.J.},
  \bibinfo{author}{Shum, C.K.}, \bibinfo{author}{Watkins, M.M.},
  \bibinfo{year}{1992}.
\newblock \bibinfo{title}{{Progress in the determination of the gravitational
  coefficient of the Earth}}.
\newblock \bibinfo{journal}{Geophys. Res. Lett.} \bibinfo{volume}{19},
  \bibinfo{pages}{529--531}.
\newblock \DOIprefix\doi{10.1029/92GL00259}.
\bibitem[{Rivoldini et~al.(2011)Rivoldini, Van~Hoolst, Verhoeven, Mocquet and
  Dehant}]{Rivo:etal11}
\bibinfo{author}{Rivoldini, A.}, \bibinfo{author}{Van~Hoolst, T.},
  \bibinfo{author}{Verhoeven, O.}, \bibinfo{author}{Mocquet, A.},
  \bibinfo{author}{Dehant, V.}, \bibinfo{year}{2011}.
\newblock \bibinfo{title}{{Geodesy constraints on the interior structure and
  composition of Mars}}.
\newblock \bibinfo{journal}{Icarus} \bibinfo{volume}{213},
  \bibinfo{pages}{451--472}.
\newblock \DOIprefix\doi{10.1016/j.icarus.2011.03.024}.
\bibitem[{Robbins and Hynek(2012)}]{RoHy12b}
\bibinfo{author}{Robbins, S.J.}, \bibinfo{author}{Hynek, B.M.},
  \bibinfo{year}{2012}.
\newblock \bibinfo{title}{{A new global database of Mars impact craters $\geq$1
  km: 2. Global crater properties and regional variations of the
  simple-to-complex transition diameter}}.
\newblock \bibinfo{journal}{J. Geophys. Res.} \bibinfo{volume}{117}.
\newblock \DOIprefix\doi{10.1029/2011JE003967}.
\bibitem[{Roberts and Barnouin(2012)}]{JHRoBa12}
\bibinfo{author}{Roberts, J.H.}, \bibinfo{author}{Barnouin, O.S.},
  \bibinfo{year}{2012}.
\newblock \bibinfo{title}{{The effect of the Caloris impact on the mantle
  dynamics and volcanism of Mercury}}.
\newblock \bibinfo{journal}{J. Geophys. Res.} \bibinfo{volume}{117}.
\newblock \DOIprefix\doi{10.1029/2011JE003876}.
\bibitem[{Roberts et~al.(2009)Roberts, Lillis and Manga}]{JHRobe:etal09}
\bibinfo{author}{Roberts, J.H.}, \bibinfo{author}{Lillis, R.J.},
  \bibinfo{author}{Manga, M.}, \bibinfo{year}{2009}.
\newblock \bibinfo{title}{{Giant impacts on early Mars and the cessation of the
  Martian dynamo}}.
\newblock \bibinfo{journal}{J. Geophys. Res.} \bibinfo{volume}{114}.
\newblock \DOIprefix\doi{10.1029/2008JE003287}.
\bibitem[{Robinson and Taylor(2001)}]{MSRoTa01}
\bibinfo{author}{Robinson, M.S.}, \bibinfo{author}{Taylor, G.J.},
  \bibinfo{year}{2001}.
\newblock \bibinfo{title}{{Ferrous oxide in Mercury's crust and mantle}}.
\newblock \bibinfo{journal}{Meteorit. Planet. Sci.} \bibinfo{volume}{36},
  \bibinfo{pages}{841--847}.
\newblock \DOIprefix\doi{10.1111/j.1945-5100.2001.tb01921.x}.
\bibitem[{Ruedas(2017)}]{Ruedas17a}
\bibinfo{author}{Ruedas, T.}, \bibinfo{year}{2017}.
\newblock \bibinfo{title}{Globally smooth approximations for shock pressure
  decay in impacts}.
\newblock \bibinfo{journal}{Icarus} \bibinfo{volume}{289},
  \bibinfo{pages}{22--33}.
\newblock \DOIprefix\doi{10.1016/j.icarus.2017.02.008}.
\bibitem[{Ruedas and Breuer(2017)}]{RuBr17c}
\bibinfo{author}{Ruedas, T.}, \bibinfo{author}{Breuer, D.},
  \bibinfo{year}{2017}.
\newblock \bibinfo{title}{{On the relative importance of thermal and chemical
  buoyancy in regular and impact-induced melting in a Mars-like planet}}.
\newblock \bibinfo{journal}{J. Geophys. Res.} \bibinfo{volume}{122},
  \bibinfo{pages}{1554--1579}.
\newblock \DOIprefix\doi{10.1002/2016JE005221}.
\bibitem[{Schultz(2017)}]{PHSchultz17}
\bibinfo{author}{Schultz, P.H.}, \bibinfo{year}{2017}.
\newblock \bibinfo{title}{{The size and nature of basin impactors on Mercury
  and the Moon}}, in: \bibinfo{booktitle}{Lunar Planet. Sci.}
\newblock \URLprefix \url{www.hou.usra.edu/meetings/lpsc2017/pdf/2704.pdf}.
\bibitem[{Schultz and Crawford(2016)}]{PHScCr16}
\bibinfo{author}{Schultz, P.H.}, \bibinfo{author}{Crawford, D.A.},
  \bibinfo{year}{2016}.
\newblock \bibinfo{title}{{Origin and implications of non-radial Imbrium
  Sculpture on the Moon}}.
\newblock \bibinfo{journal}{Nature} \bibinfo{volume}{535},
  \bibinfo{pages}{391--394}.
\newblock \DOIprefix\doi{10.1038/nature18278}.
\bibitem[{Shaw(2000)}]{DMShaw00}
\bibinfo{author}{Shaw, D.M.}, \bibinfo{year}{2000}.
\newblock \bibinfo{title}{Continuous (dynamic) melting theory revisited}.
\newblock \bibinfo{journal}{Can. Mineral.} \bibinfo{volume}{38},
  \bibinfo{pages}{1041--1063}.
\bibitem[{Shevchenko et~al.(2007)Shevchenko, Chikmachev and
  Pugacheva}]{Shev:etal07}
\bibinfo{author}{Shevchenko, V.V.}, \bibinfo{author}{Chikmachev, V.I.},
  \bibinfo{author}{Pugacheva, S.G.}, \bibinfo{year}{2007}.
\newblock \bibinfo{title}{{Structure of the South Pole-Aitken lunar basin}}.
\newblock \bibinfo{journal}{Sol. Syst. Res.} \bibinfo{volume}{41},
  \bibinfo{pages}{447--462}.
\newblock \DOIprefix\doi{10.1134/S0038094607060019}.
\bibitem[{Shoemaker and Wolfe(1987)}]{ShWo87}
\bibinfo{author}{Shoemaker, E.M.}, \bibinfo{author}{Wolfe, R.F.},
  \bibinfo{year}{1987}.
\newblock \bibinfo{title}{{Crater production on Venus and Earth by asteroid and
  comet impact}}.
\newblock \bibinfo{journal}{Lunar Planet. Sci.} \bibinfo{volume}{XVIII},
  \bibinfo{pages}{918}.
\newblock \URLprefix
  \url{http://articles.adsabs.harvard.edu//full/1987LPI....18..918S/0000918.000.html}.
\bibitem[{Shuvalov(2009)}]{Shuvalov09}
\bibinfo{author}{Shuvalov, V.}, \bibinfo{year}{2009}.
\newblock \bibinfo{title}{Atmospheric erosion induced by oblique impacts}.
\newblock \bibinfo{journal}{Meteorit. Planet. Sci.} \bibinfo{volume}{44},
  \bibinfo{pages}{1095--1105}.
\newblock \DOIprefix\doi{10.1111/j.1945-5100.2009.tb01209.x}.
\bibitem[{Shuvalov et~al.(2014)Shuvalov, K{\"u}hrt, de~Niem and
  W{\"u}nnemann}]{Shuv:etal14}
\bibinfo{author}{Shuvalov, V.}, \bibinfo{author}{K{\"u}hrt, E.},
  \bibinfo{author}{de~Niem, D.}, \bibinfo{author}{W{\"u}nnemann, K.},
  \bibinfo{year}{2014}.
\newblock \bibinfo{title}{Impact induced erosion of hot and dense atmospheres}.
\newblock \bibinfo{journal}{Planet. Space Sci.} \bibinfo{volume}{98},
  \bibinfo{pages}{120--127}.
\newblock \DOIprefix\doi{10.1016/j.pss.2013.08.018}.
\bibitem[{Smith et~al.(2017)Smith, Zuber, Neumann, Mazarico, Lemoine, Head,
  Lucey, Aharonson, Robinson, Sun, Torrence, Barker, Oberst, Duxbury, Mao,
  Barnouin, Jha, Rowlands, Goossens, Baker, Bauer, Gl{\"a}ser, Lemelin,
  Rosenburg, Sori, Whitten and Mcclanahan}]{DESmit:etal17}
\bibinfo{author}{Smith, D.E.}, \bibinfo{author}{Zuber, M.T.},
  \bibinfo{author}{Neumann, G.A.}, \bibinfo{author}{Mazarico, E.},
  \bibinfo{author}{Lemoine, F.G.}, \bibinfo{author}{Head, III, J.W.},
  \bibinfo{author}{Lucey, P.G.}, \bibinfo{author}{Aharonson, O.},
  \bibinfo{author}{Robinson, M.S.}, \bibinfo{author}{Sun, X.},
  \bibinfo{author}{Torrence, M.H.}, \bibinfo{author}{Barker, M.K.},
  \bibinfo{author}{Oberst, J.}, \bibinfo{author}{Duxbury, T.C.},
  \bibinfo{author}{Mao, D.}, \bibinfo{author}{Barnouin, O.S.},
  \bibinfo{author}{Jha, K.}, \bibinfo{author}{Rowlands, D.D.},
  \bibinfo{author}{Goossens, S.}, \bibinfo{author}{Baker, D.},
  \bibinfo{author}{Bauer, S.}, \bibinfo{author}{Gl{\"a}ser, P.},
  \bibinfo{author}{Lemelin, M.}, \bibinfo{author}{Rosenburg, M.},
  \bibinfo{author}{Sori, M.M.}, \bibinfo{author}{Whitten, J.},
  \bibinfo{author}{Mcclanahan, T.}, \bibinfo{year}{2017}.
\newblock \bibinfo{title}{Summary of the results from the lunar orbiter laser
  altimeter after seven years in lunar orbit}.
\newblock \bibinfo{journal}{Icarus} \bibinfo{volume}{283},
  \bibinfo{pages}{70--91}.
\newblock \DOIprefix\doi{10.1016/j.icarus.2016.06.006}.
\bibitem[{Steel(1998)}]{Steel98}
\bibinfo{author}{Steel, D.}, \bibinfo{year}{1998}.
\newblock \bibinfo{title}{{Distributions and moments of asteroid and comet
  impact speeds upon the Earth and Mars}}.
\newblock \bibinfo{journal}{Planet. Space Sci.} \bibinfo{volume}{46},
  \bibinfo{pages}{473--478}.
\newblock \DOIprefix\doi{10.1016/S0032-0633(97)00232-8}.
\bibitem[{Steinberger et~al.(2010)Steinberger, Werner and
  Torsvik}]{Stei:etal10}
\bibinfo{author}{Steinberger, B.}, \bibinfo{author}{Werner, S.C.},
  \bibinfo{author}{Torsvik, T.H.}, \bibinfo{year}{2010}.
\newblock \bibinfo{title}{{Deep vs. shallow origin of gravity anomalies,
  topography and volcanism on Earth, Venus and Mars}}.
\newblock \bibinfo{journal}{Icarus} \bibinfo{volume}{207},
  \bibinfo{pages}{564--577}.
\bibitem[{Stockstill-Cahill et~al.(2012)Stockstill-Cahill, McCoy, Nittler,
  Weider and Hauck~II}]{St-Ca:etal12}
\bibinfo{author}{Stockstill-Cahill, K.R.}, \bibinfo{author}{McCoy, T.J.},
  \bibinfo{author}{Nittler, L.R.}, \bibinfo{author}{Weider, S.Z.},
  \bibinfo{author}{Hauck~II, S.A.}, \bibinfo{year}{2012}.
\newblock \bibinfo{title}{{Magnesium-rich crustal compositions on Mercury:
  Implications for magmatism from petrologic modeling}}.
\newblock \bibinfo{journal}{J. Geophys. Res.} \bibinfo{volume}{117}.
\newblock \DOIprefix\doi{10.1029/2012JE004140}.
\bibitem[{Strom et~al.(2005)Strom, Malhotra, Ito, Yoshida and
  Kring}]{Stro:etal05}
\bibinfo{author}{Strom, R.G.}, \bibinfo{author}{Malhotra, R.},
  \bibinfo{author}{Ito, T.}, \bibinfo{author}{Yoshida, F.},
  \bibinfo{author}{Kring, D.A.}, \bibinfo{year}{2005}.
\newblock \bibinfo{title}{{The origin of planetary impactors in the inner Solar
  System}}.
\newblock \bibinfo{journal}{Science} \bibinfo{volume}{309},
  \bibinfo{pages}{1847--1850}.
\newblock \DOIprefix\doi{10.1126/science.1113544}.
\bibitem[{Susorney et~al.(2016)Susorney, Barnouin, Ernst and
  Johnson}]{Suso:etal16}
\bibinfo{author}{Susorney, H.C.M.}, \bibinfo{author}{Barnouin, O.S.},
  \bibinfo{author}{Ernst, C.M.}, \bibinfo{author}{Johnson, C.L.},
  \bibinfo{year}{2016}.
\newblock \bibinfo{title}{{Morphometry of impact craters on Mercury from
  MESSENGER altimetry and imaging}}.
\newblock \bibinfo{journal}{Icarus} \bibinfo{volume}{271},
  \bibinfo{pages}{180--193}.
\newblock \DOIprefix\doi{10.1016/j.icarus.2016.01.022}.
\bibitem[{Svetsov and Shuvalov(2015)}]{SvSh15}
\bibinfo{author}{Svetsov, V.V.}, \bibinfo{author}{Shuvalov, V.V.},
  \bibinfo{year}{2015}.
\newblock \bibinfo{title}{{Water delivery to the Moon by asteroidal and
  cometary impacts}}.
\newblock \bibinfo{journal}{Planet. Space Sci.} \bibinfo{volume}{117},
  \bibinfo{pages}{444--452}.
\newblock \DOIprefix\doi{10.1016/j.pss.2015.09.011}.
\bibitem[{Tackley(1996)}]{Tackley96a}
\bibinfo{author}{Tackley, P.J.}, \bibinfo{year}{1996}.
\newblock \bibinfo{title}{Effects of strongly variable viscosity on
  three-dimensional compressible convection in planetary mantles}.
\newblock \bibinfo{journal}{J. Geophys. Res.} \bibinfo{volume}{101},
  \bibinfo{pages}{3311--3332}.
\bibitem[{Tackley(2008)}]{Tackley08}
\bibinfo{author}{Tackley, P.J.}, \bibinfo{year}{2008}.
\newblock \bibinfo{title}{Modelling compressible mantle convection with large
  viscosity contrasts in a three-dimensional spherical shell using the yin-yang
  grid}.
\newblock \bibinfo{journal}{Phys. Earth Planet. Inter.} \bibinfo{volume}{171},
  \bibinfo{pages}{7--18}.
\bibitem[{Taylor and McLennan(2009)}]{SRTaMcLe09}
\bibinfo{author}{Taylor, S.R.}, \bibinfo{author}{McLennan, S.M.},
  \bibinfo{year}{2009}.
\newblock \bibinfo{title}{Planetary Crusts}.
\newblock \bibinfo{publisher}{Cambridge University Press}.
\bibitem[{Wang and Takahashi(2000)}]{WWaTa00}
\bibinfo{author}{Wang, W.}, \bibinfo{author}{Takahashi, E.},
  \bibinfo{year}{2000}.
\newblock \bibinfo{title}{{Subsolidus and melting experiments of K-doped
  peridotite KLB-1 to 27\,GPa: Its geophysical and geochemical implications}}.
\newblock \bibinfo{journal}{J. Geophys. Res.} \bibinfo{volume}{105},
  \bibinfo{pages}{2855--2868}.
\newblock \DOIprefix\doi{10.1029/1999JB900366}.
\bibitem[{W{\"a}nke and Dreibus(1994)}]{WaDr94}
\bibinfo{author}{W{\"a}nke, H.}, \bibinfo{author}{Dreibus, G.},
  \bibinfo{year}{1994}.
\newblock \bibinfo{title}{{Chemistry and accretion history of Mars}}.
\newblock \bibinfo{journal}{Phil. Trans. R. Soc. Lond.} \bibinfo{volume}{A
  349}, \bibinfo{pages}{285--293}.
\bibitem[{Warren and Taylor(2014)}]{PHWaTa14}
\bibinfo{author}{Warren, P.H.}, \bibinfo{author}{Taylor, G.J.},
  \bibinfo{year}{2014}.
\newblock \bibinfo{title}{The moon}, in: \bibinfo{editor}{Davis, A.M.} (Ed.),
  \bibinfo{booktitle}{Planets, Asteroids, Comets and The Solar System}.
  \bibinfo{edition}{2nd} ed.. \bibinfo{publisher}{Elsevier}.
  volume~\bibinfo{volume}{2} of \textit{\bibinfo{series}{Treatise on
  Geochemistry}}. chapter \bibinfo{chapter}{2.9}, pp.
  \bibinfo{pages}{213--250}.
\newblock \DOIprefix\doi{10.1016/B978-0-08-095975-7.00124-8}.
\bibitem[{Weber et~al.(2011)Weber, Lin, Garnero, Williams and
  Lognonn{\'e}}]{RCWebe:etal11}
\bibinfo{author}{Weber, R.C.}, \bibinfo{author}{Lin, P.Y.},
  \bibinfo{author}{Garnero, E.J.}, \bibinfo{author}{Williams, Q.},
  \bibinfo{author}{Lognonn{\'e}, P.}, \bibinfo{year}{2011}.
\newblock \bibinfo{title}{Seismic detection of the lunar core}.
\newblock \bibinfo{journal}{Science} \bibinfo{volume}{331},
  \bibinfo{pages}{309--312}.
\newblock \DOIprefix\doi{10.1126/science.1199375}.
\bibitem[{Weider et~al.(2015)Weider, Nittler, Starr, Crapster-Pregont,
  Peplowski, Denevi, Head, Byrne, Hauck, Ebel and Solomon}]{Weid:etal15}
\bibinfo{author}{Weider, S.Z.}, \bibinfo{author}{Nittler, L.R.},
  \bibinfo{author}{Starr, R.D.}, \bibinfo{author}{Crapster-Pregont, E.J.},
  \bibinfo{author}{Peplowski, P.N.}, \bibinfo{author}{Denevi, B.W.},
  \bibinfo{author}{Head, J.W.}, \bibinfo{author}{Byrne, P.K.},
  \bibinfo{author}{Hauck, II, S.A.}, \bibinfo{author}{Ebel, D.S.},
  \bibinfo{author}{Solomon, S.C.}, \bibinfo{year}{2015}.
\newblock \bibinfo{title}{{Evidence for geochemical terranes on Mercury: Global
  mapping of major elements with MESSENGER's X-Ray Spectrometer}}.
\newblock \bibinfo{journal}{Earth Planet. Sci. Lett.} \bibinfo{volume}{416},
  \bibinfo{pages}{109--120}.
\newblock \DOIprefix\doi{10.1016/j.epsl.2015.01.023}.
\bibitem[{Werner and Ivanov(2015)}]{WeIv15}
\bibinfo{author}{Werner, S.C.}, \bibinfo{author}{Ivanov, B.A.},
  \bibinfo{year}{2015}.
\newblock \bibinfo{title}{Exogenic dynamics, cratering, and surface ages}, in:
  \bibinfo{editor}{Spohn, T.} (Ed.), \bibinfo{booktitle}{Physics of Terrestrial
  Planets and Moons}. \bibinfo{edition}{2nd} ed..
  \bibinfo{publisher}{Elsevier}. volume~\bibinfo{volume}{10} of
  \textit{\bibinfo{series}{Treatise on Geophysics}}. chapter
  \bibinfo{chapter}{10.10}, pp. \bibinfo{pages}{327--365}.
\newblock \DOIprefix\doi{10.1016/B978-0-444-53802-4.00170-6}.
\bibitem[{Wetherill(1967)}]{Wetherill67}
\bibinfo{author}{Wetherill, G.W.}, \bibinfo{year}{1967}.
\newblock \bibinfo{title}{Collisions in the asteroid belt}.
\newblock \bibinfo{journal}{J. Geophys. Res.} \bibinfo{volume}{72},
  \bibinfo{pages}{2429--2444}.
\newblock \DOIprefix\doi{10.1029/JZ072i009p02429}.
\bibitem[{Whipple(1963)}]{Whipple63}
\bibinfo{author}{Whipple, F.L.}, \bibinfo{year}{1963}.
\newblock \bibinfo{title}{On meteoroids and penetration}.
\newblock \bibinfo{journal}{J. Geophys. Res.} \bibinfo{volume}{68},
  \bibinfo{pages}{4929--4939}.
\newblock \DOIprefix\doi{10.1029/JZ068i017p04929}.
\bibitem[{Williams and Nimmo(2004)}]{JPWiNi04}
\bibinfo{author}{Williams, J.P.}, \bibinfo{author}{Nimmo, F.},
  \bibinfo{year}{2004}.
\newblock \bibinfo{title}{{Thermal evolution of the Martian core: Implications
  for an early dynamo}}.
\newblock \bibinfo{journal}{Geology} \bibinfo{volume}{32},
  \bibinfo{pages}{97--100}.
\bibitem[{Yue et~al.(2013)Yue, Johnson, Minton, Melosh, Di, Hu and
  Liu}]{ZYue:etal13}
\bibinfo{author}{Yue, Z.}, \bibinfo{author}{Johnson, B.C.},
  \bibinfo{author}{Minton, D.A.}, \bibinfo{author}{Melosh, H.J.},
  \bibinfo{author}{Di, K.}, \bibinfo{author}{Hu, W.}, \bibinfo{author}{Liu,
  Y.}, \bibinfo{year}{2013}.
\newblock \bibinfo{title}{Projectile remnants in central peaks of lunar impact
  craters}.
\newblock \bibinfo{journal}{Nature Geosci.} \bibinfo{volume}{6},
  \bibinfo{pages}{435--437}.
\newblock \DOIprefix\doi{10.1038/ngeo1828}.
\bibitem[{Zhao et~al.(2009)Zhao, Zimmerman and Kohlstedt}]{YHZhao:etal09}
\bibinfo{author}{Zhao, Y.}, \bibinfo{author}{Zimmerman, M.E.},
  \bibinfo{author}{Kohlstedt, D.L.}, \bibinfo{year}{2009}.
\newblock \bibinfo{title}{{Effect of iron content on the creep behavior of
  olivine: 1. Anhydrous conditions}}.
\newblock \bibinfo{journal}{Earth Planet. Sci. Lett.} \bibinfo{volume}{287},
  \bibinfo{pages}{229--240}.
\newblock \DOIprefix\doi{10.1016/j.epsl.2009.08.006}.

\end{thebibliography}
\end{document}